\RequirePackage{rotating}
\documentclass[fleqn,usenatbib]{mnras}
\usepackage{newtxtext,newtxmath}

\usepackage[T1]{fontenc}

\DeclareRobustCommand{\VAN}[3]{#2}
\let\VANthebibliography\thebibliography
\def\thebibliography{\DeclareRobustCommand{\VAN}[3]{##3}\VANthebibliography}

\usepackage{graphicx}	
\usepackage{amsmath}	
\usepackage{xcolor,float}
\usepackage{orcidlink}
\usepackage{placeins}
\usepackage{caption}
\usepackage{siunitx}
\newcommand{\pccm}{pc\,cm$^{-3}$}
\newcommand{\Msun}{M$_\odot$}
\newcommand{\Msunyr}{M$_\odot$\,yr$^{-1}$}
\newcommand{\met}{$\log(Z/Z_\odot)$}

\newcommand{\Halpha}{H$\alpha$}
\newcommand{\Hbeta}{H$\beta$}
\newcommand{\Nii}{[N~{\sc ii}]}
\newcommand{\Oii}{[O~{\sc ii}]}
\newcommand{\Oiii}{[O~{\sc iii}]}
\newcommand{\Sii}{[S~{\sc ii}]}

\newcommand{\Nant}{$N_{\text{ant}}$}
\newcommand{\coordserr}[8]{\mbox{#1$^{\mathrm{h}}$#2$^{\mathrm{m}}$#3$^{\mathrm{s}}\pm$#4\arcsec} \linebreak[1] \mbox{#5\degr#6\arcmin#7\arcsec$\pm$#8\arcsec}}

\newcommand{\logMstar}{$\log(\text{M}_*/\text{M}_\odot$)}
\newcommand{\zphot}{$z_{\text{phot}}$}

\title[MeerTRAP FRB localisation]{Localisation and host galaxy identification of new Fast Radio Bursts with MeerKAT}

\author[I. Pastor-Marazuela et al.]{In\'{e}s Pastor-Marazuela$^{\orcidlink{0000-0002-4357-8027} 1,2}$\thanks{E-mail: ines.pastor-marazuela@manchester.ac.uk},
Alexa C. Gordon$^{\orcidlink{0000-0002-5025-4645} 3}$,
Ben Stappers$^{\orcidlink{0000-0001-9242-7041} 1}$,
Ilya S. Khrykin$^{\orcidlink{0000-0003-0574-7421} 4}$,
Nicolas Tejos$^{\orcidlink{0000-0002-1883-4252} 4}$,
\newauthor 
Kaustubh Rajwade$^{\orcidlink{0000-0002-8043-6909} 5}$,
Manisha Caleb$^{\orcidlink{0000-0002-4079-4648} 6,7}$,
Mayuresh~P.~Surnis$^{\orcidlink{0000-0002-9507-6985} 8}$,
Laura N. Driessen$^{\orcidlink{0000-0002-4405-3273} 6}$,
Sunil Simha$^{\orcidlink{0000-0003-3801-1496} 3,9,10}$,
\newauthor
Jun Tian$^{\orcidlink{0000-0003-1914-877X} 1}$,
J.~Xavier Prochaska$^{\orcidlink{0000-0002-7738-6875} 10,11,12}$,
Ewan Barr$^{13}$,
Sarah Buchner$^{\orcidlink{0000-0002-1691-0215} 14}$,
Wen-Fai Fong$^{\orcidlink{0000-0002-7374-935X} 3}$,
\newauthor
Fabian Jankowski$^{\orcidlink{0000-0002-6658-2811} 15}$,
Lordrick Kahinga$^{\orcidlink{0009-0007-5296-4046} 10,16}$,
Charles~D. Kilpatrick$^{\orcidlink{0000-0002-5740-7747} 3}$,
Michael Kramer$^{13,1}$,
\newauthor
Lluis Mas-Ribas$^{\orcidlink{0000-0003-4584-8841} 10}$,
Joseph Hennawi$^{\orcidlink{0000-0002-7054-4332} 17, 18}$.
\\
$^{1}$Jodrell Bank Centre for Astrophysics, University of Manchester, Oxford Road, Manchester M13 9PL, UK\\
$^{2}$ ASTRON, the Netherlands Institute for Radio Astronomy, Oude Hoogeveensedijk 4,7991 PD Dwingeloo, The Netherlands\\
$^{3}$ Center for Interdisciplinary Exploration and Research in Astrophysics (CIERA) and Department of Physics and Astronomy, Northwestern University,\\Evanston, IL 60208, USA\\
$^{4}$ Instituto de F\'isica, Pontificia Universidad Cat\'olica de Valpara\'iso, Casilla 4059, Valpara\'iso, Chile\\
$^{5}$ Astrophysics, Denys Wilkinson Building, University of Oxford, Keble Road, Oxford OX1 3RH, UK\\
$^{6}$ Sydney Institute for Astronomy, School of Physics, The University of Sydney, NSW 2006, Australia\\
$^{7}$ ARC Centre of Excellence for Gravitational Wave Discovery (OzGrav), Hawthorn, VIC 3122, Australia\\
$^{8}$ Department of Physics, IISER Bhopal, Bhauri Bypass Road, Bhopal, 462066, India\\
$^{9}$ Department of Astronomy and Astrophysics, University of Chicago, William Eckhardt Research Center, 5640 S Ellis Ave, Chicago, IL 60637\\
$^{10}$ Department of Astronomy and Astrophysics, University of California Santa Cruz, 1156 High Street, Santa Cruz, CA 95060, USA\\
$^{11}$ Institute for the Physics and Mathematics of the Universe (Kavli IPMU), 5-1-5 Kashiwanoha, Kashiwa, 277-8583, Japan \\
$^{12}$ Division of Science, National Astronomical Observatory of Japan,2-21-1 Osawa, Mitaka, Tokyo 181-8588, Japan \\
$^{13}$ Max-Planck-Institut für Radioastronomie, Bonn, Germany\\
$^{14}$ SARAO, Liesbeek House, River Park, Gloucester Road, Mowbray, Cape Town, 7700, South Africa\\
$^{15}$ LPC2E, OSUC, Univ Orleans, CNRS, CNES, Observatoire de Paris, F-45071 Orleans, France\\
$^{16}$ University of Dodoma, College of Natural and Mathematical Sciences, Department of Physics, 1 Benjamin Mkapa Road, 41218 Iyumbu, Dodoma, Tanzania\\
$^{17}$ University of California, Santa Barbara, Department of Physics, University of California, Santa Barbara, CA 93106, USA\\
$^{18}$ Leiden Observatory, Leiden University, P.O. Box 9513, 2300 RA Leiden, The Netherlands
}

\date{Accepted XXX. Received YYY; in original form ZZZ}

\pubyear{2025}

\begin{document}
\label{firstpage}
\pagerange{\pageref{firstpage}--\pageref{lastpage}}
\maketitle

\begin{abstract}
Accurately localising fast radio bursts (FRBs) is essential for understanding their birth environments and for their use as cosmological probes. Recent advances in radio interferometry, particularly with MeerKAT, have enabled the localisation of individual bursts with arcsecond precision. 
In this work, we present the localisation of 15 apparently non-repeating FRBs detected with MeerKAT.
Two of the FRBs, discovered in 2022, were localised in 8 second images from the projects which MeerTRAP was commensal to, while eight were localised using the transient buffer (TB) pipeline, and another one through SeeKAT, all with arcsecond precision. Four additional FRBs lacked TB triggers and sufficient signal, limiting their localisation only to arcminute precision. 
For eight of the FRBs in our sample, we identify host galaxies with greater than 90\% confidence, and one with 80\% confidence, while two FRBs have ambiguous associations. 
We measured spectroscopic redshifts for six host galaxies, ranging from 0.33 to 0.85, demonstrating MeerKAT's sensitivity to high redshift FRBs.
We modelled the spectral energy distributions of host galaxies with sufficient photometric coverage to derive their stellar population and star formation properties.
This work represents one of the largest uniform samples of well-localised distant FRBs to date, laying the groundwork for using MeerKAT FRBs as cosmological probes and understand how FRB hosts evolve at high redshift.
\end{abstract}




\section{Introduction}

Fast Radio Bursts (FRBs) are radio transient signals lasting from micro to milliseconds, with luminosities high enough to be seen from galaxies beyond the Milky Way (MW) \citep{petroff_fast_2019, petroff_fast_2022, cordes_fast_2019}. 
The first FRBs were discovered using the 64~m single-dish Parkes radio telescope, and their extragalactic origin was inferred from their unusually large dispersion measures (DMs), which far exceeded the predictions from MW models, as well as from their isotropic distribution across the sky \citep{lorimer_bright_2007, thornton_population_2013}.
Early FRB discoveries were made with single-dish instruments, including both non-repeaters and the first repeater, FRB~20121102A, discovered with the Arecibo radio telescope \citep{spitler_fast_2014, spitler_repeating_2016}
However, interferometric observations with radio telescope arrays became crucial for precisely localising FRBs to their host galaxies and thereby confirming their extragalactic origin. 
The repetitions from FRB~20121102A enabled follow-up observations with the Karl G. Jansky Very Large Array (VLA), leading to its localisation to a dwarf star-forming galaxy at $z\sim0.2$ \citep{chatterjee_direct_2017, tendulkar_host_2017}, and definitively establishing the cosmological origin of at least some FRBs.

In the past decade, the advent of new radio telescope arrays with dedicated FRB surveys has rapidly pushed the field forward. 
This results from a combination of increased collecting area which improves instantaneous sensitivity, multi-beam receivers that increase the field of view (FoV), and longer baselines which facilitate precise localisation. Consequently, 
thousands of FRBs have been detected to date \citep{chimefrb_collaboration_first_2021}, more than fifty have been found to repeat \citep{chimefrb_collaboration_chimefrb_2023}, and nearly a hundred FRBs have been localised to their host galaxies \citep{heintz_demographics_2020, bhandari_characterizing_2022, gordon_demographics_2023, shannon_commensal_2024, sharma_preferential_2024, bhardwaj_selection_2024, driessen_frb_2024, caleb_subarcsec_2023, rajwade_study_2024}.

Recent discoveries have further revealed crucial information about the FRB nature, which strengthen the link between FRBs and young magnetised neutron stars. \citep[e.g.][]{bochenek_fast_2020, chimefrb_collaboration_bright_2020}.
Other properties of both one-off and repeating FRBs, such as polarisation \citep{mckinven_pulsar-like_2024} and intrinsic structure \citep{pastor-marazuela_comprehensive_2025}, further resemble those observed in the Galactic pulsar and magnetar population.

Studying the host galaxy properties from which FRBs are emitted can further expose the environments where they reside, their age, and the nature of their progenitors. 
Repeating sources facilitate the follow-up with very long baseline interferometry (VLBI) to obtain milliarcsecond precision localisations. This has shown, for instance, that while some repeaters are located right outside star-forming regions \citep{marcote_repeating_2020}, others are located in globular clusters \citep{kirsten_repeating_2022}, or are associated with very old stellar populations \citep{shah_repeater_2025,eftekhari_repeater_2025}, and thus are likely to have very different ages or formation channels.
However, only $\sim2.6$\% of the currently detected FRBs seem to be repeaters \citep{chimefrb_collaboration_chimefrb_2023}, thus emphasising the importance of instantaneous localisation of any detected FRBs. For this purpose, many FRB surveys have the capacity to store voltage data when a new FRB is detected, which is then imaged in order to determine the FRB position. With this method, ASKAP \citep{scott_celebi_2023}, DSA-110 \citep{ravi_deep_2022}, and MeerTRAP \citep{rajwade_study_2024} can currently achieve sub-arcsecond localisation, and while CHIME/FRB can only attain sub-arcminute localisation on its own \citep{michilli_sub-arcminute_2022}, the CHIME/FRB Outriggers has recently started to localise FRBs with milliarcsecond precision \citep{lanman_chimefrb_2024,chime_kko_2025, chimefrb_collaboration_frb_2025}.

The identification of FRB host galaxies via optical follow-up observations has enabled the study of host galaxy demographics. These studies have revealed that
FRB host galaxies are generally star-forming, with a wide range of stellar masses and star formation rates \citep{heintz_demographics_2020,bhandari_characterizing_2022,gordon_demographics_2023}.
While there is as yet no significant difference between the hosts of repeating and one-off FRBs, repeaters may occur in galaxies with lower stellar masses \citep[e.g.][]{bhandari_characterizing_2022, gordon_demographics_2023}. Additionally, some studies have suggested that FRBs primarily trace star formation in galaxies \citep{loudas_demographics_2025} with a possible preference toward metal-rich, massive star-forming galaxies \citep{sharma_preferential_2024}, suggesting a bias toward environments favourable to magnetar formation through core-collapse supernovae (CCSN) or stellar mergers. However, other studies have posited that FRBs do not trace star formation alone, and stellar mass must also play a role, indicative of more delayed progenitor formation channels relative to star formation \citep{horowicz_demographics_2025}. Current interpretations are limited by small-number statistics, particularly at higher redshifts where only a handful of FRBs have been accurately localised. Increasing the sample of FRB hosts is hence essential for constraining their progenitor channels.

In addition to providing key insights into the nature of FRBs, accurate localisation and redshift determination facilitates their use as cosmological probes. 
One of the first such applications was using FRBs to measure the baryonic content of the intergalactic medium (IGM) through the Macquart relation \citep{macquart_census_2020}. Beyond this, other proposed applications include new measurements of the Hubble constant and other cosmological parameters \citep{hagstotz_new_2022, james_measurement_2022}, the tomographic reconstruction of the cosmic web \citep{simha_disentangling_2020, lee_constraining_2022, khrykin_flimflam_2024}, and probing the epoch of reionisation \citep[EoR, see][and references therein]{bhandari_probing_2021}. Gravitationally lensed FRBs could also provide powerful cosmological probes \citep{pastor-marazuela_fast_2025}.
These applications, however, rely on the detection and localisation of FRBs at relatively high redshift, a population that remains limited.

In this paper, we present the localisation of 15 new FRBs discovered with the MeerKAT radio telescope. Of those, 11 have been localised with arcsecond precision, thus allowing for host galaxy identification. Given their moderate to high dispersion measures (DMs), these FRBs hold significant potential for the aforementioned cosmological applications, which will be presented in Caleb, Rajwade et al. (\textit{in prep.}). The burst properties will be presented in Pastor-Marazuela et al. (\textit{in prep.}).
The paper is divided as follows:
in Section~\ref{sec:observations}, we present the MeerKAT observations in which the bursts were detected. In Section~\ref{sec:data_analysis}, we describe our host galaxy data analysis methods and new observations. In Section~\ref{sec:results}, we detail the localisation of our FRB sample and their host galaxy properties. Finally, we conclude in Section~\ref{sec:results}.

\section{Observations} \label{sec:observations}

\subsection{MeerTRAP real-time pipeline}

The MeerKAT (More (meer) Karoo Array Telescope) radio telescope, located in the Northern Cape Province in South Africa, is an array of 64 13.5\,m diameter antennas with a maximum baseline of 7.7\,km \citep{jonas_meerkat_2018}.
The antennas have three receivers that allow MeerKAT to observe in the UHF (544--1088\,MHz), L-band (856--1712\,MHz), or S-band (five different 875\,MHz bandwidths from 1750 to 3500\,MHz).
The MeerTRAP (More (meer) TRAnsients and Pulsars) project performs a commensal survey for short radio transients, simultaneous to other ongoing MeerKAT observations \citep{sanidas_meertrap_2017, rajwade_meertrap_2021}. 
To search for radio transient signals, the Filterbanking Beamformer User Supplied Equipment \citep[FBFUSE;][]{chen_wide_2021, barr_s-band_2017} generates one incoherent beam (IB) and up to 768 coherent beams (CBs) formed by coherently adding the signals from up to 40 of the inner MeerKAT antennas (\Nant). This CB tiling significantly increases the sensitivity, which scales as \Nant, though at the cost of a smaller FoV. In contrast, the IB, which sums the signal from up to 64 antennas without phasing them, offers a larger FoV but with a sensitivity scaling as $\sqrt{N_{\text{ant}}}$.
In the L-band, the CBs cover $\sim0.4$\,deg$^2$, whereas the IB spans a larger area of $\sim1$\,deg$^2$.
This setup effectively results in two complementary FRB surveys; one using the CBs that is roughly five times more sensitive than the IB but has a smaller FoV, and the other using the IB with a broader FoV but lower sensitivity \citep{jankowski_sample_2023}. 
The number of CBs ($\leq768$) is determined by the computational resources of the FBFUSE and the Transient User Supplied Equipment cluster \citep[TUSE;][]{rajwade_meertrap_2021, rajwade_first_2022}, which searches for dispersed single pulses in the FBFUSE output in near-real time.
To mitigate false detections due to radio frequency interference (RFI) occurring during the observations, we applied the outlier detection algorithm named inter-quartile range mitigation \citep[IQRM;][]{morello_iqrm_2021}, developed for and tested on data from the Lovell radio telescope (UK) and MeerKAT.
After RFI mitigation, dispersed single pulse searches are carried out with \texttt{AstroAccelerate} \citep{armour_astroaccelerate_2020, adamek_single-pulse_2020}, a GPU software that incoherently dedisperses the signal up to a dispersion measure (DM) of $\sim4000$\,\pccm\ in the L-band, with a sampling interval of 306\,$\mu$s and searching up to a maximum boxcar width of 0.67\,s. 
The burst search parameters, including DM ranges and boxcar widths, are optimised in a way that allows for the data to be searched in real-time with the available computing resources, and to have enough time to trigger the transient buffer (TB) data while the burst is held in memory. 
Although some highly dispersed FRBs could be missed due to this DM cut-off, the most dispersed MeerTRAP FRB found so far is $>1000$\,\pccm\ away from this limit, making the bias against such bursts small.
We have included Table~\ref{tab:dm_lims} in Appendix~\ref{app:dm_lims} detailing the limits that were used for each band at different times.

An overview of the MeerTRAP FRB population, including detection rates in the CBs and IB, was presented in \citet{jankowski_sample_2023}. While we have briefly discussed relevant aspects of the pulse search pipeline and configuration, a full analysis of the selection effects, detection biases (e.g. between CB and IB detections), and the impact of RFI on false positives and trigger recovery rates is beyond the scope of this paper and will be addressed in future work.

\subsection{Transient Buffer Data} \label{sec:tb}

If a radio transient candidate is identified by the real-time search pipeline within 45~s of its detection time, it triggers the storage of the TB data via a custom VOEvent \citep{petroff_voevent_2017} emission software \citep{jankowski_real-time_2022}.
The TB data that are stored contain 300\,ms of channelised complex data from all MeerKAT antennas that were available at the time of detection (up to 64), centred at the time of the burst candidate. 
To ensure the full burst is recovered, even in the case of large dispersion delays, the TB data are incoherently dedispersed following the dispersion curve of the burst. This allows us to reconstruct the full burst profile within the available TB duration and prevents temporal truncation \citep{rajwade_study_2024}.
The full observing bandwidth of the TB data is divided into 64 frequency subbands of equal bandwidth, and each subband is written to a different storage node. Some subbands might be missing due to a malfunction of the storage nodes, but this does not usually exceed 10\% of the full bandwidth.
Because the TB data contain the signal from all available antennas that we add coherently, the resulting S/N of the burst is significantly higher than in the original CB or IB detection. A typical loss of 10\% of the bandwidth due to missing subbands does not have a substantial impact on the detection of the bursts in the image domain or on the astrometric accuracy. In practice, additional frequency subbands are flagged due to RFI, which commonly exceeds the fraction due to storage node malfunctions. The loss of subbands due to node failures is hence not the dominant limitation on burst signal recovery or localisation precision.
The TB data are used to localise the FRBs and to obtain high spectro-temporal resolution, polarimetric data. While this paper focuses only on the localisation achieved with the TB data, the full TB pipeline has been detailed in \citet{rajwade_study_2024}, and the pulse properties of the FRBs in this in work will be presented in Pastor-Marazuela et al. (\textit{in prep.}).

\section{Data analysis} \label{sec:data_analysis}

\subsection{Localisation of FRBs with TB data}

The FRBs which triggered TB dumps upon detection were localised following the method described in \cite{rajwade_study_2024}. To summarise, the TB data are incoherently dedispersed to the detection DM and written to disk. Next they are correlated with \texttt{xGPU} \citep{clark_accelerating_2011}, and converted into visibility files with the appropriate metadata in a FITS-Interferometry Data Interchange (IDI) format, and then we use the Common Astronomy Software Applications package \citep[CASA,][]{casa_team_casa_2022} to convert the visibility files into measurement sets (MSs). Since the TB data divides the full observing bandwidth into 64 frequency subbands that follow the dispersion delay, we obtain up to 64 MSs.

Once the MSs are obtained, we create images with \texttt{WSClean} \citep{offringa_wsclean_2014}. Initially, we perform a simple, dirty clean step to quickly inspect the images and identify when and where the FRB appears. We create an image with the full 300\,ms integration time for each of the 64 frequency subbands, visually inspecting them to flag out those affected by RFI. We then produce a frequency averaged image from the remaining subbands. 
Next, we generate frequency averaged images of 11 time bins of each dataset around the expected burst arrival time, with each time bin having an integration time of the order of the burst duration as seen in the detection filterbank file. 
By subtracting the full integration time images, we generate difference images, in which the FRB should appear as a new source in one or more of the central time bins.
After identifying the time bins where the FRB is detected, we produce new images with more advanced cleaning parameters \citep[see][section 2.4.2]{rajwade_study_2024}. These new images include a full integration time image, an ``on'' image corresponding to the time intervals where the FRB is detected, and an ``off'' image from intervals where the source is not detected (usually the first time intervals of the MSs) with the same duration as the ``on'' image. The images are generally produced with 8192 pixels per side, and 1\,arcsec$^2$ pixel size.

The next step involves performing an astrometric correction of the resulting images and source positions, following the procedure outlined in \citet{driessen_21_2022} and \citet{driessen_frb_2024}. First, we run the Python Blob Detector and Source Finder \citep[\texttt{PyBDSF}][]{mohan_pybdsf_2015} algorithm on the full integration, ``on'' and ``off'' images to identify the image source positions and associated errors.
\texttt{PyBDSF} determines the positions and associated errors by fitting 2D Gaussian models to the radio images. The positional errors hence reflect the precision of the fitted centroid, which depends on the S/N of the source and the beam size, rather than the apparent angular extent of the source. As a result, the positional errors can be significantly smaller than the observed sources.

Next, we query various radio source catalogues to find reference sources for the astrometric correction. Although catalogues with high positional accuracy are preferred, such as the Long Baseline Array (LBA) Calibrator Survey \citep[LCS1;][]{petrov_lba_2011}, or the Australian Telescope Compact Array Parkes-MIT-NRAO \citep[ATPMN;][]{mcconnell_atpmn_2012}, these are often sparsely populated and may lack sufficient sources within the MeerKAT FoV for a reliable correction. 
The Very Large Array Sky Survey \citep[VLASS;][]{lacy_karl_2020} Epoch 2 source catalogue goes down to declinations of $-40$\degr, it usually contains tens of sources within the MeerKAT FoV, and the sources have a typical positional accuracy of 0.2\arcsec. We thus use VLASS unresolved sources as our default reference for FRBs at declinations $>-40$\degr, and assume a systematic error of 0.2\arcsec.

For declinations $\leq-40$\degr, the Rapid ASKAP Continuum Survey \citep[RACS;][]{hale_rapid_2021} typically contains tens to hundreds of sources within the FoV. However, due to insufficient VLBI reference positions in the Southern Hemisphere, the RACS-mid \citep[1367.5 MHz;][]{duchesne_rapid_2023} source positions have systematic astrometric offsets of $\sim1-2$\,\arcsec.
To overcome this, we use the Radio Fundamental Catalog (RFC\footnote{RFC: \url{https://astrogeo.org/rfc/}}) as a reference to correct the RACS-mid source positions. The RFC provides milliarcsecond accuracy, and often contains more sources within the FoV than LCS1 or ATPMN, though not enough for standalone use. Therefore, we first use RFC to correct the RACS-mid source positions locally over a large FoV of 6\degr\ in radius, then apply these corrected positions to perform the astrometric correction of the MeerTRAP images. We select RACS-mid sources classified as point-like with positional uncertainties $<0.5\arcsec$ in both right ascension (RA) and declination (Dec), and we perform the astrometric correction with the Python module \texttt{astroalign} \citep{beroiz_astroalign_2020}.

In order to avoid astrometric distortions from sources that are far from the phase centre of the MeerKAT image, we restrict our reference source selection to a certain radius. If the FRB position is contained within the half-power beamwidth (HPBW) at the central frequency of the detection bandwidth, we use the sources contained within the HPBW \citep[See][for HPBW measurements]{de_villiers_meerkat_2022, de_villiers_meerkat_2023}. If the source is outside the HPBW, we use sources contained within 4/3 of the angular separation between the FRB and the phase centre.

Once we have selected our reference and MeerKAT sources, we match them using a 5\arcsec\ tolerance. We then compute an affine transformation matrix by performing a least squares minimisation of the positional offsets between the matched MeerKAT and reference sources. We weight the minimisation by the inverse of the positional variance, where we compute the variance as the quadrature sum of the positional errors from both the reference and the MeerKAT catalogues. Finally, we apply the resulting transformation matrix to the MeerKAT sources and images using the the python module \texttt{scikit-image} \citep{walt_scikit-image_2014}.

\begin{figure*}
    \raggedright
    \includegraphics[width=0.49\textwidth]{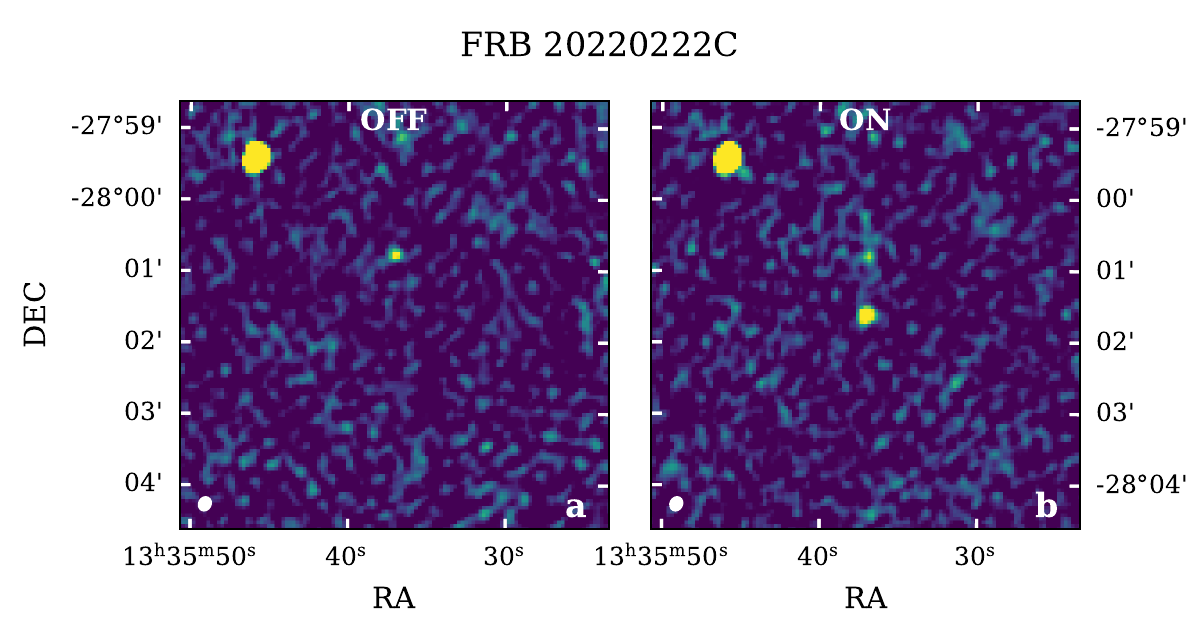}
    \includegraphics[width=0.49\textwidth]{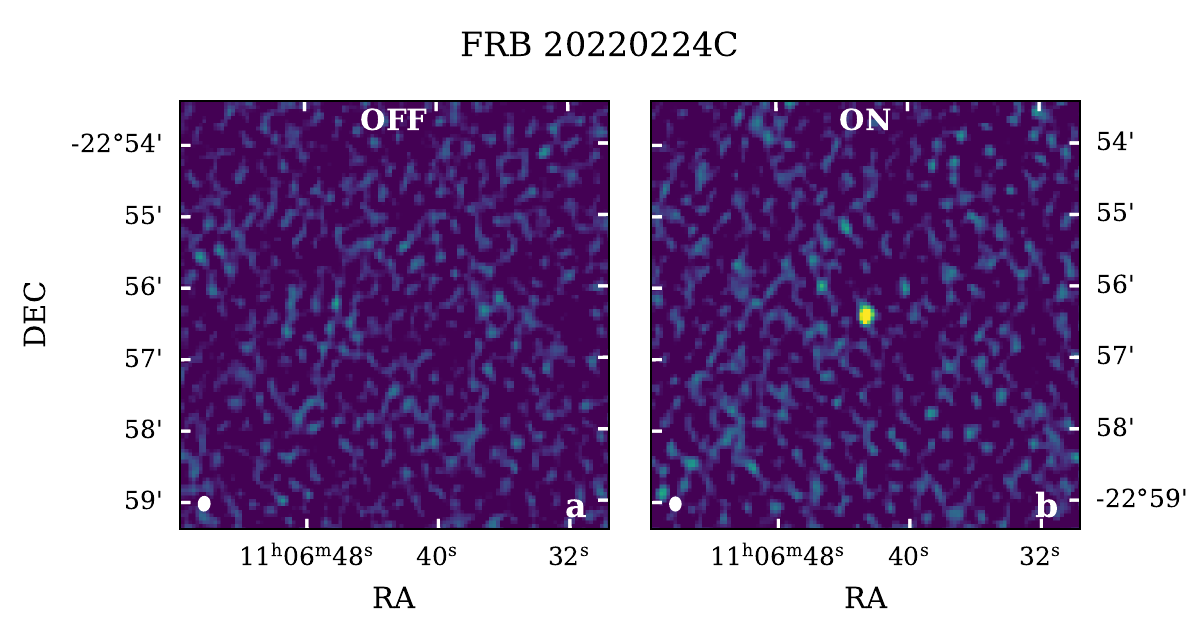}\\
    \includegraphics[width=0.49\textwidth]{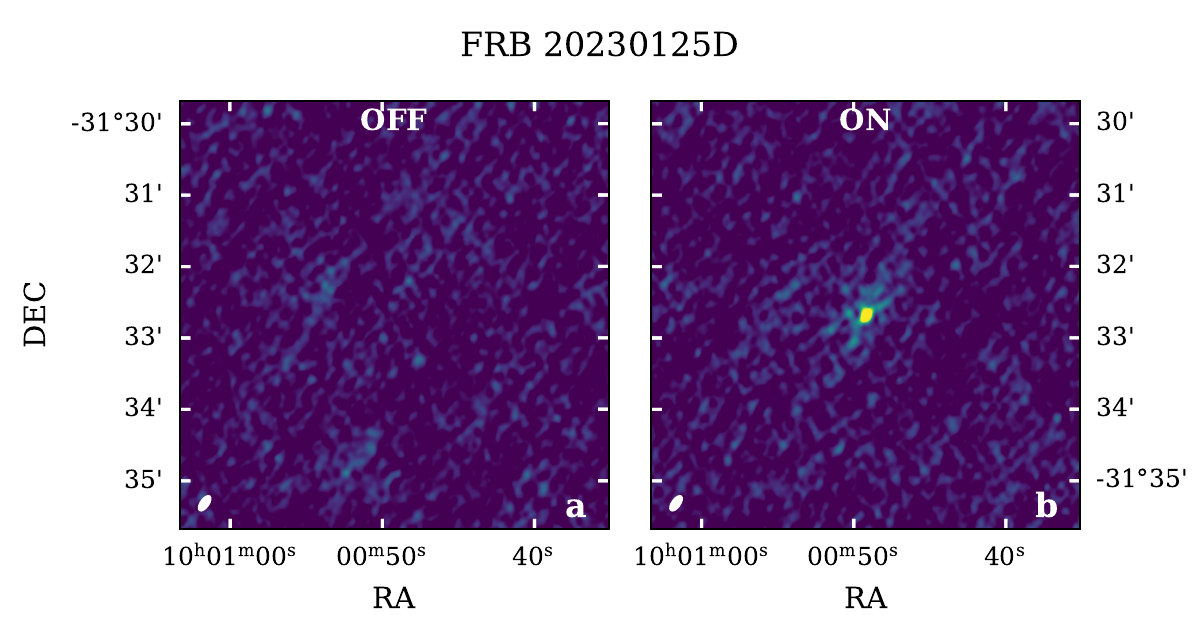}
    \includegraphics[width=0.49\textwidth]{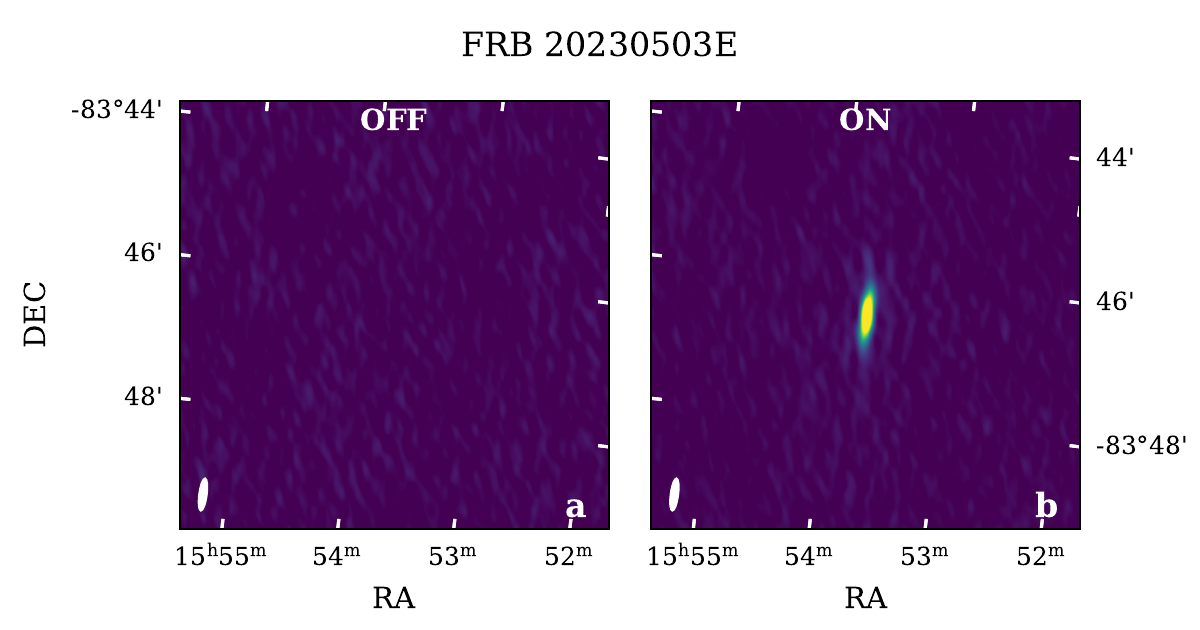}\\
    \includegraphics[width=0.49\textwidth]{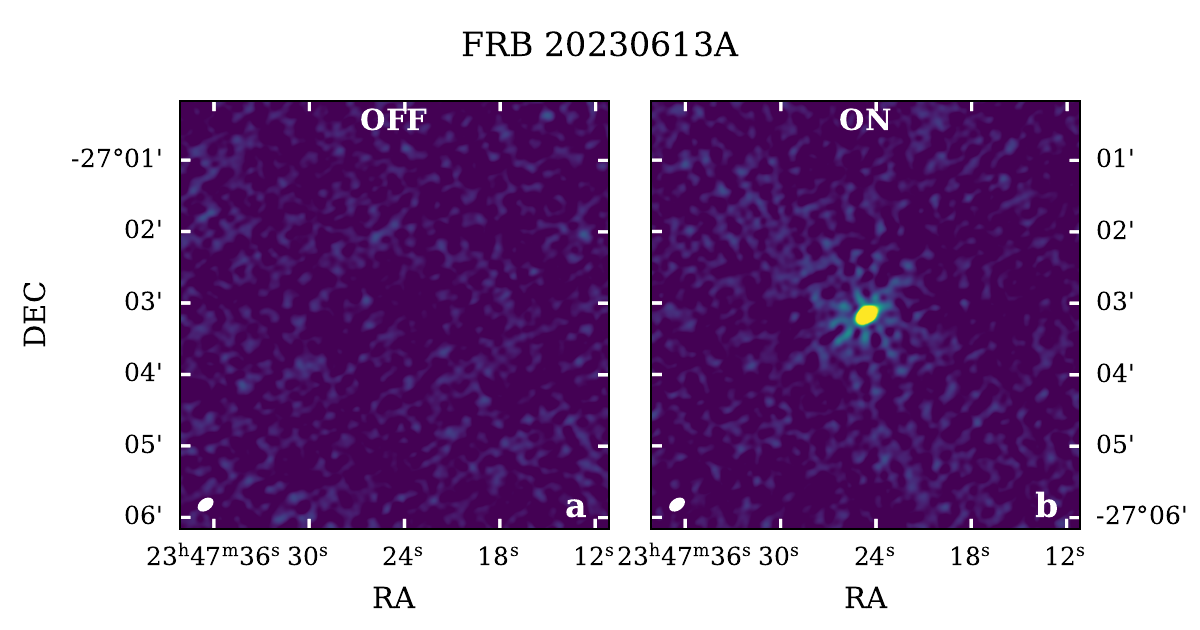}
    \includegraphics[width=0.49\textwidth]{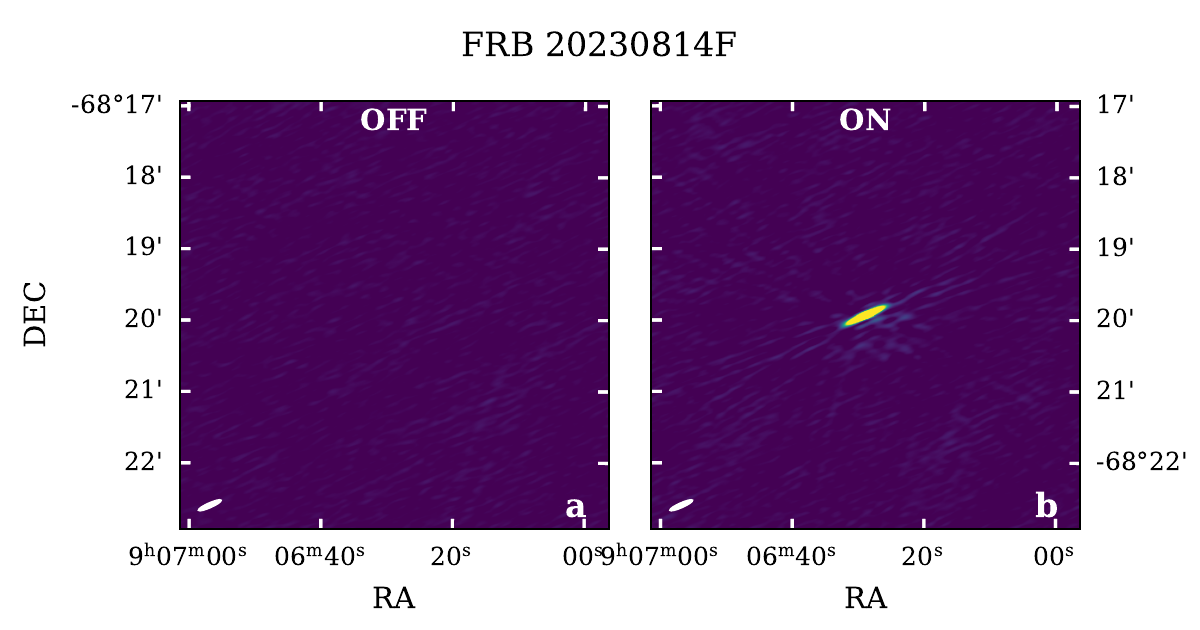}\\
    \includegraphics[width=0.49\textwidth]{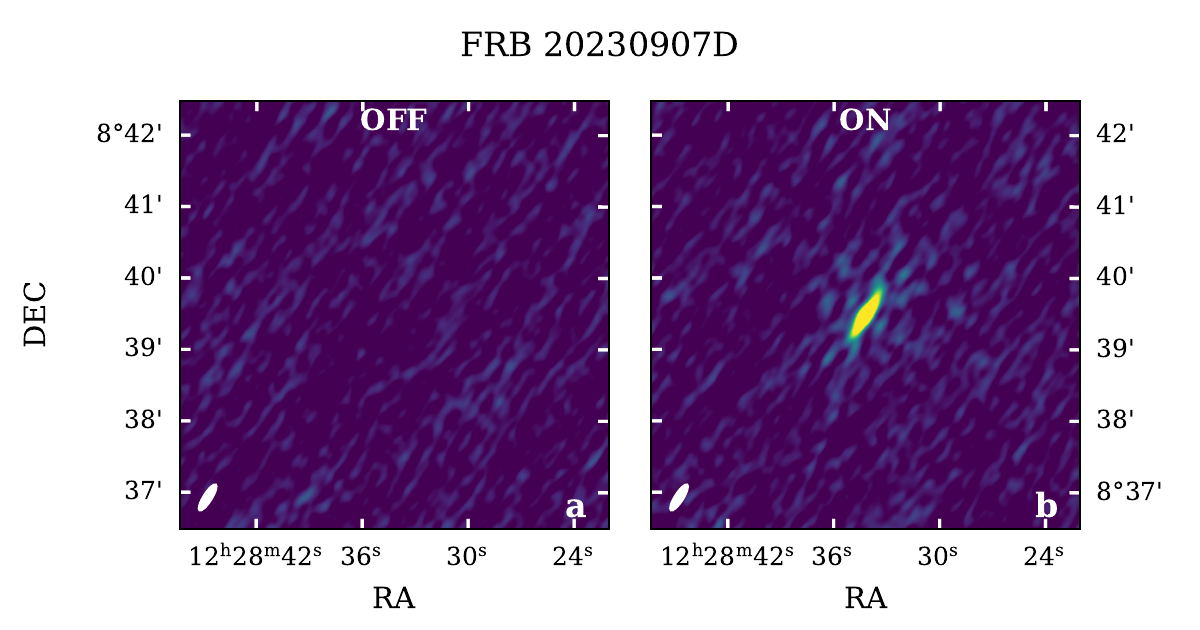}
    \includegraphics[width=0.49\textwidth]{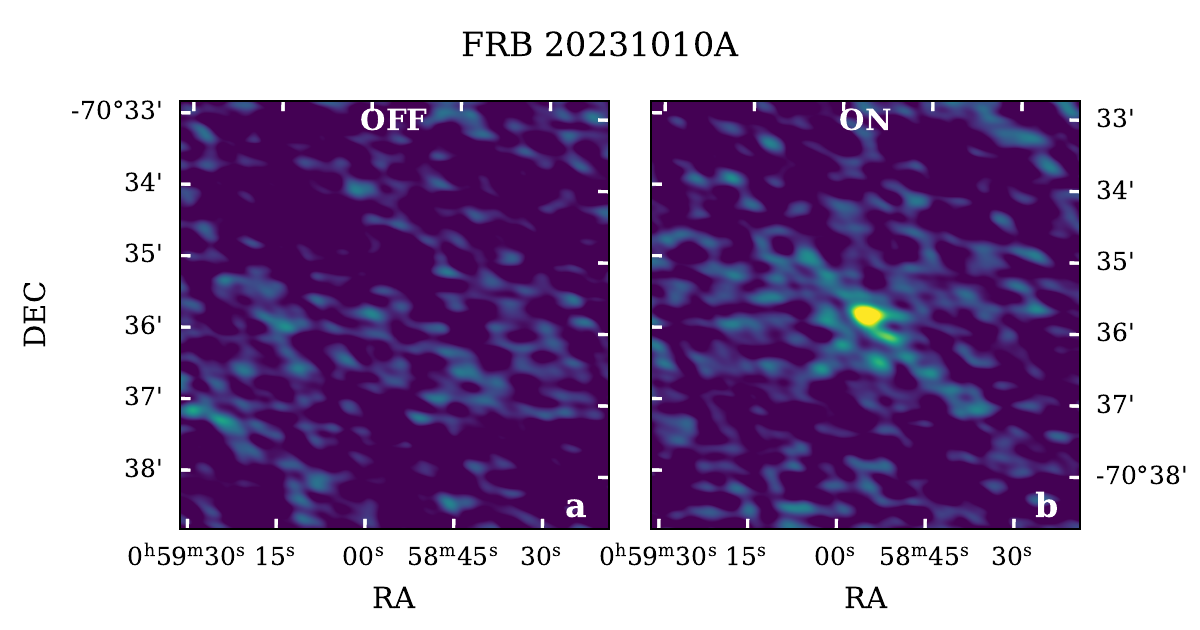}\\
    \includegraphics[width=0.49\textwidth]{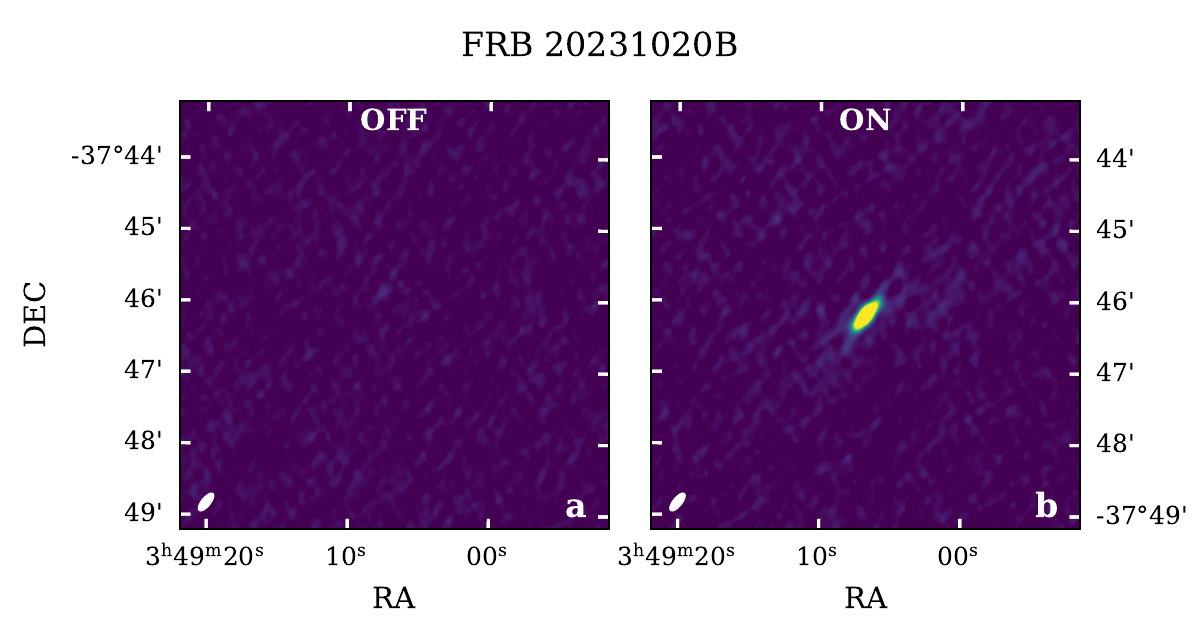}
    \includegraphics[width=0.49\textwidth]{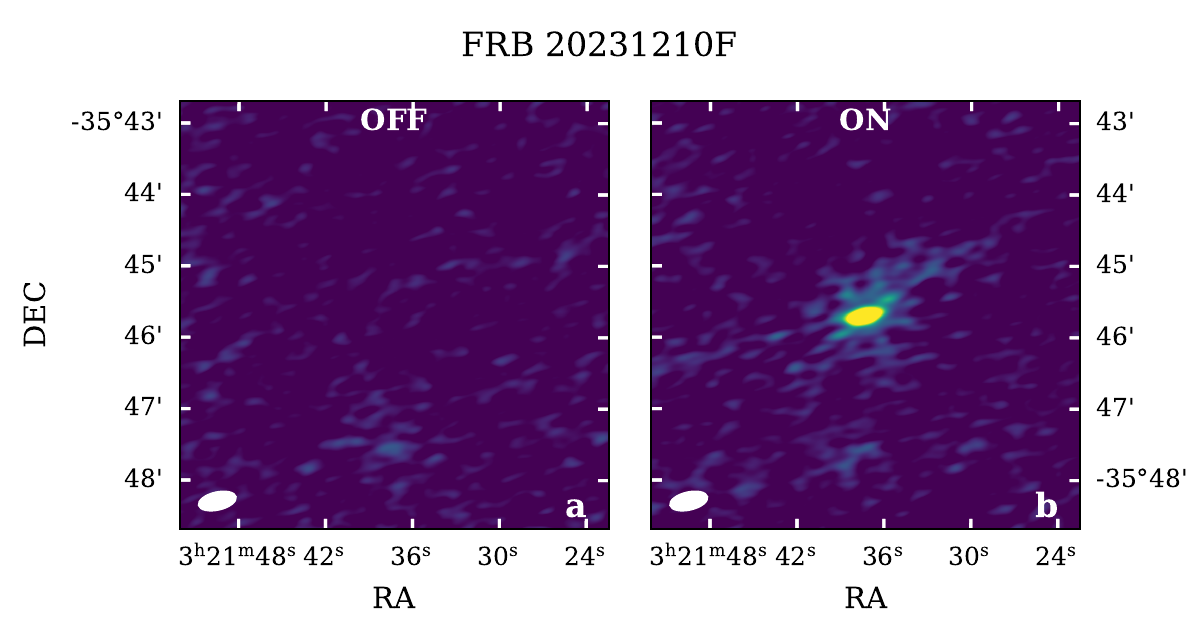}
    \caption{Localisation images of the FRBs presented in this work. In each row, the left panel shows an image before the arrival time of the burst (off), and the right panel the image during the burst (on). The images are centred at the FRB location, and have a FoV of 6\arcmin$\times$6\arcmin. The beam size of each observation is shown on the bottom left corner of each panel as a white ellipse.}
    \label{fig:on-off}
\end{figure*}

The resulting source position thus has several sources of uncertainty, and the final error we report is obtained by summing these in quadrature. The first source of error is the positional uncertainty obtained from \texttt{PyBDSF}, and the second one is from the astrometric correction. We estimate the astrometric error by computing the average separation in RA and DEC between the corrected and reference sources after each alignment, and add them in quadrature. For VLASS, we add an additional 0.2\arcsec systematic error in RA and DEC to account for the typical positional source accuracy. This error budget explicitly accounts for potential residual systematics in the astrometry, ensuring that the reported uncertainties are conservative and that our host associations are robust.
The final ``on" and ``off" images are shown in Fig.~\ref{fig:on-off}, while the $1\sigma$ localisation regions overlaid on the optical background are shown in Fig.~\ref{fig:optical}.

\subsection{Localisation of FRBs without TB data} \label{sec:tb_loc}

Although the MeerTRAP search pipeline runs in near-real-time \citep{rajwade_meertrap_2021, rajwade_first_2022}, only the bursts identified within 45\,s of the time they occurred will trigger a TB dump. 
In some cases the identification can be delayed beyond this limit, most commonly when an excess of candidate transients slows the clustering process, although other factors may also contribute. A full investigation will be presented in future work.
The bursts that did not produce a TB trigger thus have to be localised through different techniques. 
The python package \texttt{SeeKAT}\footnote{\texttt{SeeKAT}: \url{https://github.com/BezuidenhoutMC/SeeKAT}} was developed to localise MeerTRAP and TRAPUM sources by combining the pointing direction of the beams where they were detected and the S/N values in each one through a maximum likelihood estimation (MLE) approach. This method is described in detail in \citet{bezuidenhout_tied-array_2023}, and it has already been tested for previous MeerTRAP localisations \citep[e.g.][]{rajwade_first_2022, jankowski_sample_2023}. While a bright source detected in many coherent beams can be localised with (sub-)arcsecond accuracy, the position of a source detected in fewer than three CBs or in the IB, in most cases, will not be constrained enough to determine its most likely host galaxy.

Some FRBs do not produce enough CB detections to achieve a good localisation with \texttt{SeeKAT}. For bright FRBs, an alternative method is to localise them using imaging data from the observations running commensally to MeerTRAP, where the shortest available integration times are either 8\,s or 2\,s. 
This technique was successfully used to localise the MeerTRAP FRBs 20210405I \citep[8\,s integration time,][]{driessen_frb_2024} and 20210410D \citep[2\,s integration time,][]{caleb_subarcsec_2023}. However, since the typical millisecond duration of FRBs is orders of magnitude shorter than the available integration times, and the dispersion delay can smear the signal across multiple time intervals, which have fixed start times, only sufficiently bright FRBs can be localised through this method. To mitigate the signal loss, we select the snapshots containing the burst, and restrict the frequency range when appropriate to maximise S/N, though the reduced signal due to temporal averaging and dispersion cannot be fully corrected.

\subsection{Host galaxy observations \& analysis}

Once we have localised an FRB with (sub-)arcsecond precision, we search for archival optical images at the burst location to identify any host galaxy candidates. 
The tenth data release (DR10) of the DESI Legacy Survey \citep[DESI hereafter,][]{dey_overview_2019} covers the coordinates of most MeerKAT pointings. Since it can reach magnitude depths $\lesssim25$ in the $g$ and $r$ bands, it is likely to contain the host galaxies of most of the FRBs presented in this paper. 
The DESI DR10 source catalogues provide source fluxes in the g, r, i, and z bands from the dark energy camera \citep[DECam,][]{flaugher_dark_2015}, as well as in the W1-4 infrared filters from the Near-Earth Object Wide-field Infrared Survey Explorer \citep[NEOWISE,][]{mainzer_initial_2014} when available. 

\subsubsection{Optical data acquisition} \label{sec:optical_obs}

Although archival optical images are generally available, for some FRB coordinates there were no observations deep enough to identify the most likely host galaxy. In those cases, we observed the FRB locations with ground-based optical telescopes, with observing time granted through various proposals to use Gemini/GMOS, Gemini/Flamingos2, Keck/DEIMOS, Keck/LRIS, and SOAR/Goodman; the program details of specific observations are listed in Table~\ref{tab:observations}. 
After identifying the most likely FRB host galaxies using the Probabilistic Association of Transients to their Hosts \citep[\texttt{PATH},][see Section~\ref{sec:host_gal_id}]{aggarwal_probabilistic_2021}, we obtained spectra to identify their redshifts, using the same observatories as for the deep optical images.

The Gemini observatory consists of two optical/NIR telescopes with a diameter of 8.1\,m, with Gemini-North located in Hawaii and Gemini-South in Chile, which together cover most of the Northern and Southern sky. Gemini-South is especially well suited to follow up any potential MeerTRAP source. 
Both telescopes are equipped with the Gemini Multi-Object Spectrograph (GMOS) instrument, which provides imaging over a $5.5'\times5.5'$ FoV with a spatial sampling of 0.0807"/pixel \citep{hook_gemini-north_2004, gimeno_-sky_2016}. Imaging can be performed using the Sloan Digital Sky Survey (SDSS) filter system (u', g', r', i', and z'). By default, we used the r-band filter to perform deep imaging observations and identify any potential host galaxy at the FRB location. GMOS further supplies 360 to 1030\,nm spectroscopy, either in long-slit mode with a 5.5' long slit, or in multi-object mode with up to 60 custom-made slits.

Additionally, Gemini-South is currently equipped with the FLoridA Multi-object Imaging Near-infrared Grism Observational Spectrometer 2 (FLAMINGOS-2), a near-infrared imager and spectrograph \citep{eikenberry_flamingos-2_2006}. 
In imaging mode, it covers a circular 6.1\arcmin\ FoV with a spatial sampling of 0.179\arcsec/ pixel, with the J, H, Ks, K-blue and K-red filters available. 
For spectroscopic observations, it has both a long-slit mode and a multi-object spectroscopy mode, covering a wavelength range from 950 to 2500\,nm with a spectral resolution $R = 1000-3000$\footnote{Gemini/Flamingos2 specifications: \url{https://noirlab.edu/public/programs/gemini-observatory/gemini-north/flamingos-2/}}. 

The W. M. Keck observatory (Keck hereafter) consists of two telescopes (I and II) with a 10\,m diameter each, located in Hawaii. 
The Low Resolution Imaging Spectrometer (LRIS) is an instrument to perform optical imaging and spectroscopy installed on Keck I. The light received by the telescope is split into two and sent separately to a red and a blue camera \citep{oke_keck_1995, rockosi_low-resolution_2010,kassis+2022}. The imaging mode on Keck~I includes the UBVGRI filters; the FoV is $6'\times7.8'$, and the pixels are 0.135" wide. We typically obtain images with the R-band filter.
Spectra are obtained with a grating on the red side and a grism on the blue side, which together cover the wavelength range between 320 and 1000\,nm. The different gratings and grisms that can be used result in different wavelength resolutions, roughly from 0.16 to 0.92\,nm. The spectra are produced either with 175" long slits of different widths, or with up to 100 slits in the multi-object mode using custom-made slit masks\footnote{Keck/LRIS specifications: \url{https://www2.keck.hawaii.edu/inst/lris/pre_observing.html}}.

The DEep Imaging Multi-Object Spectrograph (DEIMOS) instrument is a  visible-wavelength, faint-object, multi-slit imaging spectrograph installed on the Keck II telescope. In imaging mode, DEIMOS has a FoV of $16.7'\times5.0'$, with a pixel size of 0.1185". It utilises BVRIZ broadband filters\footnote{Keck/DEIMOS filters: \url{www2.keck.hawaii.edu/realpublic/observing/public_instrument_info/deimos/filter_list.html}} to obtain deep images from $\sim$400\,nm to $\sim$1000\,nm. To identify any potential host galaxy at the FRB location, we used the R-band filter by default \citep{faber_deimos_2003}.
In spectroscopic mode, DEIMOS uses either custom-made slit masks to observe multiple objects simultaneously, with up to 130 slits per mask, or 20" long slits with various different widths. It covers a wavelength range between 410 and 1100\,nm, and a spectral resolution between 0.11 and 0.35\,nm depending on the selected grating\footnote{Keck/DEIMOS specifications: \url{https://www2.keck.hawaii.edu/inst/deimos/pre_observing.html}}. 

The Southern Astrophysical Research (SOAR) Telescope with a 4.1\,m mirror is located in Cerro Pach\'on, Chile. The Goodman high throughput spectrograph, mounted on SOAR, can be used both for imaging and spectroscopy \citep{clemens_goodman_2004}. In imaging mode, it is equipped with the SDSS filters u, g, r, i, and z, as well as the Bessell filters U, B, V, R, and I, and an FoV of $7.2\arcmin\times7.2\arcmin$ with a spatial sampling of 0.15\arcsec/pixel. In spectroscopic mode, both a long slit mode and multi-slit object masks over a field of $3\arcmin\times5\arcmin$ arcminutes are available, and it covers a wavelength range from 320 to 900\,nm with a spectral resolution $R=1400-14000$\footnote{SOAR/Goodman specifications: \url{https://noirlab.edu/public/programs/ctio/soar-telescope/goodman/}}.

All instruments thus possess similar characteristics, and the selection of the instruments will mainly be dependent on the declination of the target and on the available observing time.

We reduced the imaging data acquired with Gemini and Keck using the Pipeline for Optical/infrared Telescopes in Python for Reducing Images (\texttt{POTPyRI})\footnote{\texttt{POTPyRI}: \url{https://github.com/CIERA-Transients/POTPyRI}}. 
The pipeline first creates master bias, dark, and flat frames from the available calibration files, which are then applied to the science frames before aligning and stacking them. Next, it applies an astrometric correction using the Gaia DR3 catalogue \citep{gaia_collaboration_gaia_2023} for optical observations, or the Two Micron All Sky Survey \citep[2MASS][]{skrutskie_two_2006} for near-infrared observations. In cases where the positional discrepancy between the sources in the image and the reference catalogue is $\geq0.5\arcsec$, we first apply a manual WCS alignment. Once the astrometric solution is successful, we calculate the zero-point magnitude of the image and perform aperture photometry.
This is done with the Python packages \texttt{photutils} \citep{bradley_astropyphotutils_2024} and \texttt{petrofit} \citep{geda_petrofit_2022} to detect sources, perform aperture photometry, and determine their angular radii. We then cross-match these sources with reference catalogues that provide calibrated magnitudes in the FoV, such as Pan-STARRS \citep{chambers_pan-starrs1_2019} or SkyMapper \citep{onken_skymapper_2024}. These matches are used to compute the zero-point magnitude for each matched source, from which we derive the magnitudes of all detected objects using the median zero-point. 
The magnitudes reported in these two surveys are consistent to within $\lesssim0.02$ mag in overlapping filters, with Pan-STARRS showing systematics of only 7–12\,millimag across the sky\footnote{Pan-STARRS photometric quality: \url{https://skymapper.anu.edu.au/data-release/dr1/?utm_source=chatgpt.com\#dr1p1\_photometric_quality}}. Any systematic offsets between the calibrations are hence negligible compared to our measurement uncertainties.

To reduce the SOAR imaging observations, we used the \texttt{Photpipe} pipeline\footnote{\texttt{Photpipe}: \url{https://photpipe-docs.readthedocs.io/en/latest/}} instead \citep{rest_cosmological_2014}. \texttt{Photpipe} performs image calibration, including bias subtraction and flat-fielding, followed by astrometric calibration using Gaia DR3 as a reference. The fits images are next sky-subtracted, stacked, and re-gridded to the same pixel scale and field centre using \texttt{SWarp}\footnote{\texttt{SWarp}: \url{https://www.astromatic.net/software/swarp/}} \citep{bertin_terapix_2002}. Finally, it performs a photometric calibration, computing the zero-point magnitude using again either Pan-STARRS or SkyMapper as a reference, which is applied to the aperture photometry.

The uncertainty in the calibrated magnitude for each source combines the photometric measurement error with the zero-point uncertainty. The measurement error is computed from the flux uncertainty reported by \texttt{photutils}, while the zero-point uncertainty is estimated from the scatter of matched reference catalogue sources. These two contributions are added in quadrature to obtain the total magnitude uncertainty. The contribution from the reference catalogues themselves is negligible compared to the other sources of uncertainty.

\subsubsection{Host galaxy identification} \label{sec:host_gal_id}

In order to identify the FRB hosts, we apply \texttt{PATH} \citep{aggarwal_probabilistic_2021} in our optical/NIR images around the FRB positions. \texttt{PATH} is a Bayesian framework for identifying the most likely host galaxy of the FRB by computing priors using the magnitudes and angular separations of all candidate host galaxies, along with the probability that the host galaxy is not visible (`unseen') in the image. These priors are then combined with the FRB localisation region to calculate the posterior probabilities, identifying the most probable host.

If the FRB localisation is covered by DESI, the survey's optical depth is typically enough to identify a sample of the nearest galaxies. 
In cases where we obtained deeper optical imaging, we apply a source fitting technique to extract the properties of sources near the FRB location. Although the DESI photometric catalogues provide the magnitudes and angular radii in up to eight filters (griz, W1-4), along with photometric redshifts in regions outside of the Galactic plane \citep{dey_overview_2019}, we process the $r$-band DESI images with the same pipeline used for the targeted FRB fields to ensure homogeneity across the whole sample.
Whether we use the DESI catalogue or the sources identified in a targeted optical observation, we select the positions, half-light radii and magnitudes of the extended sources within a 10" separation from the FRB localisation to generate the input source catalogue for the \texttt{PATH} analysis. We exclude from the candidate list the sources identified as stars in the DESI catalogue, or identified as point-like by the fitting technique.

The \texttt{PATH} algorithm requires the unseen host prior, $P(U)$, as an input. We estimate this prior independently for each FRB, following an approach similar to that described by \citet[][Appendix B2]{marnoch_unseen_2023}. 
To define a plausible redshift range for each FRB, we begin by estimating the extragalactic dispersion measure (DM). We assess the Galactic interstellar medium (ISM) contribution by averaging the expected values from the NE2001 \citep{cordes_ne2001i_2003} and YMW16 \citep{yao_new_2017} models, and then add the Galactic halo contribution from the \citet{yamasaki_galactic_2020} model. Subtracting the combined Galactic contribution from the observed DM yields the extragalactic DM. We then apply the Macquart relation \citep{macquart_census_2020} using the \texttt{FRBs} package \citep{prochaska_frbsfrb_2025}, to obtain the 95\%\ confidence redshift range expected for each FRB. We assume a host galaxy contribution to the DM in its rest frame of 100\,\pccm, which is consistent with empirical estimates and simulations \citep{mo_dispersion_2023, kovacs_dispersion_2024, bernales-cortes_empirical_2025}.
Using a reference sample of 56 securely localised FRBs with known $r$-band magnitudes and spectroscopic redshifts, we compute how their apparent magnitudes would evolve if they were located at different redshifts. For each trial redshift,  we construct a probability density function using a Gaussian kernel density estimate (KDE) based on the projected magnitudes. Then, for each FRB, we determine the limiting magnitude of the optical observation used to identify host galaxy candidates. Using the KDE corresponding to the Macquart redshift upper limit of each FRB, we compute the fraction of reference galaxies that would fall below the detection threshold. We adopt this fraction as the unseen prior $P(U)$.
After running \texttt{PATH}, we identify any galaxy with a posterior probability of association $P(O_i|x)>0.9$ as the putative host.

The estimate of $P(U)$ based on a DM-derived redshift upper limit, magnitude completeness, and the luminosity of known FRB hosts, provides a conservative prior compared to adopting a fixed 5-10\% value, reducing arbitrary assumptions in cosmological applications. Our association probabilities are largely insensitive to moderate changes in the assumed host DM, because of DM dilution with redshift, simulations suggesting that host DMs tend to be larger at higher redshift \citep{kovacs_dispersion_2024}, and our conservative unseen prior. However, $P(U)$ might be biased if FRB hosts are systematically fainter than the known sample, or if the Macquart relation underestimated redshifts along certain lines of sight.

Certain FRB sightlines, especially for FRBs with high DMs, may intersect foreground galaxy clusters that contribute significantly to the observed DM. In order to estimate whether these could affect the inferred redshift, we searched for clusters within 2\degr\ of each FRB location. For our search, we selected the galaxy cluster catalogue from the DESI Legacy Imaging Surveys \citep{zou_galaxy_2021}, which offers the most extensive optical coverage available, and the second release of the meta-catalogue of X-ray detected clusters of galaxies \citep[MCXC-II;][]{sadibekova_mcxc-ii_2024}, which provides the most complete compilation of X-ray clusters. In Section~\ref{sec:results}, these searches are mentioned wherever we find clusters intersecting the FRB sightlines.

For each cluster intersecting the FRB sightline within twice its characteristic radius, $R_{500}$, we estimated the DM contribution by modelling their gas density profiles with a $\beta$-model \citep[e.g.][]{arnaud__2009}. We assume typical values of $\beta\sim2/3$, a core radius $r_c\sim0.15 R_{500}$, and a gas fraction $f_{\text{gas}}\sim0.1$ of the total mass. Next, we integrate the electron density at the projected impact parameter between the cluster centre and the FRB, up to $2 R_{500}$. Finally, we apply a redshift correction to the resulting DM. While the values above are typical for X-ray detected clusters, their large uncertainties result in errors of up to 30\% of the reported values. This simple calculation is only intended to verify that the observed host redshift falls within the expected Macquart range; a detailed analysis of the foreground cluster contributions will be reported in an upcoming publication.

\subsubsection{Analysing spectroscopic and photometric data}

The spectra are obtained once the putative host has been identified. If more than one host galaxy candidate remains after running PATH, we use either the long-slit or the multi-object spectroscopy slit masks available with Gemini-S/GMOS, Keck/LRIS, or Keck/DEIMOS to obtain simultaneous spectra of those sources.
Once the spectra have been obtained, we reduce them with \texttt{PypeIt} using standard practices \citep{prochaska_pypeit_2020}, and determine the redshift.

For some FRBs, the observation constraints have not allowed us to obtain a spectrum yet. In those cases, we obtain a photometric redshift if enough photometry is available in at least four bands.
The DESI-DR10 catalogue provides photometric redshifts computed through a random forest algorithm using DECam and WISE fluxes for all galaxies with more than one exposure in the required filters, and contained within unmasked regions\footnote{Legacy survey files: \url{https://www.legacysurvey.org/dr10/files/}}. 

To infer the physical properties of the most likely host galaxies of our FRB sample, we use the Code Investigating GALaxy Emission \citep[\texttt{CIGALE},][]{boquien_cigale_2019}. 
Using the redshift and the photometric data as input, \texttt{CIGALE} fits the spectral energy distributions (SEDs) to estimate the stellar mass, star formation rate (SFR), metallicity, the age of the oldest stars, and other properties that characterise the host galaxies. We initialise \texttt{CIGALE} adopting the following models: delayed-exponential star-formation history without burst population, a synthetic stellar population described by \citet{bruzal2003}, the initial mass function given by \citet{chabrier2003}, dust attenuation models from \citet{calzetti2001}, and dust emission template from \citet{dale2014}. This upper limit follows the prescription used in previous FRB host studies \citep{khrykin_flimflam_2024} and is consistent with the range expected from the \citet{dale2014} templates for typical star-forming galaxies. In cases where SED residuals or mid-IR photometry suggest a different AGN contribution, we adjust this fraction to obtain more reliable estimates on the stellar population parameters.

\texttt{CIGALE} provides Bayesian estimates of the physical parameters and their associated uncertainties based on the likelihood-weighted average and standard deviation of their probability distribution function (PDF) derived from a model grid. 
However, these estimates rely on a good photometric coverage, with 10 to 25 photometric measurements across a broad wavelength range typically required to get robust constraints on the galactic properties. 
When spectroscopic redshifts are unavailable, \texttt{CIGALE} can also infer photometric redshifts. Since we verified that the spectroscopic redshifts in our sample are consistent with the photometric redshifts reported on the DESI DR10 catalogue, we limited the redshift range that \texttt{CIGALE} explores to within those uncertainties when possible (see Appendix~\ref{app:photoz}). In cases where no photometric redshift is available, we fit for a photometric redshift between 0 and 1.
We note however that the uncertainties on the derived physical parameters are significantly larger for galaxies lacking spectroscopic redshifts. 
While other tools, such as \texttt{Prospector} \citep{johnson_stellar_2021}, can perform joint photometric and spectroscopic SED fits, such analysis is beyond the scope of this work. Moreover, recent comparisons between the properties derived by \texttt{CIGALE} and \texttt{Prospector} for FRB host galaxies \citep{gordon_demographics_2023} show that the resulting SFR estimates are in good agreement (private comm.), justifying the reliability of our \texttt{CIGALE} results.
The photometric information and \texttt{CIGALE} results are summarised in Table~\ref{tab:sed_results}.

\section{Results} \label{sec:results}

During all of the MeerTRAP observations that were carried out throughout the year 2023, we discovered fourteen new, so far non-repeating, FRBs. One of the bursts, FRB\,20230808F, has already been presented in \cite{hanmer_contemporaneous_2025}. In this work, we present the localisation of the remaining thirteen 2023 FRBs, as well as the localisation of two FRBs discovered in February 2022 before the TB data became available. 
Of the total 15 localised FRBs, two lie too close to the Galactic plane to be covered by DESI-DR10 imaging. Among the 11 FRBs with arcsecond localisations and DESI coverage, two have no visible host galaxy candidates in the DESI data, consistent with the survey’s imaging depth. The remaining four FRBs have arcminute localisations; although these regions are included in DESI-DR10, their large sizes prevent the identification of a single host galaxy.
General FRB and localisation properties are listed in Table~\ref{tab:frb_loc}, while detailed properties of the host galaxy candidates are shown in Table~\ref{tab:frb_hosts}.

\subsection{Arcsecond localised FRBs} \label{sec:new_loc_frbs}

\begin{table*}
	\centering
	\caption{FRB localisation general properties.}
	\label{tab:frb_loc}
	\begin{tabular}{l
                    S[table-format=4.2(2)]
                    c
                    l
                    l
                    c
                    c
                    c} 
		\hline
		\multicolumn{1}{c}{ID} & \multicolumn{1}{c}{DM} & DM$_{\text{NE2001}}$ & \multicolumn{1}{c}{RA} & \multicolumn{1}{c}{DEC} & $z_{\text{spec}}$ & \zphot & Localisation \\
         & \multicolumn{1}{c}{(\pccm)} & (\pccm) & \multicolumn{1}{c}{(hh:mm:ss)} & \multicolumn{1}{c}{(dd:mm:ss)} & & & method \\
		\hline
20220222C & 1071.2(0.8) & 56 & 13:35:37.08$\pm$0.54\arcsec & -28:01:36.93$\pm$0.55\arcsec & 0.853 & -- & 8-s image\\
20220224C & 1140.2(1.8) & 52 & 11:06:42.61$\pm$0.42\arcsec & -22:56:23.48$\pm$0.64\arcsec & 0.6271 & $0.57^{+0.06}_{-0.06}$ & 8-s image\\
20230125D & 640.08(0.03) & 88 & 10:00:49.21$\pm$0.24\arcsec & -31:32:40.77$\pm$0.26\arcsec & 0.3265 & -- & TB \\
20230306F & 689.5(0.9) & 23 & 12:24:01$\pm$60\arcsec & +14:54:17$\pm$50\arcsec & -- & -- & SeeKAT \\
20230413C & 1532.2(0.05) & 45 & 05:13:00$\pm$63\arcsec & -39:50:00$\pm$57\arcsec & -- & -- & SeeKAT \\
20230503E & 483.74(0.04) & 88 & 15:53:43.19$\pm$0.69\arcsec & -83:46:30.93$\pm$0.87\arcsec & -- & $0.32^{+0.15}_{-0.15}$ & TB \\
20230613A & 483.51(0.01) & 30 & 23:47:24.65$\pm$0.4\arcsec & -27:03:10.01$\pm$0.52\arcsec & 0.3923 & $0.42^{+0.03}_{-0.03}$ & TB\\
20230814F & 471.44(0.03) & 134 & 09:06:28.88$\pm$0.35\arcsec & -68:19:55.70$\pm$0.31\arcsec & -- & -- & TB \\
20230827E & 1433.7(0.1) & 38 & 04:08:28.242$\pm$0.75\arcsec & -18:16:58.55$\pm$1.5\arcsec & -- & -- & SeeKAT \\
20230907D & 1030.79(0.04) & 29 & 12:28:34.20$\pm$0.4\arcsec & +08:39:29.13$\pm$0.57\arcsec & 0.4638 & $0.44^{+0.03}_{-0.02}$ & TB \\
20231007C & 2660.4(1.9) & 42 & 23:38:18$\pm$50\arcsec & +21:52:34$\pm$55\arcsec & -- & -- & SeeKAT \\
20231010A & 442.59(0.02) & 41 & 00:58:55.67$\pm$0.52\arcsec & -70:35:46.93$\pm$0.3\arcsec & -- & $0.61^{+0.18}_{-0.18}$ & TB \\
20231020B & 952.2(0.3) & 34 & 03:49:06.77$\pm$0.37\arcsec & -37:46:11.56$\pm$0.4\arcsec & 0.4775 & $0.46^{+0.07}_{-0.08}$ & TB\\
20231204B & 1772.1(0.3) & 41 & 01:05:07$\pm$64\arcsec & -70:37:16.5$\pm$68\arcsec & -- & -- & SeeKAT \\
20231210F & 720.6(0.2) & 32 & 03:21:37.28$\pm$0.25\arcsec & -35:45:41.13$\pm$0.25\arcsec & -- & $0.50^{+0.08}_{-0.08}$ & TB\\
		\hline
	\end{tabular}
\begin{minipage}{\linewidth}
    \textbf{Notes.} 
    additional details about the FRB host galaxies and their physical prperties are listed in the Appendix Tables~\ref{tab:frb_hosts}, \ref{tab:sed_results}.
    DM$_{\text{NE2001}}$ is the dispersion measure from the NE2001 model evaluated at 30\,kpc, $z_{\text{spec}}$ is the spectroscopic redshift, and \zphot\ the photometric redshift.
\end{minipage}
\end{table*}

\subsubsection{FRB\,20220222C}

\begin{figure}
    \centering
    \includegraphics[width=\linewidth]{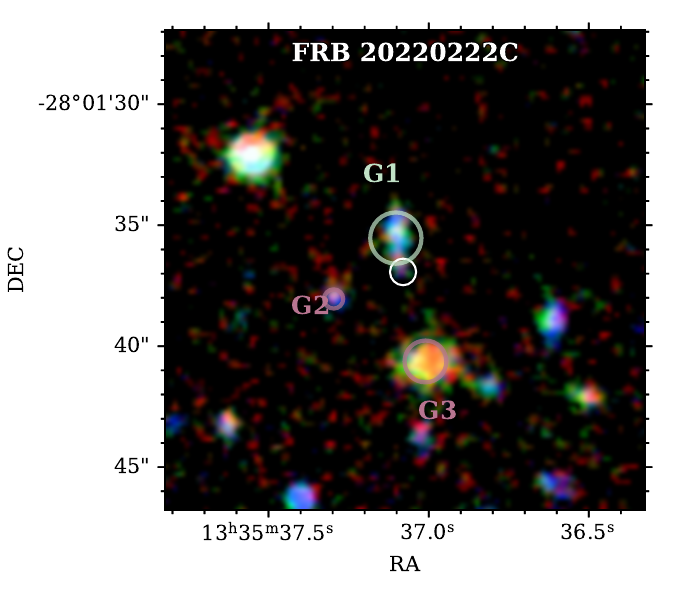}
    \caption{Localisation of FRB~20220222C. The error region is shown as a white ellipse, and the background image shows a composite image with the $r$, $z$, and $J$ band observations. The candidate hosts are indicated by coloured circles of half-light radii labelled G1-3.}
    \label{fig:frb20220222}
\end{figure}

FRB\,20220222C was discovered while MeerTRAP was observing commensally in the L-band with the MHONGOOSE \citep[MeerKAT HI Observations of Nearby Galactic Objects -- Observing Southern Emitters,][]{de_blok_mhongoose_2024} large survey project (Proposal ID: SCI-20180516-EB-01), with a DM of 1071.2$\pm$0.8\,\pccm. The dynamic spectrum showed the burst to be present only in the lower half of the frequency band. It was detected prior to the successful implementation of the TB system, and there is thus no available TB trigger. However, it was detected in seven CBs and the IB with a S/N of 28 in the IB, and was hence reasonably well localised. The burst was identified in the 8-s images generated from the MHONGOOSE observation, by searching for new sources appearing at the burst time of arrival (ToA) within the coordinates the detection CBs were pointing at\footnote{This search was undertaken under the auspices of the ThunderKAT project \cite{thunderkat_2016}.}. 
Since the burst was only visible at low frequencies, we created images of the bottom half of the band. For these data, we found a new source in the difference images at the coordinates \coordserr{13}{35}{37.08}{0.54}{-28}{01}{36.93}{0.55}, where the errors include the \texttt{PyBDSF} and the astrometric error on the source position, for which we used VLASS as a reference. The off/on images are shown in Fig.~\ref{fig:on-off}.

A search in the DESI images at the source position revealed no host galaxy candidate within the error region, and we thus acquired deep images in the r, z, J, and Ks bands with Gemini-South, with total exposure times of 2500, 2500, 600, and 1800\,s respectively. The observation details are summarised in Table~\ref{tab:observations}.

In the resulting images, we identified a galaxy (G1) centred 1.43" away from the FRB coordinates with a magnitude $r=23.86\pm0.04$. This galaxy has a \texttt{PATH} probability of 94.4\% of being associated with the FRB after assuming an unseen prior of 5\%, and we therefore identify it as the putative host. A second source with magnitude $r=25.01\pm0.07$ is located 3.08" away from the FRB coordinates (G2), and a third brighter galaxy with $r=23.32\pm0.04$ is 3.81" away (G3). Both those galaxies have association probabilities $<0.01$\%. A composite RGB image from the r, z, and J images, the FRB error region, and candidate host galaxies are shown in Fig.~\ref{fig:frb20220222}.

To measure the redshift of the two brightest galaxies, G1 and G3, we obtained a spectrum with Keck/LRIS in long-slit mode on 2023 April 17 (Program ID: U173; PI: X. Prochaska). The red-side grating was centred at 808.2\,nm with the 600/10000 grating, with a total exposure time of 6450\,s, while the blue-side grism used the 300/5000 configuration and the exposure time was 6600\,s.
No spectral lines were identified in the blue grism for the putative host. On the red grating, the \Hbeta\ line and the \Oiii\ doublet were identified in the spectrum, which allowed us to determine a redshift of 0.853. This is in agreement with the redshift range expected from the Macquart relation, $z_{\text{Macquart}}=1.17^{+0.26}_{-0.58}$. The projected distance between the galaxy centre and the FRB location would thus be $11.3\pm6.1$\,kpc.
For G3, we identified a feature in the blue-side spectrum which might be the \Oii\ doublet, and would correspond to a redshift of 0.795, but since no lines were identified in the red-side spectrum, the measurement is not convincing.

Based on the magnitudes of G1, that was detected in the r, z, and J filters, but not in the Ks filter, we performed an SED fit. We find the value of the stellar mass to be \logMstar=$10.1\pm0.2$. The best SFR value is 7.2\,\Msunyr, with a 0.3 dex error, and the metallicity \met$<-0.13$. The oldest stars in the galaxy have an age of $1.8\pm0.7$\,Gyr. The CIGALE fit is shown in Fig.~\ref{fig:host_sed}.

\subsubsection{FRB\,20220224C}

\begin{figure}
    \centering
    \includegraphics[width=\linewidth]{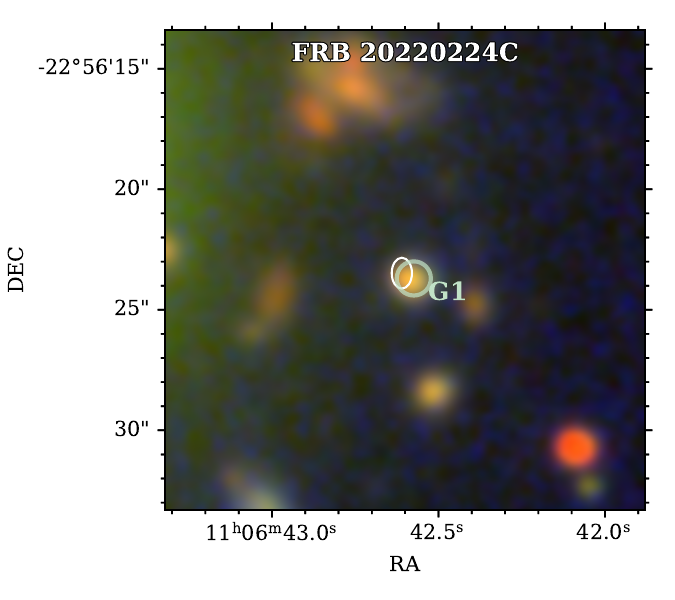}
    \caption{Localisation of FRB~20220224C. The error region is shown by a white ellipse, and the background is a composite image from the $r$ and $i$ observations and the $g$ image from DESI-DR10. The candidate host galaxy is indicated by a coloured circle labelled G1.}
    \label{fig:frb20220224}
\end{figure}

This FRB was detected with a DM of 1140.2$\pm$1.8\,\pccm\ in another MHONGOOSE observation (Proposal ID: SCI-20180516-EB-01), prior to the successful implementation of the TB system. Since it was found only in the IB, SeeKAT could not constrain the position well. However, the burst signal was strong enough to be localised in the 8-s images of the commensal MHONGOOSE observation\footnotemark[\value{footnote}]. Due to the dispersion delay, the burst signal was divided across two consecutive 8-s images.
The on image, which adds together the two 8-s images where it was detected, is shown together with an off image in Fig.~\ref{fig:on-off}.

We performed an astrometric correction of the final on image using VLASS as a reference, and obtained the final coordinates \coordserr{11}{06}{42.61}{0.42}{$-$22}{56}{23.48}{0.64}. We identified a host galaxy candidate in DESI DR10 located 0.52\arcsec away from the position and with an r-band magnitude of $21.8\pm0.3$. 
Furthermore, we acquired deep $r$ and $i$-band images with GMOS, with total exposure times of 2700\,s each, on 2023 January 29 and 30 (Proposal ID: GS-2022B-Q-123; PI: A. Gordon). A composite colour image is shown in Fig.~\ref{fig:frb20220224}.

Based on the DM of this FRB, after removing the MW contribution and 100\,\pccm\ from the host, the expected Macquart redshift would be $z_{\text{Macquart}}=1.27^{+0.28}_{-0.62}$. 
Assuming an unseen prior of 67\%, we find its \texttt{PATH} association probability to be 93.8\%, with the probability of association of all surrounding sources being negligible. In order to measure the redshift, we obtained LRIS spectra of the putative host as well as two other nearby sources. The spectra were acquired on 24 January 2023 (Program ID: U055; PI: J. Hennawi) with the 300/5000 blue grism and an exposure of 2780\,s, and with the 600/10000 red grating centred at 892.7\,nm and an exposure time of 2740\,s.
In the spectra, we identified the \Oii\ and \Oiii\ doublets, as well as the \Hbeta\ line, which we used to measure the FRB host redshift to be 0.6271. At this redshift, the projected physical separation between the FRB and the galaxy centre is $3.8\pm5.4$\,kpc.

The spectroscopic redshift hence falls below the 95\% lower limit. To assess whether any structure in the foreground could be significantly contributing to the observed DM, we searched for galaxy clusters in the foreground.
From the DESI Legacy Imaging Surveys, \cite{zou_galaxy_2021} produced a galaxy cluster catalogue, and there we found the galaxy cluster ACO~S~651 to be located 550\arcsec away from the FRB location, at a redshift of 0.0639. The cluster is also detected in X-rays \citep{piffaretti_mcxc_2011}, and it has a characteristic radius $R_{500}=830$\,kpc. The impact parameter of the FRB, $\sim700$\,kpc, falls within its radius.
From a cluster with a typical density profile, as described in Section~\ref{sec:host_gal_id}, the expected DM contribution could be up to $800\pm100$\,\pccm\ in the cluster frame, or $\sim750$\,\pccm\ after correcting for redshift. This contribution likely explains the large DM observed.

Through the DESI-DR10 source catalogues\footnote{DESI-DR10: \url{https://www.legacysurvey.org/dr10/description/}}, we had access to photometric measurements of the host galaxy candidate in a broad wavelength range. These include the \textit{griz} filters from DECam and the NEOWISE filters W1-4. This extensive wavelength coverage enables us to perform SED fitting of the putative host galaxy with CIGALE.
The SED parameters derived for the putative host are as follows: a stellar mass \logMstar$=10.3\pm0.1$, a star formation rate SFR$=18.9$\,\Msunyr, with a 0.2\,dex error, a metallicity $\log(Z/Z_\odot)<-0.7$, and an age of $1.4\pm0.5$\,Gyr. The CIGALE fit is shown in Fig.~\ref{fig:host_sed}.

\subsubsection{FRB\,20230125D}

\begin{figure}
    \centering
    \includegraphics[width=\linewidth]{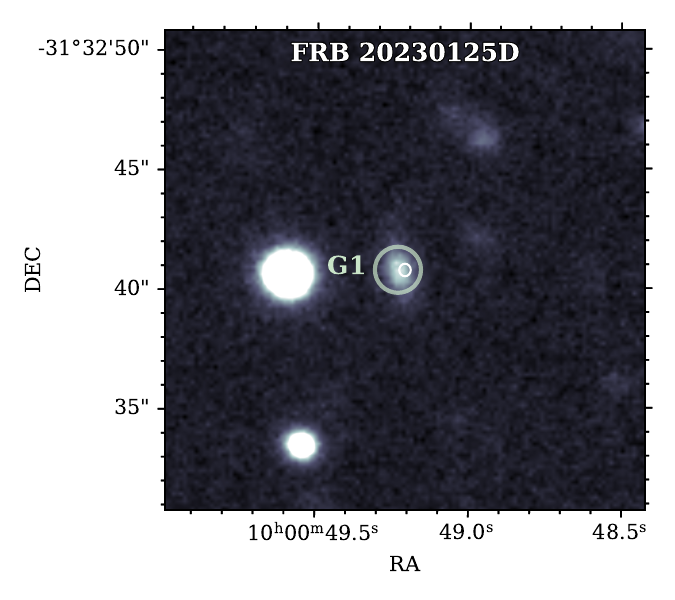}
    \caption{Localisation of FRB~20230125D. The error region is shown by a white ellipse, and the background image is a Gemini-S observation. The candidate host galaxy is indicated by a coloured circle labelled G1.}
    \label{fig:frb20230125}
\end{figure}

This FRB, with a DM of 640.08$\pm$0.03\,\pccm, was detected in the L-band during an open time observation (Proposal ID: SCI-20220822-IP-01). It was detected in the IB and the CB 261, and its detection triggered the storage of the TB data. We localised the burst in the TB data as described in Section~\ref{sec:tb_loc}, and performed an astrometric correction with VLASS as the reference catalogue. The on and off images where the burst is found are shown in Fig.~\ref{fig:on-off}. We localised the burst to the coordinates \coordserr{10}{00}{49.21}{0.24}{$-$31}{32}{40.77}{0.26}. The FRB coordinates were not covered by DESI, and it only had been observed in the i-band with Pan-STARRS. We identified a potential source at the location of the FRB in the i-band, and followed up with a deep r-band image using the GMOS with a total 2700~s exposure time (see observation details in Table~\ref{tab:observations}). In the deep image, we identified a galaxy 0.29" away from the FRB localisation centre with a magnitude $r=22.16\pm0.02$, as shown in Fig.~\ref{fig:frb20230125}. Based on the image magnitude limit and the Macquart redshift upper limit, we assume an unseen prior of 36\% and obtain a posterior \texttt{PATH} probability of 97.81\%, and we hence identify this as the host galaxy. The deep r-band image is shown in Fig.~\ref{fig:optical}. Subsequently, we obtained a spectrum of the putative host using Keck/DEIMOS (see Table~\ref{tab:observations}).
In the spectrum, we identified the \Halpha\ and \Nii\ spectral lines (see Fig.~\ref{fig:host_spectra}), from which we measured a redshift of $z=0.3265$, which falls on the lower end of what is expected from the Macquart relation ($z_{\text{Macquart}}=0.62^{+0.14}_{-0.34}$). 
To explain its apparent DM excess, we searched for known galaxy clusters in the foreground, but found no matches in MCXC-II. For the same reason why no DESI images are available at this location, we do not have information about galaxy clusters identified in the optical. Future observations might explain whether the excess DM can be attributed to foreground clusters or to the host galaxy instead.
At the spectroscopic redshift, the projected distance between the galaxy centre and the FRB location is $1.4\pm1.7$\,kpc. Because we only have the host galaxy magnitude in one filter, we are not able to perform a CIGALE SED fitting at this stage.

\subsubsection{FRB\,20230503E}

\begin{figure}
    \centering
    \includegraphics[width=\linewidth]{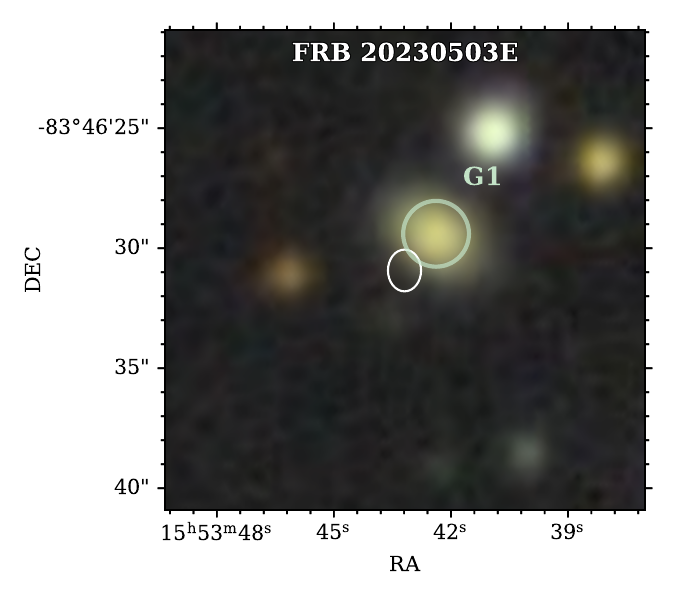}
    \caption{Localisation of FRB~20230503E. The error region is shown by a white ellipse, and the background image is from DESI-DR10. The candidate host galaxies are indicated by coloured circles labelled G1-2. The source inside the FRB error region is a star.}
    \label{fig:frb20230503}
\end{figure}

FRB\,20230503E was discovered during a gain calibration observation in the L-band of the calibrator J1619-8418 (Proposal ID: SCI-20180516-PW-03), and it has a DM of 483.74$\pm$0.04\,\pccm. The burst was only detected in the IB, and it triggered the storage of the TB data, that we used to localise it. 
After obtaining the original coordinates, we performed an astrometric correction using the ATPMN sources in a 6\degr\ radius to correct the RACS-mid source coordinates. 

We determined the final localisation region to be \coordserr{15}{53}{43.19}{0.69}{$-$83}{46}{30.93}{0.87}. 
We identify a galaxy located 2.02\arcsec away from the FRB localisation, with an r-band magnitude of $20.1\pm0.2$, as can be seen in Fig.~\ref{fig:frb20230503}. Assuming an unseen prior of 5\%, we obtain a \texttt{PATH} probability of association of 97.3\%, and we thus identify this galaxy as the putative host. 

Although the DESI images of these coordinates exist, this region was masked on the DESI source detection pipeline, and thus the host galaxy candidates are not listed in the source catalogue.
We used the computed DECam griz magnitudes and the WISE W1 and W2 magnitudes to perform an SED fit of the host galaxy with CIGALE. Because we do not have a spectroscopic or photometric redshift constraints for this galaxy, its physical properties are poorly constrained. 
\texttt{CIGALE} finds a loosely constrained redshift \zphot$=0.29\pm0.26$, and subsequent parameters \logMstar$<10.6$, SFR$<26$\,\Msunyr, \met$<0.2$, and age $2\pm1$\,Gyr.
Once a spectroscopic redshift is obtained for this FRB, refined SED fit results will be reported in a future publication.

\subsubsection{FRB\,20230613A}

\begin{figure}
    \centering
    \includegraphics[width=\linewidth]{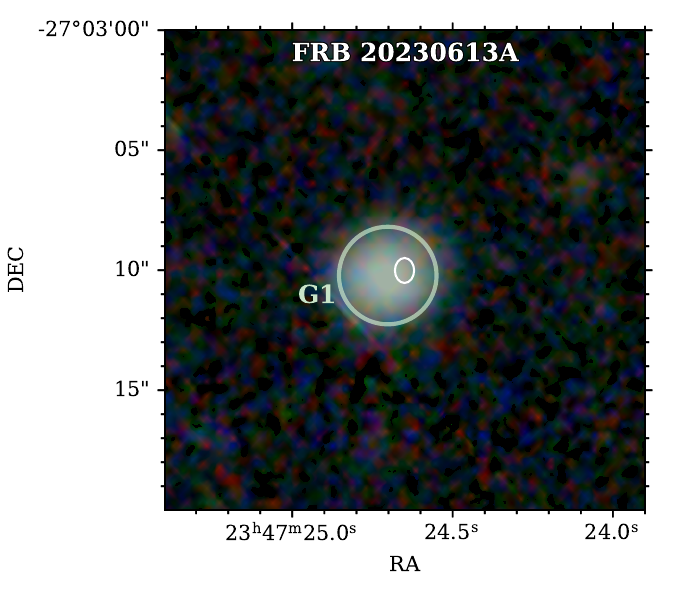}
    \caption{Localisation of FRB~20230613A. The error region is shown by a white ellipse, and the background image is from DESI-DR10. The candidate host galaxy is indicated by the coloured circle labelled G1.}
    \label{fig:frb20230613}
\end{figure}

This FRB was detected with a DM of 483.51$\pm$0.01\,\pccm\ in CB~33 (observation with Proposal ID: SCI-20220822-MC-01), which provides us with an initial position estimate. Its detection triggered the storage of the TB data, which allowed us to localise the FRB with difference images.
At the ToA of the FRB, one of the \textit{Galileo} satellites, part of the global navigation satellite system (GNSS)\footnote{\textit{Galileo}: \url{https://defence-industry-space.ec.europa.eu/eu-space/galileo-satellite-navigation_en}}, was passing approximately $1.6^{\circ}$ away from the phase centre of the observation. Despite this relatively large distance, it appeared as an extremely bright source, producing strong artifacts that were difficult to clean in images with such short integration times. Given the significant distance from the phase centre, the satellite position was located outside of the main lobe of the observation. However, because the power profile is frequency-dependent, the satellite's position fell within the sidelobes in some frequency subbands, while it coincided with the nulls between them in other subbands \citep{de_villiers_meerkat_2022}. As a result, we had to flag out 36 of the 64 frequency subbands from the TB data to localise the FRB.


The FRB was fortunately bright enough to be localised in this reduced subband number. After locating it in the difference images (see Fig.~\ref{fig:on-off}), we performed an astrometric correction using VLASS, and determined its final localisation region to be \coordserr{23}{47}{24.65}{0.40}{$-$27}{03}{10.01}{0.52}. On DESI, we identified a galaxy centred 0.73\arcsec away from the FRB location, with a magnitude $r=20.132\pm0.004$, resulting in an association probability 99.9\% after assuming an unseen prior of 7\%. Hence we determine this galaxy to be the putative host, which can be seen in Fig.~\ref{fig:frb20230613}.

To determine the redshift of the FRB, we obtained a spectrum with Keck/DEIMOS on 11 August 2023 (Program ID: U051; PI: X. Prochaska). We used the 600ZD grating at 700\,nm central wavelength for a total exposure time of 2835\,s. In the spectrum, we identified \Halpha, \Nii, \Hbeta, \Oii\ and \Oiii\ lines, which we used to determine the redshift to be $z=0.3923$ (see Fig.~\ref{fig:host_spectra}), consistent with the photometric redshift of the source, \zphot$=0.42\pm0.03$, as well as the expected redshift from the Macquart relation, $z_{\text{Macquart}}=0.48^{+0.14}_{-0.26}$. Given the redshift, the projected physical offset between the galaxy centre and the FRB location is $3.4\pm3.2$\,kpc.

We noticed that the FRB was located $\sim5.5$\arcmin\ away from the galaxy cluster Abell~4038, located in the foreground of the FRB host galaxy with a redshift of $z\sim0.03$ \citep{abell_catalog_1989, lopes_optical_2018}.
The cluster has a characteristic radius $R_{500}=1.25$\,Mpc, while the impact parameter of the FRB is roughly twice that, 2.43\,Mpc. With its mass $M_{500}=5.78\times10^{10}$\,\Msun, and following the method described in Section~\ref{sec:host_gal_id}, we find that Abell~4038 could contribute 100--200\,\pccm\ to the DM of the FRB.

The host galaxy of FRB~20230613A has broad photometric coverage from optical to infrared, including the $griz$ DECam filters and the WISE W1 and W2 filters, although no W3 and W4 magnitudes are available. We used the measured magnitudes in each filter to fit the SED with CIGALE. We find the stellar mass of the galaxy to be \logMstar$=10.1\pm0.2$, the star formation rate SFR$=4.5$\,\Msunyr\ with 0.3\,dex error, the metallicity $\log(Z/Z_\odot)<-0.13$, and the age of the galaxy $2.0\pm0.8$\,Gyr.

\subsubsection{FRB\,20230814F}

\begin{figure}
    \centering
    \includegraphics[width=\linewidth]{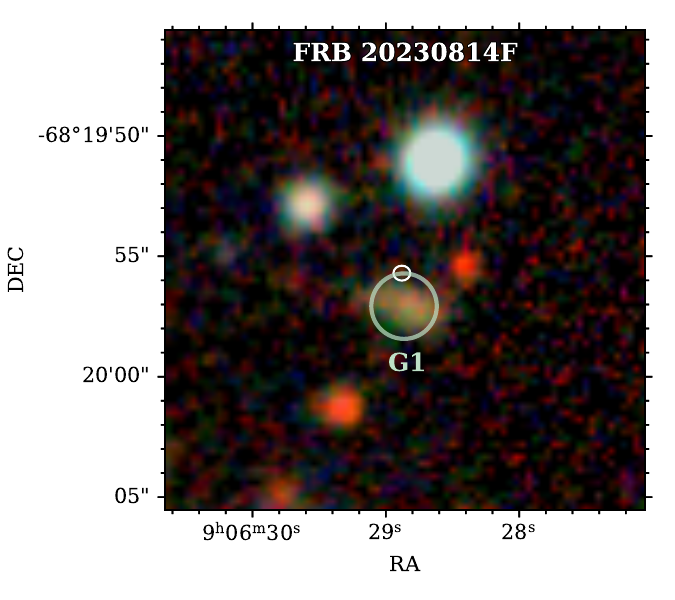}
    \caption{Localisation of FRB~20230814F. The error region is shown by a white ellipse, and the background image is a composite RGB image using DECam data. The candidate host galaxies are indicated by coloured circles labelled G1-2.}
    \label{fig:frb20230814}
\end{figure}

FRB\,20230814F is the first ever one-off FRB discovered at S-band. It has a DM of 471.44$\pm$0.03\,\pccm, and further details will be published in an accompanying paper (Pastor-Marazuela et al. \textit{in prep.}). It was found during a gain calibration observation of the calibrator J0906$-$6829\footnote{MeerKAT gain calibrators: \url{https://skaafrica.atlassian.net/wiki/spaces/ESDKB/pages/1452146701/L-band+gain+calibrators}} (Proposal ID: SCI-20200703-MK-03). The burst was detected in many CBs as well as the IB, and its detection triggered the storage of the TB data, which we used for the localisation. Given the high S/N of the burst as well as the calibrator, we used deeper cleaning parameters with \texttt{WSClean} to remove artifacts and accurately determine the position of the burst and other sources in the field. The off/on images are shown in Fig.~\ref{fig:on-off}. We used other sources in the FoV to perform the astrometric correction with the RFC and RACS-mid catalogues, and determined its final error region to be \coordserr{09}{06}{28.88}{0.35}{$-$68}{19}{55.70}{0.31}. Given this corresponds to a Galactic latitude $\sim-13\fdg95$, these coordinates are not covered by DESI, whose sky coverage is bounded by $|b|>18\degr$. However, optical imaging already existed in the NOIRLab archive from DECam for the g, r, and i filters\footnote{Astro Data Lab image cutout access: \url{https://datalab.noirlab.edu/sia.php}} \citep{fitzpatrick_noao_2014, nikutta_data_2020}. We thus ran the source finding method described in Section~\ref{sec:optical_obs} and built an RGB composite image from the $irg$ data, as shown in Fig.~\ref{fig:optical}. We identify an extended source at the FRB position with a magnitude $r=22.11\pm0.09$, and a separation of 1.37\arcsec. By assuming an unseen prior of 1\%, we find the \texttt{PATH} association probability to be 92.0\%, and we thus identify it as the putative host. The FRB localisation and optical background are shown in Fig.~\ref{fig:frb20230814}.
A spectrum of the host galaxy, required to determine its redshift, has not been obtained yet.
Because the existing photometric coverage is limited, we do not attempt to perform an SED fit.

\subsubsection{FRB\,20230827E}

\begin{figure}
    \centering
    \includegraphics[width=\linewidth]{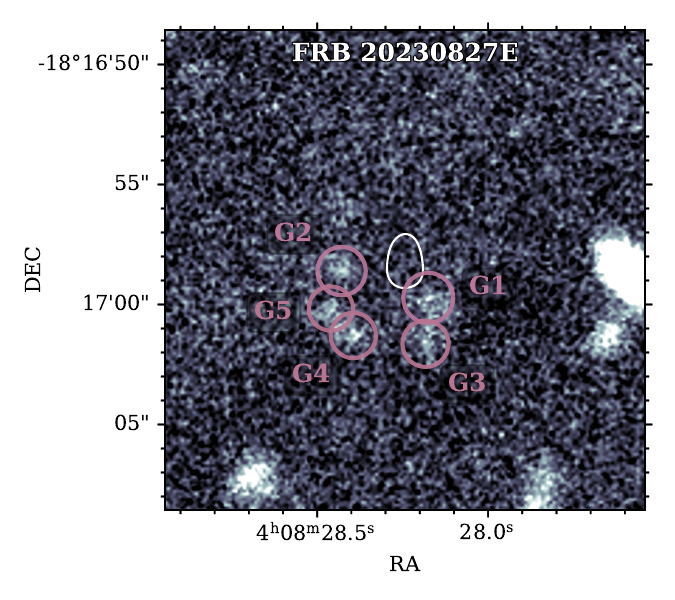}
    \caption{Localisation of FRB~20230827E. The error region is shown by a white ellipse, and the background image is from a Keck/DEIMOS R-band observation. The candidate host galaxies are indicated by coloured circles labelled G1-5.}
    \label{fig:frb20230827}
\end{figure}

This bright FRB was detected with a DM of 1433.7$\pm$0.1\,\pccm, in a gain calibration observation of J0409-1757 (Proposal ID: EXT-20210823-RT-02). It was detected in the IB and many CBs, but unfortunately the clustering of radio transient candidates in the MeerTRAP pipeline took longer than the allowed 45\,s after the burst ToA (see Section~\ref{sec:tb}), and hence the TB data were not triggered. The large number of beams where it was detected, however, was sufficient to achieve a sub-arcsecond localisation with SeeKAT. The coordinates of the FRB we determined are \coordserr{04}{08}{28.24}{0.75}{$-$18}{16}{58.55}{$^{1.50}_{0.75}$}.

The DESI DR10 archival images did not reveal any potential host galaxy coincident with the FRB location at the magnitude limit of 25.02. This is unsurprising given the high DM of the burst. Because of the high redshift expected for this source from the Macquart relation, the emission from the host galaxy is expected to be redshifted to the near-infrared (NIR) band. We thus obtained two 45 minute Keck/DEIMOS observations, one in the R-band on 2023 September 6, and the other in the Z-band on 2023 October 7 (Program ID: O438; PI: A. Gordon). These observations revealed three to five faint sources coincident with the FRB location in the R-band, as indicated in Fig.~\ref{fig:frb20230827} by G1 to G5, but they were not detected in the Z-band.
The five main host galaxy candidates, located from 1.81\arcsec to 3.78\arcsec away from the best FRB position, have magnitudes ranging from 25.6 to 26.4. 
Given these magnitudes, obtaining sufficient signal to determine the redshift of the potential host galaxy is challenging with ground-based telescopes.

In the direction of this FRB, the MW ISM contribution to the DM ranges from 134\,\pccm\ to 210\,\pccm\ according to the NE2001 and YMW16 models respectively.  The MW halo is expected to contribute $\sim50$ additional DM units, resulting in an excess DM $\sim1212$\,\pccm. This corresponds to an expected redshift $z_{\text{Macquart}}=1.45^{+0.32}_{-0.68}$, which could make this FRB one of the most distant localised yet \citep{ryder_luminous_2023, connor_gas_2024, caleb_fast_2025}. We have not identified any galaxy cluster overlapping with this FRB's sightline within twice its critical radius in the MCXC-II catalogue or the DESI galaxy cluster catalogue that could significantly contribute to the observed DM.

After performing a \texttt{PATH} analysis assuming an unseen prior of 10\%, we could not identify any of the candidate host galaxies as the host, with their posterior probabilities of association ranging from 0.8\% for G1, 0.7\% for G2, and <0.01\% for G3 to G5. The resulting unseen posterior is 98.4\%.

The configuration of the galaxy candidates is reminiscent of FRB~20220610A \citep{ryder_luminous_2023}, which was localised to a compact galaxy group at $z\sim1$ \citep{gordon_fast_2023}.
If the galaxies in this field are physically associated, the FRB may also originate in a dense or interacting group that could enhance both recent star formation and local dispersion. Spectroscopic or integral field unit (IFU) observations will be crucial to determine the redshifts of the candidates and to confirm whether they form a physically bound group.
These deeper observations could allow us to identify any faint galaxy co-located with the FRB position.

\subsubsection{FRB\,20230907D}

\begin{figure}
    \centering
    \includegraphics[width=\linewidth]{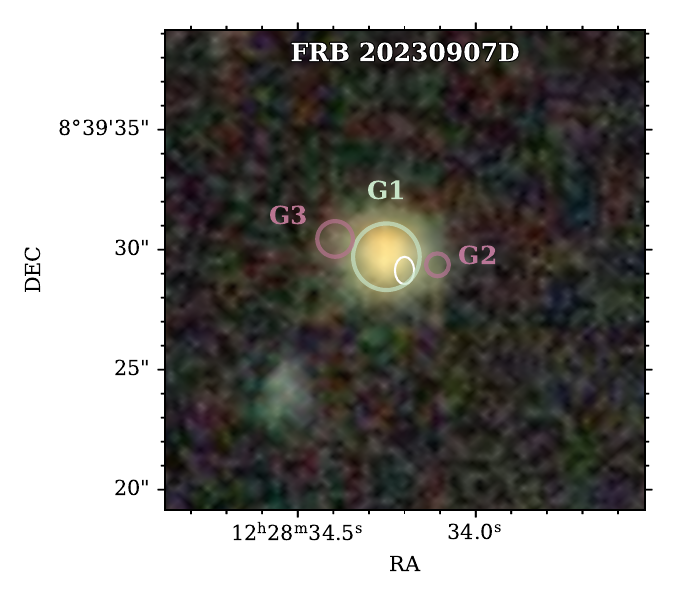}
    \caption{Localisation of FRB~20230907D. The error region is shown by a white ellipse, and the background image is from DESI-DR10. The candidate host galaxy is indicated by a coloured circle labelled G1.}
    \label{fig:frb20230907}
\end{figure}

\begin{figure}
    \centering
    \includegraphics[width=\linewidth]{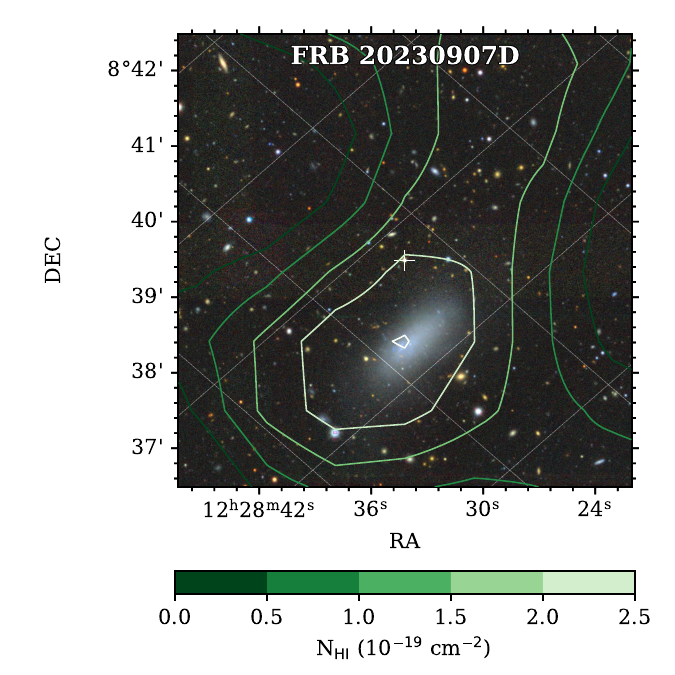}
    \caption{Localisation of FRB~20230907D near the low surface brightness galaxy UGC\,7596. The white cross indicates the FRB and host galaxy position, and the green contours the HI column density from \citet{sorgho_h_2017}.}
    \label{fig:frb20230907_hi}
\end{figure}

FRB\,20230907D was discovered during an L-band ThunderKAT observation (Proposal ID: SCI-20180516-PW-04). The burst, with a DM of 1030.79$\pm$0.04\,\pccm\ in the IB, triggered the storage of the TB data. After locating the FRB in the images generated from the TB data (see off/on images in Fig.~\ref{fig:on-off}), we performed an astrometric correction with VLASS. We determined the final coordinates of the burst to be \coordserr{12}{28}{34.20}{0.40}{+08}{39}{29.13}{0.57}. 

At the FRB location, our source finding pipeline identified a single galaxy with a magnitude $r=19.83\pm0.09$, centred 0.92\arcsec away from the FRB position, as shown in Fig.~\ref{fig:frb20230907}. However, DESI-DR10 identifies three separate sources, a brighter one G1 with an r-band magnitude of $19.50\pm0.01$, and two dimmer ones, G2 and G3, with magnitudes of $23.1\pm0.1$ and $23.0\pm0.1$, located respectively 0.92\arcsec, 1.39\arcsec, and 3.18\arcsec\ away from the FRB localisation centre.

The propagation path of the FRB traversed 2.7\,kpc away from the centre of the low surface brightness galaxy UGC\,7596\footnote{UGC\,7596: \href{https://simbad.cds.unistra.fr/simbad/sim-id?Ident=UGC+7596&NbIdent=1&Radius=2&Radius.unit=arcmin&submit=submit+id}{SIMBAD}} \citep{nilson_uppsala_1973}, with a redshift $z=0.001886$ or a distance of 4.6\,Mpc \citep{karachentsev_updated_2013}, belonging to the Southern filament of the Virgo Cluster \citep[][see Fig.~\ref{fig:frb20230907_hi}]{sorgho_h_2017}. Given the large angular extent of the Virgo Cluster in the sky, although the FRB location has an angular separation of $\sim4$\degr\ with respect to the cluster centre, its contribution to the DM can be significant. To compute the expected DM contribution from the Virgo cluster to the DM of this FRB, we followed the method described in \cite[][Eq.~4]{agarwal_fast_2019}, where they use the electron density distribution of the cluster derived from the \textit{Planck} data \citep{planck_collaboration_planck_2016}, instead of the typical cluster parameters described in Section~\ref{sec:host_gal_id}, and integrate at the FRB impact parameter (1.17\,Mpc) up to twice the Virial radius of the cluster (2.4\,Mpc). Using this method, we estimate the contribution of the Virgo cluster to the FRB DM to be $\sim230$\,\pccm. Additionally, UGC\,7596 could contribute $\sim50$\,\pccm\ to the FRB DM based on its stellar mass and optical radius \citep{diaz-garcia_characterization_2016}.

By subtracting these two contributions from the extragalactic DM of the FRB, we expect its redshift to be $z_{\text{Macquart}}=0.84^{+0.18}_{-0.44}$. Based on the Macquart redshift upper limit, we apply an unseen prior of 40\%\ to the \texttt{PATH} analysis. If we consider the three galaxies identified in the DESI catalogue, we find the association probability to be 
$P(O|x) = 94.2\%$ for G1, which we hence identify as the putative host, 2.7\% for G2, and 0.3\% for G3. If instead we only consider G1 as identified by our source finding algorithm, we find the association probability to be 98.5\%.

To determine the redshift of the host, we observed the galaxy with Keck/LRIS on 5 January 2024 (Program ID: O438; PI: A. Gordon). We observed with the 400/3400 blue grism and the 400/8500 red grating at 783\,nm for a total exposure time of 2400\,s. The spectrum revealed several bright emission lines, including \Halpha, \Hbeta, abd the \Nii, \Sii, and \Oiii\ doublets, which allowed us to determine a spectroscopic redshift $z=0.464\pm0.015$. This redshift is well in agreement with the DESI photometric redshift, \zphot$=0.44^{+0.03}_{-0.02}$, 
and it is at the lower end of what we expect from the Macquart relation after removing the Virgo cluster and UGC~7596, $z_{\text{Macquart}}=0.84^{+0.18}_{-0.44}$. 
Motivated by this large apparent DM excess, we searched for additional intervening structures and identified three galaxy clusters listed in \citet{zou_galaxy_2021} located in the foreground of the FRB. These clusters correspond to the catalogue identifiers 2074100051, 2074100062, and 2125300112. If we assume typical galaxy cluster profiles (Section~\ref{sec:host_gal_id}), their combined DM contribution would exceed the observed value. Since these clusters are not detected in X-rays, our adopted cluster parameters are likely uncertain, and the actual DM contribution may be substantially lower. A detailed analysis of the DM budget for this and other FRBs in the sample will be presented in an upcoming publication.


The putative host galaxy of FRB~20230907D has a broad photometric coverage in the optical and IR (DECam $griz$ and WISE W1-4), and we thus fitted the SED with CIGALE. We find the galaxy stellar mass to be \logMstar$=10.9\pm0.2$, the SFR$=14.9$\,\Msunyr\ with a 0.3\,dex uncertainty, the metallicity \met$<-0.11$, and the age $2\pm1$\,Gyr.
The projected physical offset between the galaxy centre and the FRB location is $5.7\pm4.2$\,kpc.

\subsubsection{FRB\,20231010A}

\begin{figure}
    \centering
    \includegraphics[width=\linewidth]{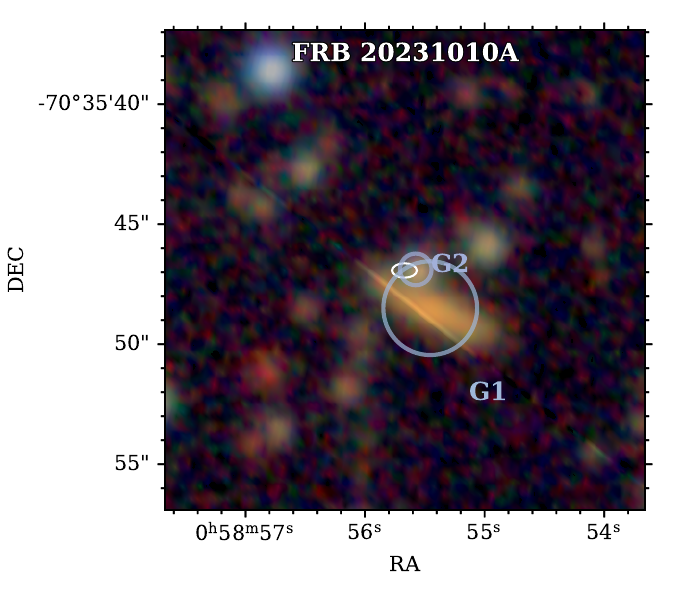}
    \caption{Localisation of FRB~20231010A. The error region is shown by a white ellipse, and the background image is from DESI-DR10. The candidate host galaxies are indicated by coloured circles labelled G1-2. The diagonal line that crosses the FRB location is a product of the overlap between two adjacent exposures. The line going through G1 is an artifact from the CCD detector edges of the DESI observations.}
    \label{fig:frb20231010}
\end{figure}

This FRB was detected during an observation of the MeerTime project for pulsar timing \citep{bailes_meerkat_2020} (Proposal ID: SCI-20180516-MB-04).
The FRB had a DM of 442.59$\pm$0.02\,\pccm, and it was found in CB 437. The burst detection triggered the storage of the TB data, which was used to localise it in the difference images, shown in Fig.~\ref{fig:on-off}. After the astrometric correction with RFC and RACS, we obtained the final position \coordserr{00}{58}{55.67}{0.52}{$-$70}{35}{46.93}{0.30}. In the DESI DR10, images, we find this position to be close to an edge-on galaxy as well as a fainter galaxy, as can be seen in Fig.~\ref{fig:frb20231010}. 
The edge-on galaxy G1 has a magnitude of $r=21.2\pm0.6$ and a half-light radius of 1.95\arcsec, and its centre has a separation of 1.9\arcsec from the FRB localisation centre. In contrast, the fainter galaxy G2 has an r-band magnitude of $22.5\pm0.8$, a half-light radius of 0.66\arcsec, and is located 0.47\arcsec away from the FRB best coordinates. Assuming an unseen prior of 5\%, \texttt{PATH} assigns a marginally higher probability of association to the brighter galaxy G1 (52.3\%) compared to the dimmer galaxy G2 (46.6\%). However, the photometric redshift from G1, $z_{\text{phot, G1}}=0.61\pm0.18$, is more in agreement with the expected redshift from the Macquart relation, $z_{\text{Macquart}}=0.42^{+0.12}_{-0.24}$, compared to G2 where $z_{\text{phot, G2}}=0.84^{+0.27}_{-0.30}$.
If the host galaxy was G2, there would be additional contributions to the DM from G1 located in the foreground, which could lead to a disagreement with the expected Macquart redshift.

The sky location where we localised this FRB has DECam $griz$ coverage, but no WISE photometry is available. We use \texttt{CIGALE} to both estimate a photometric redshift within the DESI uncertainties, and to fit the SED, but with results poorly constrained.
For G1 we find a \zphot$=0.61\pm0.09$, and the physical parameters \logMstar$<10.5$, SFR$=13.8$\,\Msunyr\ with a 0.5\,dex uncertainty, \met$<0.2$, and age $1.4\pm0.4$\,Gyr.
For G2, the initial fit obtained with the default AGN fraction of $<20\%$ produced an excess of infrared emission and a poorly constrained stellar mass. To investigate this, we tested AGN fractions of 5, 10, 20, and 30\%, and found that the best-fit solution was obtained for an AGN fraction of 5\%. For this model, we derive \zphot$=0.54\pm0.1$, \logMstar$=9.1\pm0.3$, SFR$=2.5$,\Msunyr\ with a 0.2\,dex error, \met$<-0.4$, and age$=1.1\pm0.1$\,Gyr.
The projected physical offset would be $13.2\pm4.2$\,kpc for G1 and $3.1\pm3.9$\,kpc for G2.
 
\subsubsection{FRB\,20231020B}

\begin{figure}
    \centering
    \includegraphics[width=\linewidth]{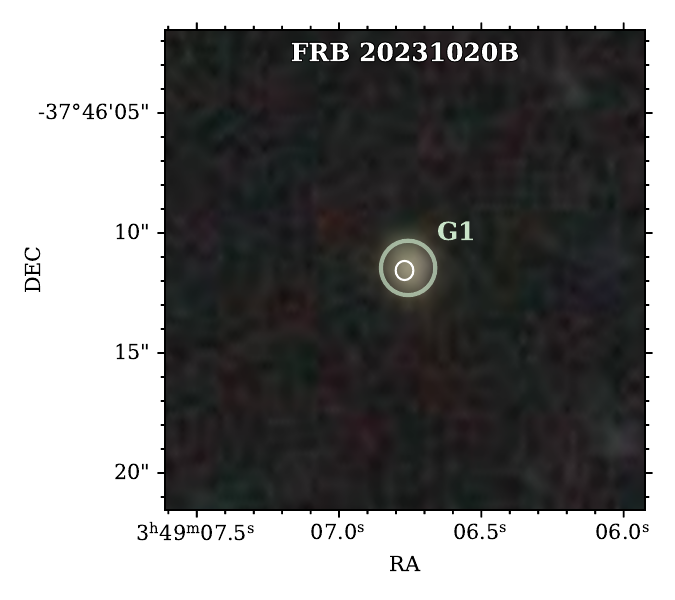}
    \caption{Localisation of FRB~20231020B. The error region is shown by a white ellipse, and the background image is from DESI-DR10. The candidate host galaxy is indicated by a coloured circle labelled G1.}
    \label{fig:frb20231020}
\end{figure}

FRB\,20231020B was detected with a DM of 952.2$\pm$0.3\,\pccm\ during a MHONGOOSE observation (Proposal ID: SCI-20180516-EB-01), a low-mass, late-type, gas-rich dwarf galaxy \citep{blok_meerkat_2020}. The burst was detected in the IB, and it triggered the storage of the TB data. We used the TB data to generate images where we localised the burst, as shown in Fig.~\ref{fig:on-off}. After performing an astrometric correction with VLASS, we determined the burst coordinates to be \coordserr{03}{49}{06.77}{0.37}{$-$37}{46}{11.56}{0.40}. In the archival DESI DR10 images, we find a single galaxy coincident with the FRB location, with a magnitude of $21.79\pm0.01$, and a photometric redshift \zphot$=0.46^{+0.07}_{-0.08}$, shown in Fig.~\ref{fig:frb20231020}. The \texttt{PATH} analysis determines the probability of association to be 99.8\%, assuming an unseen prior of 35\%.
In order to determine the spectroscopic redshift of the putative host galaxy, we obtained a Keck/DEIMOS observation (Program ID: U051; PI: X. Prochaska) on 14 December 2023. We used the 600ZD grating at 650\,nm central wavelength for a total exposure time of 730\,s. From the \Oiii, \Hbeta\ and \Oii\ spectral lines, we determined the redshift of the host to be 0.477. The redshift is in good agreement with the photometric redshift, but it is lower than what we expect from the Macquart relation, $z_{\text{Macquart}}=1.04^{+0.24}_{-0.52}$.

We searched for galaxy clusters in the catalogue from \cite{zou_galaxy_2021}, and identified the cluster with ID~3959300018 and photometric redshift $\sim0.327$ to be centred $\sim127$\arcsec\ away from the FRB location. The cluster has a characteristic radius $R_{500} = 526$\,kpc, while the impact parameter of the FRB is 620\,kpc. Its mass is $M_{500} = 5.75 \times 10^{13}$\Msun.
Following the method described in Section~\ref{sec:host_gal_id}, we compute the expected DM contribution from the foreground cluster to be up to $\sim400$\,\pccm\ in the cluster frame, or $\sim300$\,\pccm\ after correcting for redshift. This contribution from the foreground cluster could explain the observed DM.

This sky location has photometric coverage in the DECam $griz$ filters and the WISE W1-3 filters, which we used to perform an SED fitting with CIGALE. We obtain a mass \mbox{\logMstar$=10.7\pm0.3$}, an SFR$=14$\,\Msunyr\ with a 1\,dex uncertainty, a metallicity \met$<0.1$, and an age of $2.2\pm1.4$\,Gyr.
The projected physical offset between the FRB location and the host galaxy centre is $1.2\pm3.3$\,kpc.

\subsubsection{FRB\,20231210F}

\begin{figure}
    \centering
    \includegraphics[width=\linewidth]{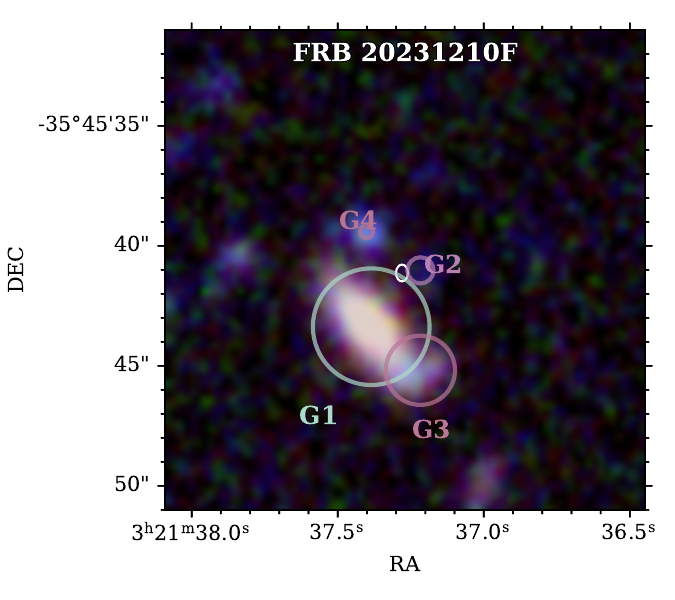}
    \caption{Localisation of FRB~20231210F. The error region is shown by a white ellipse, and the background image is a composite RGB image from DESI-DR10. The candidate host galaxies are indicated by coloured circles labelled G1-4.}
    \label{fig:frb20231210}
\end{figure}

This burst was detected in CB 306 during an observation of the MeerKAT Fornax Survey \citep[MFS,][Proposal ID: SCI-20180516-PS-01]{serra_meerkat_2023}, with a DM of 720.6$\pm$0.2\,\pccm. The detection triggered the TB data, from which we generated images to localise the burst, as shown in Fig.~\ref{fig:on-off}, and then we performed an astrometric correction of the position with VLASS. The resulting FRB coordinates are \coordserr{03}{21}{37.28}{0.25}{$-$35}{45}{41.13}{0.25}. In the DESI DR10 images, we find this position to coincide with an edge-on galaxy G1, and a fainter galaxy G2, apparently in the background based on the photometric redshift, in a similar configuration to those of FRB\,20231010A (see Fig.~\ref{fig:frb20231210}). The edge-on galaxy has a magnitude of $21.1\pm0.01$ and it is located at 2.58\arcsec from the FRB location, while the second galaxy has a magnitude of $r=26.4\pm1.0$ and a separation of 0.78\arcsec. Two further galaxies, G3 and G4, fainter than the first one but brighter than the second one, are located respectively 4.12\arcsec and 2.27\arcsec away from the FRB location.
Assuming an unseen prior of 20\%, the \texttt{PATH} analysis finds a probability of association of 81.0\% with G1, 14.8\% with G2, 1.6\% with G3, and 0.03\% with G4. Although the probability does not reach a 90\% confidence for G1, we still identify it as the putative host. A spectrum of the two brightest potential host galaxies is pending to be obtained, and although we cannot yet determine the spectroscopic redshift, DESI-DR10 reports a photometric redshift of $0.50\pm0.08$ for the G1, and $1.37\pm0.43$ for the G2. The Macquart relation predicts a redshift $z_{\text{Macquart}}=0.78^{+0.20}_{-0.40}$, which also favours G1 as the host galaxy.

The coordinates of this FRB have photometric coverage in all DECam $griz$ filters and in the WISE W1, W2, and W4 filters, which we used to fit the SED for G1. We find an optimal photometric redshift of \zphot$=0.43\pm0.01$, a mass \logMstar$=9.7\pm0.1$, which is the lowest in out host galaxy sample, SFR$<0.2$\,\Msunyr, metallicity \met$<-2.2$, and age $1.0\pm0.1$\,Gyr.
The projected physical offset would be $19.6\pm3.7$\,kpc. 

\subsection{Arcminute localised FRBs}

\begin{figure*}
    \raggedright
    \includegraphics[height=7.5cm]{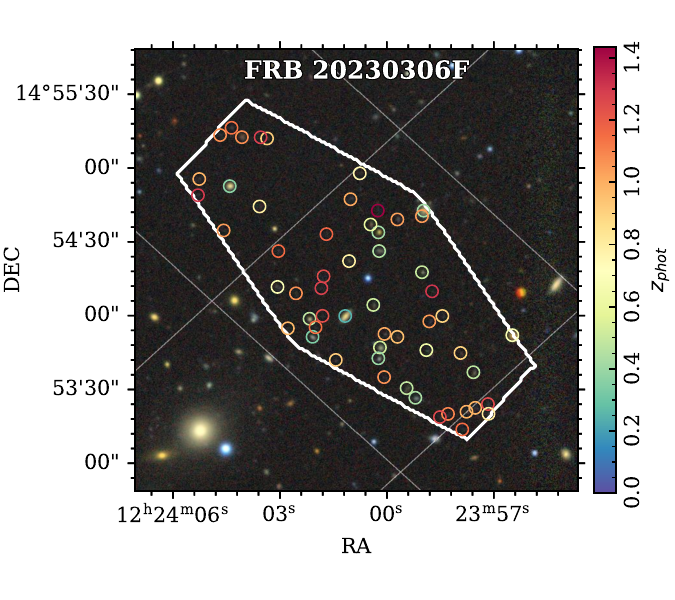}
    \includegraphics[height=7.5cm]{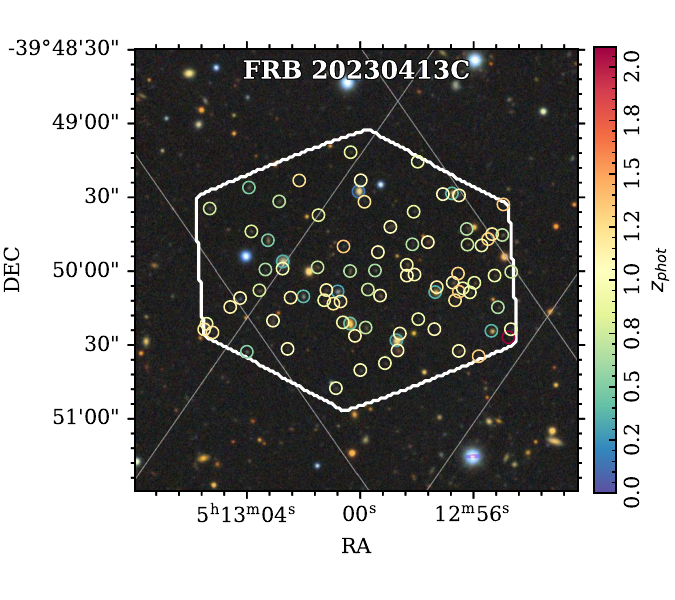}\\
    \includegraphics[height=7.5cm]{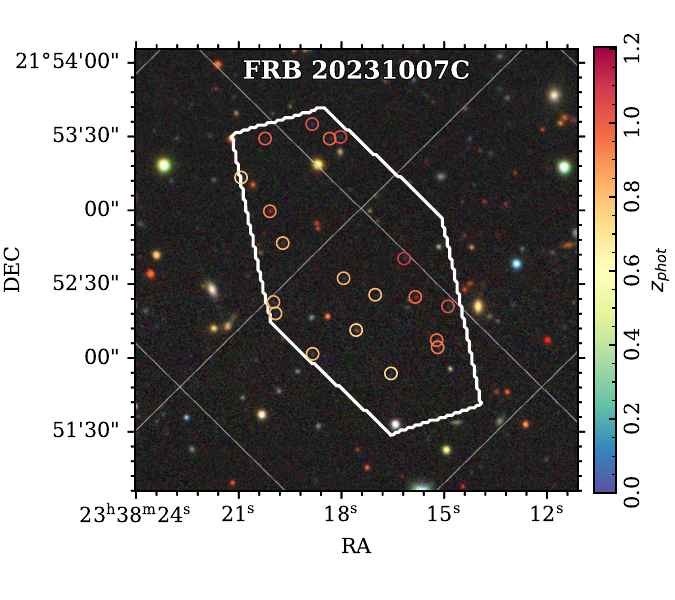}
    \includegraphics[height=7.5cm, trim=0.0cm 0.0cm 1.45cm 0.0cm, clip]{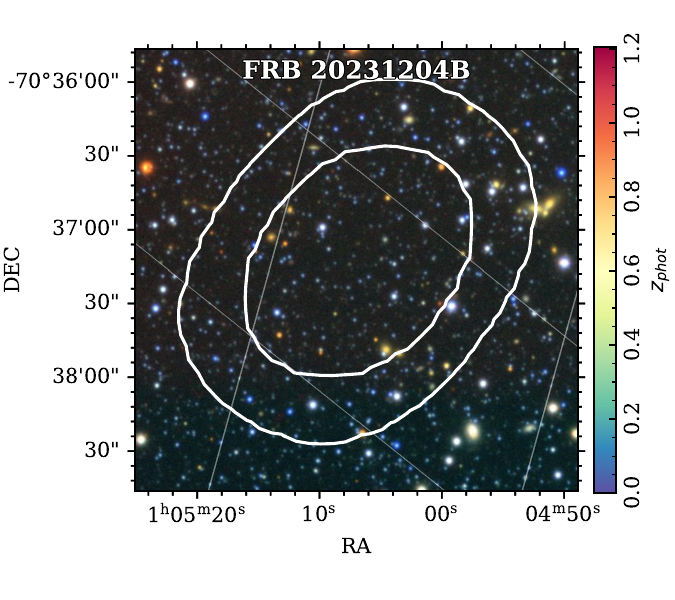}
    \caption{SeeKAT FRB localisations. The white contours show the FRB localisation regions, the background shows the DESI-Legacy image, and the coloured markers show the position of the known galaxies within the error regions, going from blue to red with increasing photometric redshift.}
    \label{fig:seekat_localisations}
\end{figure*}

Some of the FRBs that were detected in 2022 and 2023 did not trigger the storage of the TB data and were too faint to be detected in several CBs or were localised by imaging the 2-s/8-s data that were taken simultaneously to the MeerTRAP observations. In this section, we present the four FRBs detected during the 2023 observations that we could not accurately localise. We show their localisation areas and the sources identified as galaxies within those in Fig.~\ref{fig:seekat_localisations}.

\subsubsection{FRB\,20230306F}

FRB\,20230306F, with a DM of 689.5$\pm$0.9\,\pccm, was detected during an observation of the Virgo Cluster in CB 149 (Proposal ID: SCI-20220822-FD-02). The TB was triggered $<2$\,s after the maximum possible time delay, and thus the voltage data could not be saved. 
Given the low detection S/N$\sim10$, a localisation through imaging of the 8-s data was not possible. Since it was detected only in a single CB, we could only constrain its localisation to the hexagon shown in Fig.~\ref{fig:seekat_localisations} centred at the coordinates \coordserr{12}{24}{01}{60}{+14}{54}{17}{50}. The DESI-DR10 catalogue contains 56 sources identified as galaxies within this error region and the photometric redshift upper limit of 0.94 (within errors) expected from the Macquart relation.
The DESI DR10 $r$-band limiting magnitude in the direction of this FRB is 24.9, so the resulting unseen prior is $\sim23$\% at this limiting redshift.
We ran a \texttt{PATH} analysis on these galaxies, and the highest probability of association, given to the brightest galaxy with an r-band magnitude of 19.81, is 3\%. The next most likely host has an association probability of 1.3\%, and all other galaxies are lower than 1\%. Meanwhile, the probability of an unseen host is around 90\%. This confirms that with such large localisation errors it becomes unfeasible to identify the host galaxy.

\subsubsection{FRB\,20230413C}

This FRB was detected during a UHF MeerTime observation (Proposal ID: SCI-20180516-MB-04) with a DM of 1532.2$\pm$0.5\,\pccm.
Unfortunately, the burst was detected when the observation had just started, and since the buffers were still being cleared, the TB data were not triggered.
The burst was not detected in any other beams, and although its S/N of $\sim15$ was higher than that of FRB\,20230306F, localisation was still not possible through the imaging of the simultaneous 8-s resolution data. The SeeKAT localisation thus resulted in the hexagon shown in Fig.~\ref{fig:seekat_localisations}, centred around the coordinates \coordserr{05}{13}{00}{63}{$-$39}{50}{00}{57}. 
This region contains 84 sources identified as galaxies in the DESI-DR10 catalogue with photometric redshifts below the maximum of 2.11 expected from the Macquart relation, or 60 between a minimum of 0.92 and the maximum of 2.11 (including $1\sigma$ photometric errors).
In this field, the DESI DR10 limiting $r$-band magnitude is 25.1, which results in an unseen prior of $\sim68$\%.
All galaxies in this sample have \texttt{PATH} association probabilities $<1$\%, while the probability of an unseen host is 98\%, similarly to the previous FRB. It is thus not feasible to confidently identify the host galaxy of this source.

\subsubsection{FRB\,20231007C}

This burst was detected in CB 324 during an Open Time observation \citep[][Proposal ID: SCI-20230907-TD-01]{healy_a2626_2021}. 
It has a DM of 2660.4$\pm$1.9\,\pccm, the largest of the current MeerTRAP FRB sample, but unfortunately the trigger arrived 0.5\,s too late and the TB data were not stored. Since it was a single CB detection with a S/N$\sim10$, we could only localise it with SeeKAT to the hexagon shown in Fig.~\ref{fig:seekat_localisations}, centred around the coordinates \coordserr{23}{38}{18}{50}{+21}{52}{34}{55}.
The DESI-DR10 catalogue contains 19 sources identified as galaxies between photometric redshifts of 0.9 --half of the expected minimum redshift from the Macquart relation-- and 3.74, the maximum expected redshift. The source number increases to 25 if we make no redshift selection. Additionally, many more galaxies probably exist within that volume but have magnitudes above the DESI depth limit. 
The $r$-band limiting magnitude in this field from the DESI DR10 is 24.3, so we would expect $\sim99$\% of the FRB hosts to not be seen at the redshift upper limit.
The \texttt{PATH} analysis finds the probability of an unseen host to be 99.95\%; it is thus not possible to confidently identify the host galaxy candidate.

\subsubsection{FRB\,20231204B}

FRB\,20231204B was found during a UHF MeerTime observation (Proposal ID: SCI-20180516-MB-04) with a DM of 1772.1$\pm$0.3\,\pccm.
At the time of detection, this was the highest DM ever found for a MeerTRAP UHF FRB.
It was detected in a CB (335) at the edge of the tiling and in the incoherent beam. Unfortunately, the trigger arrived 2\,s after the maximum allowed delay, and the TB data were not stored. Although the S/N$\sim25$ in the CB is higher than other FRBs in this section, it is still not enough to localise the FRB in the 8-s images.
By combining the CB and IB detections, we localised the FRB with SeeKAT to an annular region centred around the coordinates \coordserr{01}{05}{07}{64}{$-$70}{37}{16.5}{68}. Although some galaxies are visible by eye in the DESI-DR10 image shown in Fig.~\ref{fig:seekat_localisations}, the DESI source finding algorithm has not been run in that region due to the proximity to the Small Magellanic Cloud (SMC). Similarly to all other FRBs presented in this section, the large localisation area is expected to contain a considerable amount of galaxies, and hence determining the most likely host is not possible.
At the $r$-band limiting magnitude of 25.5 from DESI DR10, we expect $\sim68$\% would not be visible at the redshift upper limit.

\subsection{Host galaxy sample}

\begin{figure}
    \centering
    \includegraphics[width=\linewidth]{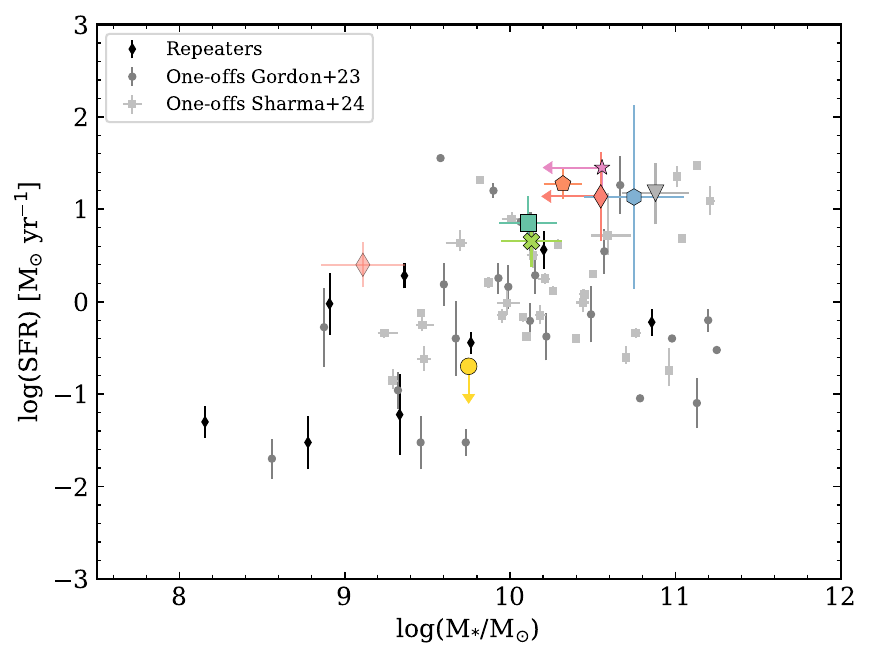}
    \includegraphics[width=\linewidth]{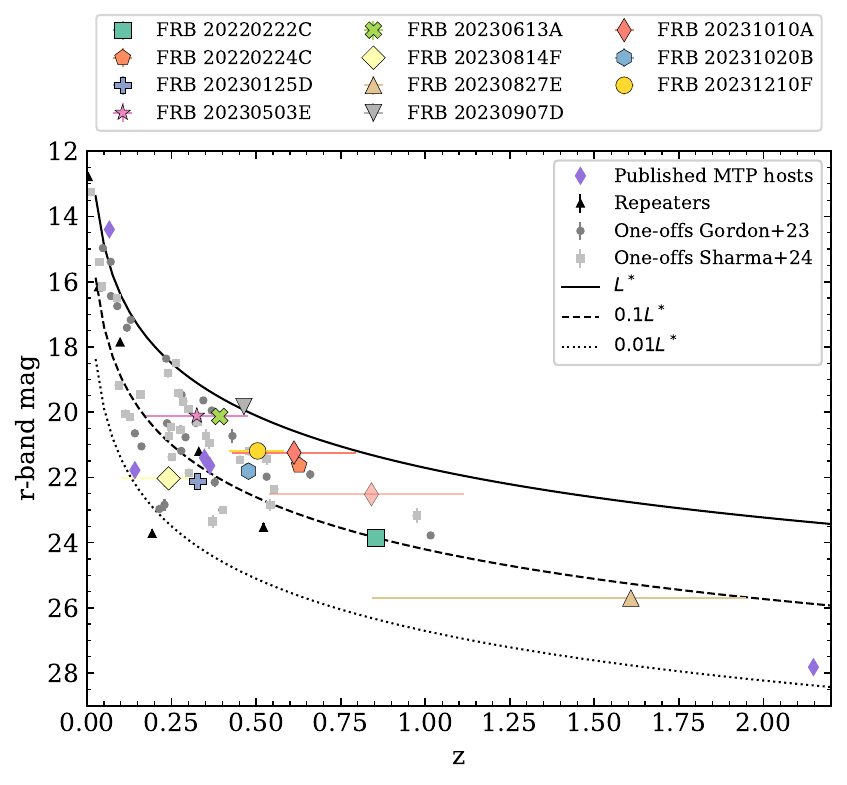}
    \caption{FRB host galaxy properties. The top panel shows the star formation rate as a function of stellar mass, and the bottom panel the r-band magnitude as a function of redshift. In each panel, the coloured markers show the FRBs presented in this work, the black triangles the host galaxy properties of repeaters, medium grey the galaxy properties of one-off FRBs presented in \citet{gordon_demographics_2023}, and light grey the galaxies presented in \citet{sharma_preferential_2024}. The purple diamonds are previously published MeerTRAP FRB hosts \citep{driessen_frb_2024, caleb_subarcsec_2023, hanmer_contemporaneous_2025, caleb_fast_2025}. Note that on the top panel, the error bars on the MeerKAT FRBs are too large to be shown.}
    \label{fig:gal_prop}
\end{figure}

For eight of the FRBs with arcsecond-scale localisation presented in this work, we had access to multi-wavelength photometric data, which we used to perform an SED fitting with \texttt{CIGALE}. This allowed us to estimate key parameters of the host galaxy candidates, including stellar mass, star formation rate, metallicity, and age. 
Figure~\ref{fig:gal_prop} provides a visual representation of some of these properties compared to previous samples of localised FRBs, including several repeaters, and those from \cite{gordon_demographics_2023}, and \cite{sharma_preferential_2024}. The top panel shows the stellar masses and star formation rates, while the bottom one shows $r$-band magnitudes as a function of redshift.

Our host galaxy sample has stellar masses that are consistent with the samples of repeaters, \cite{gordon_demographics_2023}, and \cite{sharma_preferential_2024}.
We assessed this by performing a Kolmogorov-Smirnov (KS) two-sample test to compare the underlying distributions of these galaxy properties, considering only the most likely hosts. Since all p-values exceed 0.05, we find no significant differences in the sample distributions in terms of stellar mass given the current dataset size.
However, the SFRs of the MeerTRAP FRB hosts appear to be larger than the sample of repeaters, with a p-value of 0.03. Our SFRs are consistent with those from \cite{gordon_demographics_2023} and \cite{sharma_preferential_2024}, where we find p-values $>0.05$. 
We note however that the galaxy SEDs were modelled with a different code using different assumptions from \cite{gordon_demographics_2023} and \cite{sharma_preferential_2024}, and thus our results might be subject to systematic differences.
While our characterisation of FRB host galaxies remains incomplete, in future work we will incorporate the missing hosts and host galaxy optical spectra into the SED fitting to refine our analysis.

Despite the limited sample size, MeerTRAP is producing a population of high redshift FRBs. 
A KS test comparing the redshifts (spectral or photometric) of MeerTRAP FRB hosts with those of all previously localised FRBs returns p-values $<0.01$, suggesting a statistically significant difference in their redshift distributions.
Given the high sensitivity of MeerKAT and the high DM values found in the pre-existing MeerTRAP FRB sample \citep{jankowski_sample_2023, pastor-marazuela_comprehensive_2025}, which naturally predispose it to detecting higher-redshift sources, this result is consistent with our expectations.

\section{Conclusions} \label{sec:conclusions}

In this work, we have presented the localisation of 15 newly discovered FRBs within the MeerTRAP project in 2022 and 2023. 
While MeerTRAP has previously localised FRBs with (sub-)arcsecond accuracy \citep{driessen_frb_2024, caleb_subarcsec_2023, rajwade_study_2024, hanmer_contemporaneous_2025, tian_detection_2024, tian_meerkat_2025}, this work presents the first large sample of MeerTRAP localised FRBs.
The two FRBs from 2022 were localised in the 8-s commensal imaging observations, since the transient buffer (TB) system was not yet operational, achieving arcsecond precision. In 2023, eight FRBs were localised by imaging the 300~ms of TB data stored upon their detection. While five additional FRBs were localised using SeeKAT, only one of these had sufficient beam detections to achieve sub-arcsecond accuracy. The remaining four were constrained to larger regions, making host galaxy identification unfeasible. 

Among the 11 FRBs with (sub-)arcsecond localisation accuracy, we identified host galaxies with >90\% confidence for eight of them. One additional FRB has a likely host galaxy candidate with >80\% confidence. While the host of FRB~20231010A remains ambiguous based on the \texttt{PATH} analysis, with two nearby galaxies showing comparable association probabilities, the photometric redshift of the brighter, edge-on galaxy G1 is consistent with the range we expect from the Macquart relation. Obtaining a spectroscopic redshift for both host galaxy candidates could hence help us confirm the host of this FRB.
FRB~20230827E has five host candidates and a large unseen posterior. Its host galaxy could either be too faint to be detected in ground-based optical observations given its large excess DM of $\sim$1212\,\pccm. Alternatively, it could be associated to a high-redshift compact galaxy group, akin to FRB~20220610A \citep{gordon_fast_2023}.
For four FRBs in our sample, FRB~20220224C, FRB~20230613A, FRB~20230907D, and FRB~20231020B, we identify foreground galaxy clusters which significantly contribute to the observed DM, demonstrating how FRBs can be used to probe foreground structures. 

Spectroscopic redshifts were obtained for six of these hosts using Keck and Gemini observations, ranging from 0.32 to 0.85. 
Determining redshifts for potential host galaxies remains challenging, especially for FRBs with faint or distant hosts, such as those we typically localise with MeerTRAP. Timing constraints and securing optical follow-up observations add further difficulties. As our sample grows, photometric redshift catalogues, such as DESI-DR10, are becoming essential, and they will play an even greater role when large FRB localisation projects, such as the CHIME/FRB Outriggers \citep{lanman_chimefrb_2024}, begin detecting hundreds of FRBs yearly.

We estimated the galaxy properties of eight FRB host galaxies (or candidate hosts) with sufficient photometric coverage using the \texttt{CIGALE} code. The inferred stellar masses span from $10^{9.75}$ to $10^{10.94}$\,\Msun, while the star formation rates range from 0.1 to 28.5\,\Msunyr, although some of these values represent upper limits.
These stellar masses and SFR estimates are broadly consistent with those reported for other FRB samples \citep{gordon_demographics_2023, sharma_preferential_2024}, despite potential systematic differences arising from our different SED fitting techniques.
We emphasise that the availability of spectroscopic redshifts, broader photometric coverage, and the inclusion of spectroscopic data in the SED fitting process can substantially improve constraints on the host galaxy properties of our sample. These improvements will be explored in future work.

With an increasing number of localised FRBs, MeerTRAP is uniquely assembling one of the largest uniform samples of high redshift FRBs to date, following ASKAP \citep{shannon_commensal_2024} and DSA-110 \citep{law_deep_2024}. This dataset is crucial to use FRBs as cosmological tools, particularly for measuring the baryonic content of the IGM, a topic that will be explored in an upcoming publication (Caleb et al. \textit{in prep.}). The localisation of these distant FRBs brings us closer to answering fundamental questions about the distribution of cosmic baryons, the nature of their host galaxies, and the environments that give rise to these enigmatic bursts.

\section*{Acknowledgements}

We thank Amidou Sorgho for sharing the UGC7596 HI emission map with us. We would like to thank the Fornax, MeerTime, MHONGOOSE, and ThunderKAT LSPs for allowing us to observe commensally. 

The MeerTRAP collaboration acknowledges funding from the European Research Council under the European Union’s Horizon 2020 research and innovation programme (grant agreement No 694745).
IPM acknowledges funding from an NWO Rubicon Fellowship, project number 019.221EN.019.
BWS and JT acknowledge funding through a consolidated grant from STFC.
ACG and the Fong Group at Northwestern acknowledges support by the National Science Foundation under grant Nos. AST-2206494 and CAREER grant No. AST-2047919.  ACG acknowledges support from NSF grants AST-1911140, AST-1910471 and AST-2206490 as a member of the Fast and Fortunate for FRB Follow-up team. W. M. Keck Observatory access was supported by Northwestern University and the Center for Interdisciplinary Exploration and Research in Astrophysics (CIERA). 
IK and NT would like to acknowledge the support received by the Joint Committee ESO-Government of Chile grant ORP 40/2022.
NT acknowledges support by FONDECYT grant 1252229.
MC acknowledges support of an Australian Research Council Discovery Early Career Research Award (project number DE220100819) funded by the Australian Government.
SS gratefully acknowledges support by the Brinson Foundation as a joint NU-UC Brinson Postdoctoral Fellow.
W.F. gratefully acknowledges support by the David and Lucile Packard Foundation, the Alfred P. Sloan Foundation, and the Research Corporation for Science Advancement through Cottrell Scholar Award \#28284.
 LK is supported by the National Science Foundation under grant Nos. AST1909358 and AST-2308182 and CAREER grant No. AST2047919.


The MeerKAT telescope is operated by the South African Radio Astronomy Observatory, which is a facility of the National Research Foundation, an agency of the Department of Science and Innovation. \\
This work has made use of the "MPIfR S-band receiver system" and the FBFUSE  backend, designed, constructed and maintained by funding of the MPI für Radioastronomie.\\
Some of the data presented herein were obtained at Keck Observatory, which is a private 501(c)3 non-profit organization operated as a scientific partnership among the California Institute of Technology, the University of California, and the National Aeronautics and Space Administration. The Observatory was made possible by the generous financial support of the W. M. Keck Foundation. \\
Based on observations obtained at the international Gemini Observatory, a program of NSF NOIRLab, which is managed by the Association of Universities for Research in Astronomy (AURA) under a cooperative agreement with the U.S. National Science Foundation on behalf of the Gemini Observatory partnership: the U.S. National Science Foundation (United States), National Research Council (Canada), Agencia Nacional de Investigación y Desarrollo (Chile), Ministerio de Ciencia, Tecnología e Innovación (Argentina), Ministério da Ciência, Tecnologia, Inovações e Comunicações (Brazil), and Korea Astronomy and Space Science Institute (Republic of Korea).\\
(Some of) The data presented herein were obtained at the W. M. Keck Observatory, which is operated as a scientific partnership among the California Institute of Technology, the University of California and the National Aeronautics and Space Administration. The Observatory was made possible by the generous financial support of the W. M. Keck Foundation.\\
The DESI Legacy Imaging Surveys consist of three individual and complementary projects: the Dark Energy Camera Legacy Survey (DECaLS), the Beijing-Arizona Sky Survey (BASS), and the Mayall z-band Legacy Survey (MzLS). DECaLS, BASS and MzLS together include data obtained, respectively, at the Blanco telescope, Cerro Tololo Inter-American Observatory, NSF’s NOIRLab; the Bok telescope, Steward Observatory, University of Arizona; and the Mayall telescope, Kitt Peak National Observatory, NOIRLab. NOIRLab is operated by the Association of Universities for Research in Astronomy (AURA) under a cooperative agreement with the National Science Foundation. Pipeline processing and analyses of the data were supported by NOIRLab and the Lawrence Berkeley National Laboratory (LBNL). Legacy Surveys also uses data products from the Near-Earth Object Wide-field Infrared Survey Explorer (NEOWISE), a project of the Jet Propulsion Laboratory/California Institute of Technology, funded by the National Aeronautics and Space Administration. Legacy Surveys was supported by: the Director, Office of Science, Office of High Energy Physics of the U.S. Department of Energy; the National Energy Research Scientific Computing Center, a DOE Office of Science User Facility; the U.S. National Science Foundation, Division of Astronomical Sciences; the National Astronomical Observatories of China, the Chinese Academy of Sciences and the Chinese National Natural Science Foundation. LBNL is managed by the Regents of the University of California under contract to the U.S. Department of Energy.



\section*{Data Availability}

The data underlying this article will be shared on reasonable request to the corresponding author.
 



\bibliographystyle{mnras}
\bibliography{biblio} 




\appendix

\section{Dispersion measure search limits} \label{app:dm_lims}

The DM ranged searched by the real-time MeerTRAP pipeline have varied over time for the different available observing bands. While originally the searches went up to 5118\,\pccm, when the TB system was implemented, the limits were lowered in order to reduce the trigger generation time, and improve the chances of capturing the TB data for a larger fraction of bursts.
The DM limits remain much larger than the largest DM FRB we have ever detected, and hence it should not significantly bias our observed population.
Table~\ref{tab:dm_lims} lists the DM upper limits we used for the searches at different frequencies and periods of time.

\begin{table}
    \centering
    \caption{DM upper limits (\pccm) that were used for MeerTRAP transient searches throughout the observations when the FRBs in this work were detected.}
    \begin{tabular}{l|ccc}
         \hline\hline
         Date & UHF & L & S\\
         \hline
         2022 -- 2023-05-05 & 2664 & 5241 & 5241 \\ 
         2023-05-05 -- 2023-06-01 & 2288 & 4241 & 5241 \\
         2023-06-02 -- 2023-10-22 & 2162 & 4241 & 5241 \\
         2023-10-22 onwards & 2162 & 3841 & 5241 \\
         \hline\hline
    \end{tabular}
    \label{tab:dm_lims}
\end{table}

\section{Photometric redshifts} \label{app:photoz}

To evaluate the accuracy of the photometric redshifts reported for the FRB host galaxies, we compared the performance of \texttt{CIGALE} when the redshift was left free to vary versus when it was constrained by the DESI-DR10 photometric redshift uncertainties. For each host with both spectroscopic and DESI photometric redshift measurements available, we performed two tests with \texttt{CIGALE}: first allowing the redshift to vary freely between 0 and 1, and second restricting it to the range defined by the DESI photometric redshift $1\sigma$ uncertainties.

\begin{figure}
    \centering
    \includegraphics[width=\linewidth]{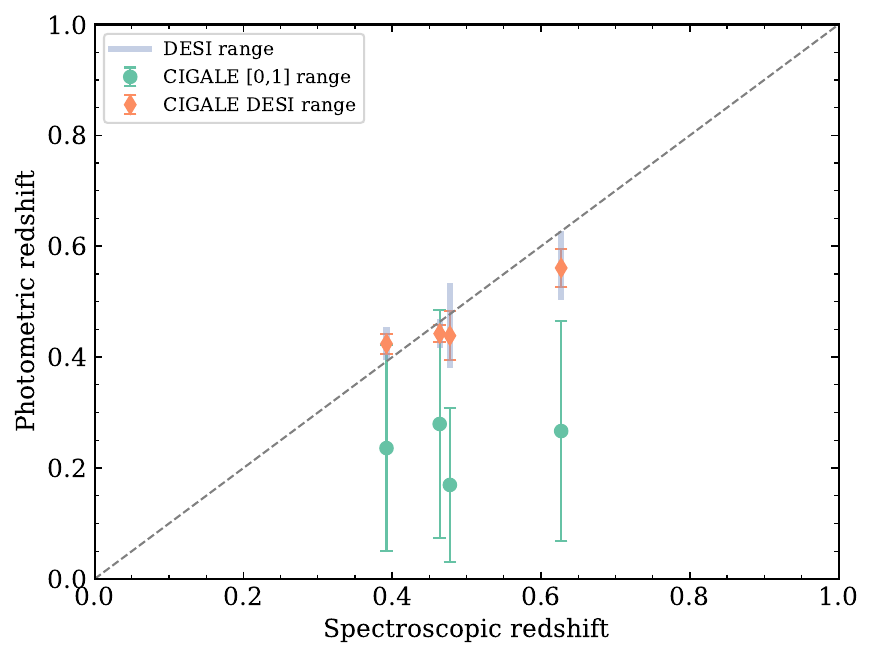}
    \caption{Comparison between spectroscopic and photometric redshifts for the FRB host galaxy sample. For each host with available spectroscopic and DESI photometric redshift (light grey bars), we show the photometric redshift derived from \texttt{CIGALE} when (i) the fitting redshift range is allowed to vary freely between 0 and 1 (green circles), and (ii) constrained to the DESI photometric redshift uncertainty range (orange diamonds). The dashed grey line indicates the one-to-one relation expected for perfect agreement.}
    \label{fig:photoz_test}
\end{figure}

Figure~\ref{fig:photoz_test} shows the resulting comparison between spectroscopic and photometric redshifts. The \texttt{CIGALE} results with an unconstrained redshift (green circles) typically cluster around 0.2, and show larger deviations from the spectroscopic values than those constrained to the DESI range (orange diamonds), which are generally closer to the ideal relation (dashed grey line). This confirms that DESI photometric redshifts provide a more accurate prior than the unconstrained \texttt{CIGALE} results, and we subsequently adopt the DESI uncertainties when fitting their SEDs with \texttt{CIGALE}.

We acknowledge that we have performed this test on a small galaxy sample. While adopting this method to analyse a large galaxy sample would introduce biases, since the models assumed by DESI and \texttt{CIGALE} are different, this appears to be a good option for our small galaxy sample with limited photometric and spectroscopic information.

\section{Optical images}

\begin{figure*}
    \raggedright
    \includegraphics[width=0.33\textwidth]{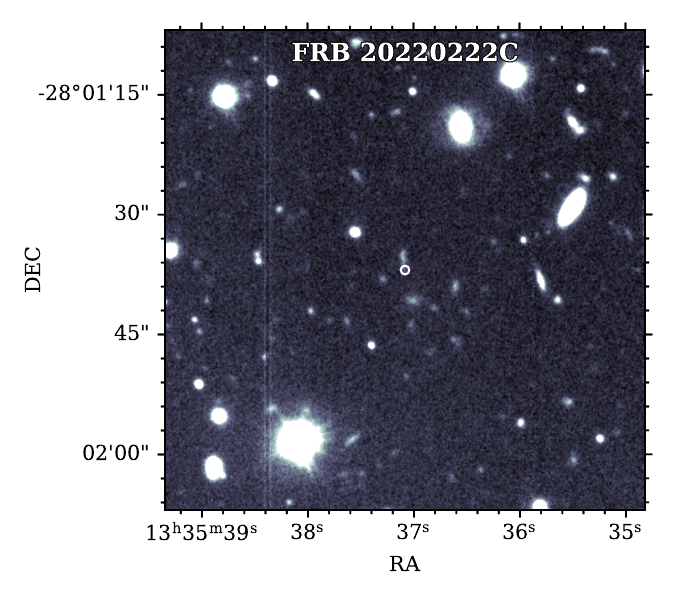}
    \includegraphics[width=0.33\textwidth]{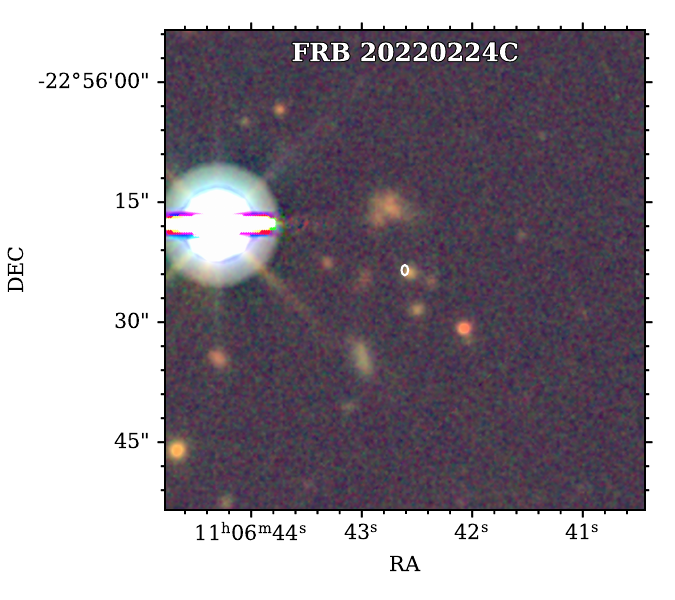}
    \includegraphics[width=0.33\textwidth]{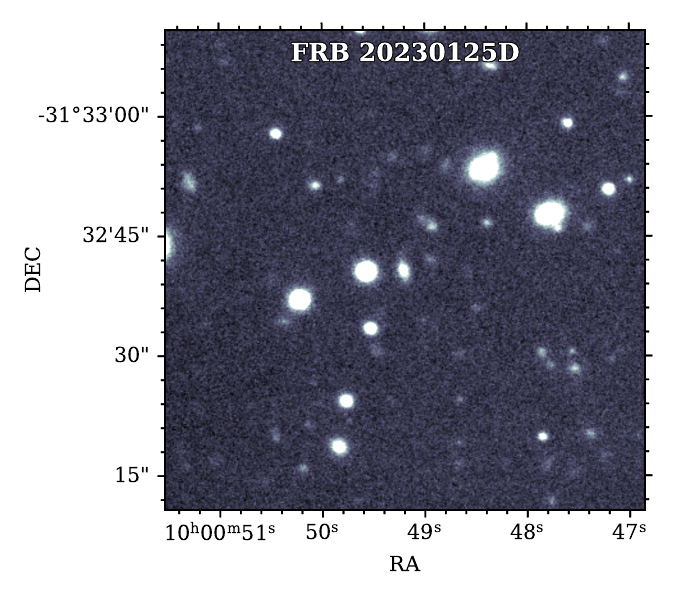}\\
    \includegraphics[width=0.33\textwidth]{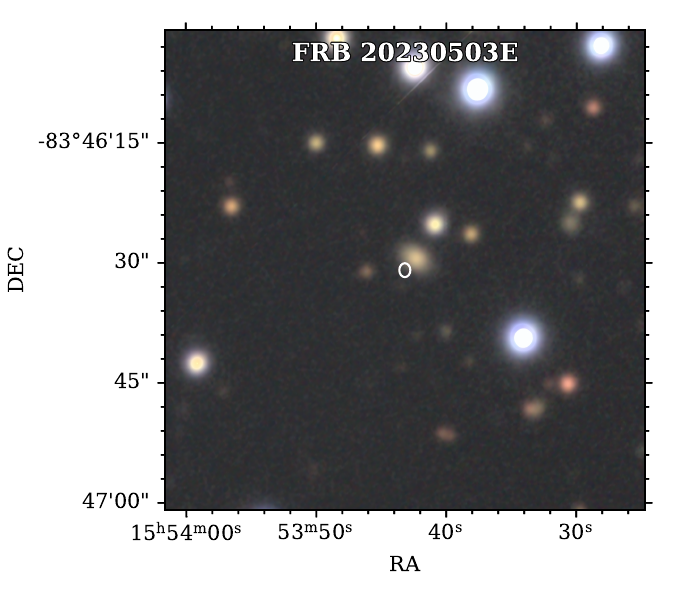}
    \includegraphics[width=0.33\textwidth]{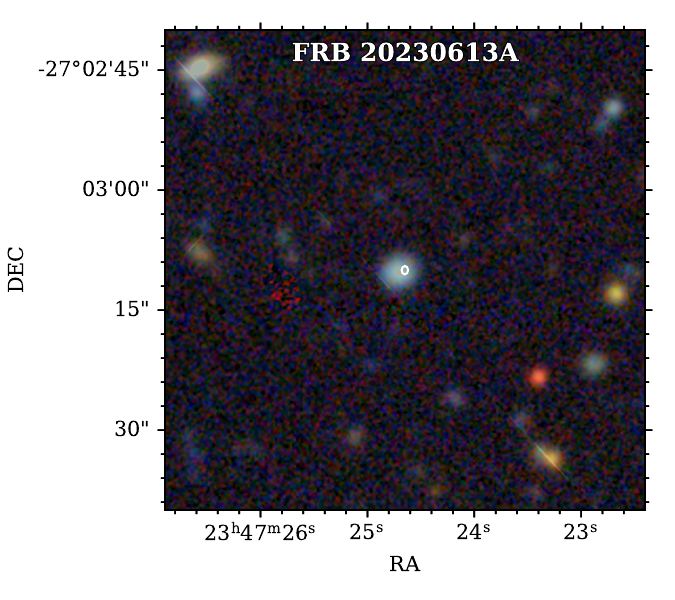}
    \includegraphics[width=0.33\textwidth]{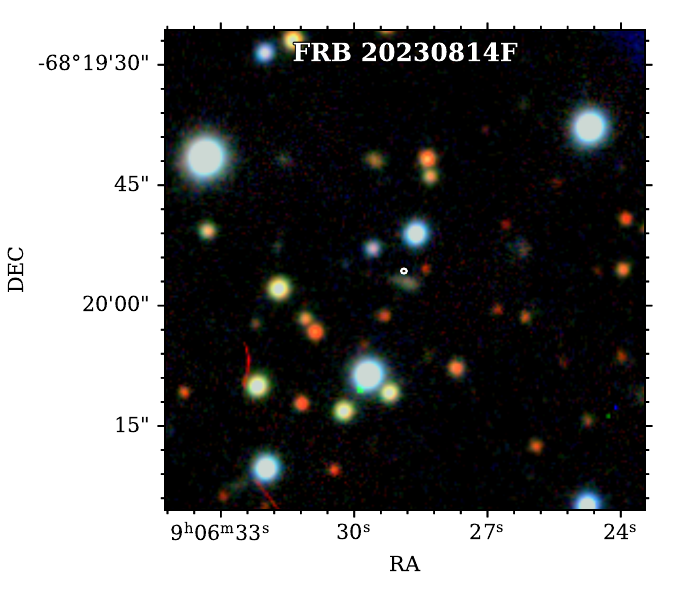}\\    
    \includegraphics[width=0.33\textwidth]{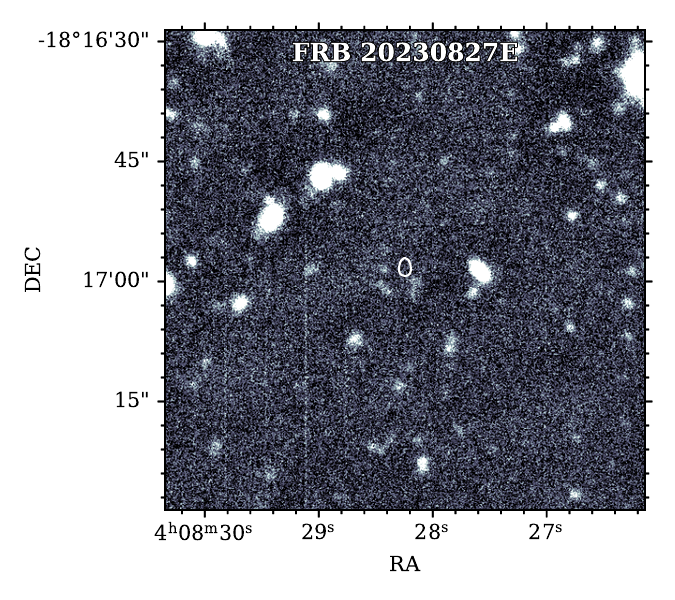}
    \includegraphics[width=0.33\textwidth]{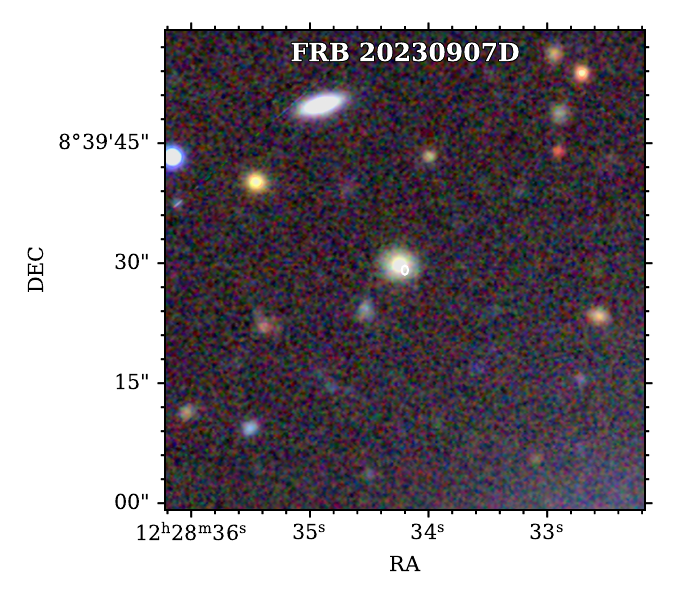}
    \includegraphics[width=0.33\textwidth]{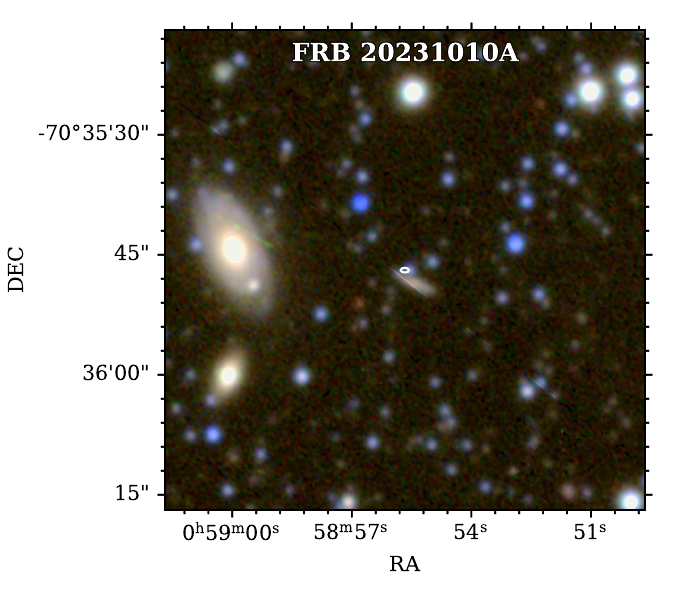}\\
    \includegraphics[width=0.33\textwidth]{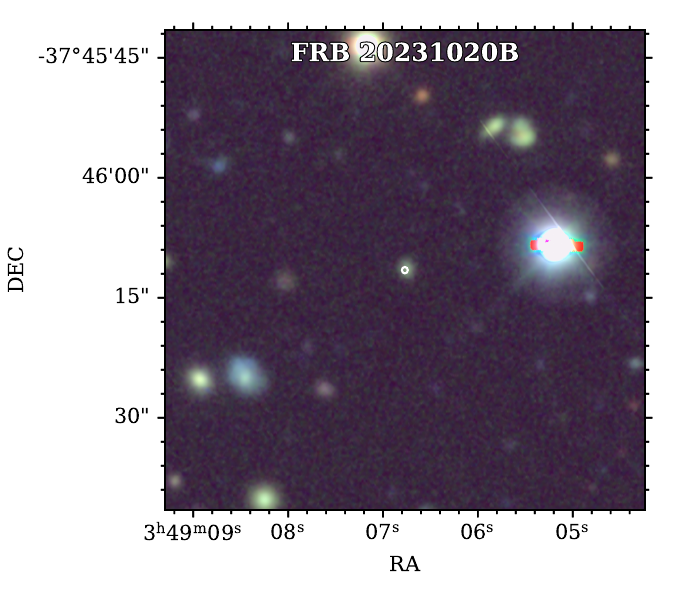}
    \includegraphics[width=0.33\textwidth]{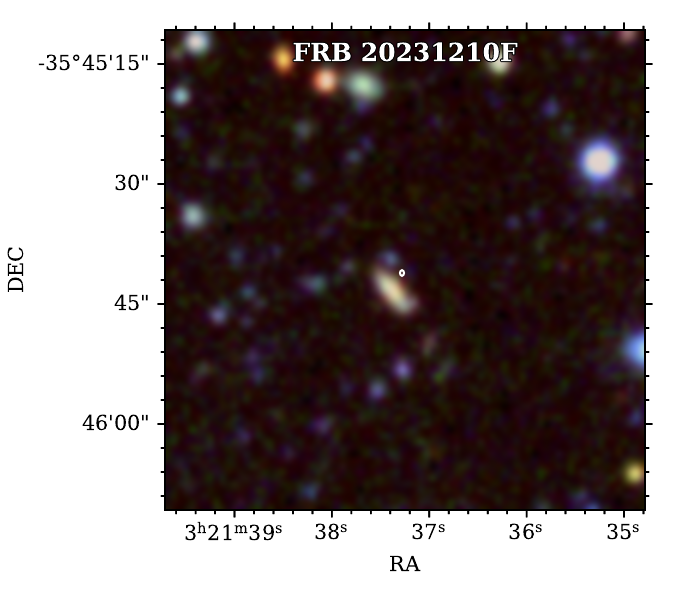}
    \caption{FRB localisations. The white ellipses at the centre of each image show the FRB localisation error regions and the background shows the optical image from either the DESI-Legacy survey or the deep r-band images obtained through the observing programmes mentioned in the text. The FRB identifier is indicated at the top of each image. Each image has an FoV of 60"$\times$60"}
    \label{fig:optical}
\end{figure*}

Figure~\ref{fig:optical} shows a collection of the optical images matching the location of the FRBs presented in this paper with (sub-)arcsecond localisation.

\section{Host galaxy spectra}

In this section, we show the optical and NIR properties of the putative host galaxies of the FRBs presented in Section~\ref{sec:new_loc_frbs} for which we were able to obtain a spectrum and/or magnitudes in several bands. The optical spectra and emission lines we identified are shown in Fig.~\ref{fig:host_spectra}. The SED fitting to the magnitudes from DECam and WISE (when available) is shown in Fig.~\ref{fig:host_sed}. The configuration of each imaging and spectroscopic observations that were taken are detailed in Table~\ref{tab:observations}.

\renewcommand{\arraystretch}{1.5}
\begin{sidewaystable*} 
  \centering
	\caption{FRB and host galaxy candidates. We have listed galaxy candidates within 5\arcsec of the FRB localisation centre with a probability of association >1\%. The galaxy labels match those in the main text and figures.}
	\label{tab:frb_hosts}
	\begin{tabular}{cccccccccccc} 
		\hline\hline
		FRB ID & DM & RA & DEC & GID & Galaxy coord. & $z_{\text{spec}}$ & \zphot & mag$_{r}$ & $R_{\text{hl}}$ (") & sep. (") & $P(O|x)$  \\
		\hline\hline
20220222C & 1070.17 & 13:35:37.08 (0.54) & -28:01:36.93 (0.55) & G1 & 13:35:37.10 -28:01:35.53 & 0.853 & -- & 23.86$\pm$0.04 & 1.06 & 1.43 & 0.9442\\
20220224C & 1145.0 & 11:06:42.61 (0.42) & -22:56:23.48 (0.64) & G1 & 11:06:42.58 -22:56:23.70 & 0.6271 & $0.57^{+0.06}_{-0.06}$ & 21.63$\pm$0.05 & 0.68 & 0.52 & 0.9964\\
20230125D & 640.37 & 10:00:49.21 (0.24) & -31:32:40.77 (0.26) & G1 & 10:00:49.23 -31:32:40.78 & 0.3265 & -- & 22.12$\pm$0.02 & 0.96 & 0.29 & 0.9781\\
20230503E & 484.03 & 15:53:43.19 (0.69) & -83:46:30.93 (0.87) & G1 & 15:53:42.38 -83:46:29.40 & -- & $0.32^{+0.15}_{-0.15}$ & 20.11$\pm$0.01 & 1.37 & 2.02 & 0.9733\\
20230613A & 483.82 & 23:47:24.65 (0.40) & -27:03:10.01 (0.52) & G1 & 23:47:24.70 -27:03:10.22 & 0.3923 & $0.42^{+0.03}_{-0.03}$ & 20.132$\pm$0.004 & 1.39 & 0.73 & 0.9993\\
20230814F & 471.73 & 09:06:28.88 (0.35) & -68:19:55.70 (0.31) & G1 & 09:06:28.86 -68:19:57.07 & -- & -- & 22.03$\pm$0.05 & 1.36 & 1.37 & 0.9199\\
20230827E & 1434.0 & 04:08:28.242 (0.75) & -18:16:58.55 (1.50) & G1 & 04:08:28.17 -18:16:59.71 & -- & -- & 26.40$\pm$0.02 & 1.04 & 1.51 & 0.0082\\
          &        &           &           & G2 & 04:08:28.43 -18:16:58.59 & -- & -- & 25.81$\pm$0.02 & 1.01 & 2.64 & 0.0074\\
          &        &           &           & G3 & 04:08:28.18 -18:17:01.63 & -- & -- & 26.40$\pm$0.02 & 0.96 & 3.19 & $3\times10^{-6}$\\
          &        &           &           & G4 & 04:08:28.39 -18:17:01.28 & -- & -- & 25.56$\pm$0.01 & 0.94 & 3.48 & $7\times10^{-6}$\\
          &        &           &           & G5 & 04:08:28.46 -18:17:00.16 & -- & -- & 26.13$\pm$0.02 & 0.92 & 3.48 & 0.0001\\
20230907D & 1031.11 & 12:28:34.20 (0.40) & +08:39:29.13 (0.57) & G1 & 12:28:34.25 +08:39:29.71 & 0.4638 & $0.44^{+0.03}_{-0.02}$ & 19.83$\pm$0.09 & 1.39 & 0.92 & 0.9585\\
20231010A & 442.81 & 00:58:55.67 (0.52) & -70:35:46.93 (0.30) & G1 & 00:58:55.45 -70:35:48.50 & -- & $0.61^{+0.18}_{-0.18}$ & 21.24$\pm$0.01 & 1.9 & 2.38 & 0.5232\\
            &        &           &           & G2 & 00:58:55.58 -70:35:46.89 & -- & $0.84^{+0.27}_{-0.30}$ & 22.52$\pm$0.01 & 0.47 & 1.61 & 0.4662\\
20231020B & 953.84 & 03:49:06.77 (0.37) & -37:46:11.56 (0.40) & G1 & 03:49:06.76 -37:46:11.45 & 0.4775 & $0.46^{+0.07}_{-0.08}$ & 21.81$\pm$0.02 & 0.88 & 0.18 & 0.9984\\
20231210F & 721.46 & 03:21:37.28 (0.25) & -35:45:41.13 (0.25) & G1 & 03:21:37.39 -35:45:43.37 & -- & $0.50^{+0.08}_{-0.08}$ & 21.19$\pm$0.02 & 1.97 & 2.58 & 0.8103\\
            &        &           &           & G2 & 03:21:37.22 -35:45:41.02 & -- & $1.37^{+0.43}_{-0.44}$ & 26.4$\pm$1.0 & 0.41 & 0.78 & 0.1483\\
            &        &           &           & G3 & 03:21:37.22 -35:45:45.18 & -- & $0.99^{+0.70}_{-0.54}$ & 24.1$\pm$0.1 & 1.08 & 4.12 & 0.0160\\
            &        &           &           & G4 & 03:21:37.40 -35:45:39.41 & -- & $0.86^{+0.27}_{-0.33}$ & 23.9$\pm$0.1 & 0.56 & 2.27 & 0.0003\\
		\hline\hline
	\end{tabular}
\end{sidewaystable*}


\begin{figure*}
    \centering
    \includegraphics[width=0.33\textwidth]{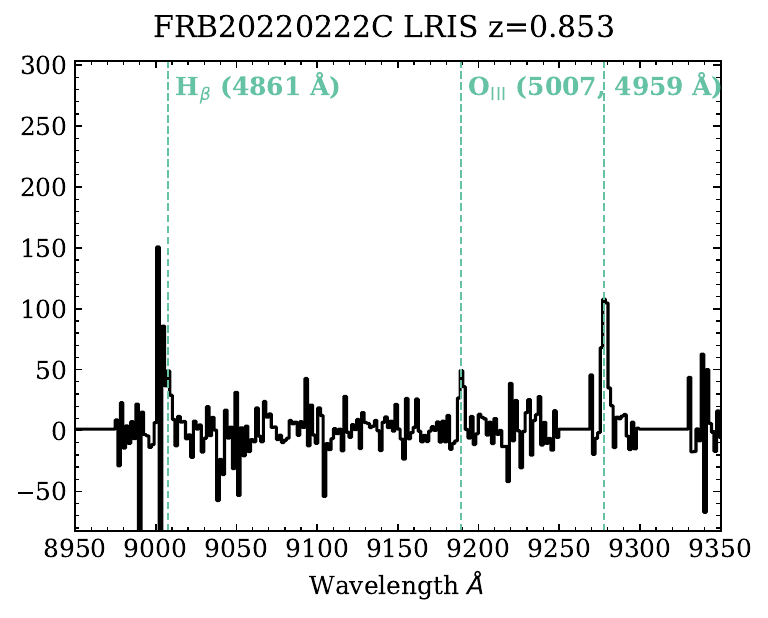}
    \includegraphics[width=0.66\textwidth]{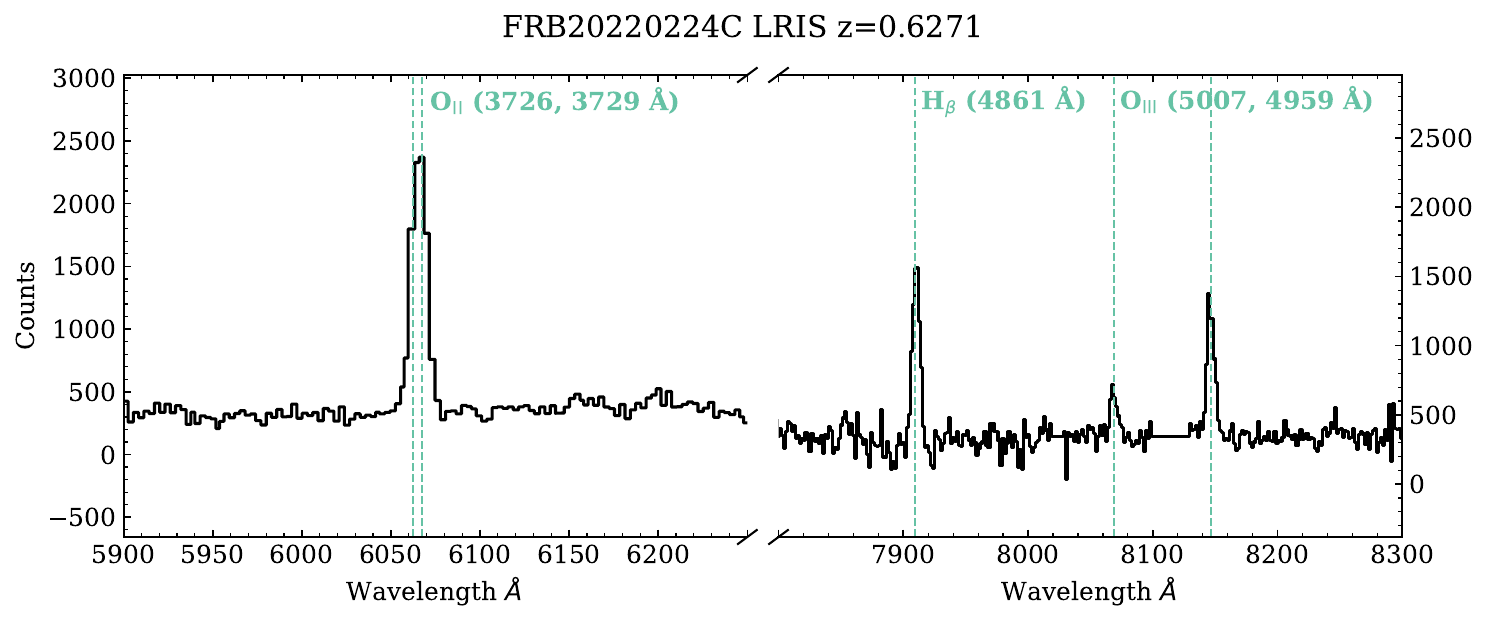}\\
    \includegraphics[width=0.99\textwidth]{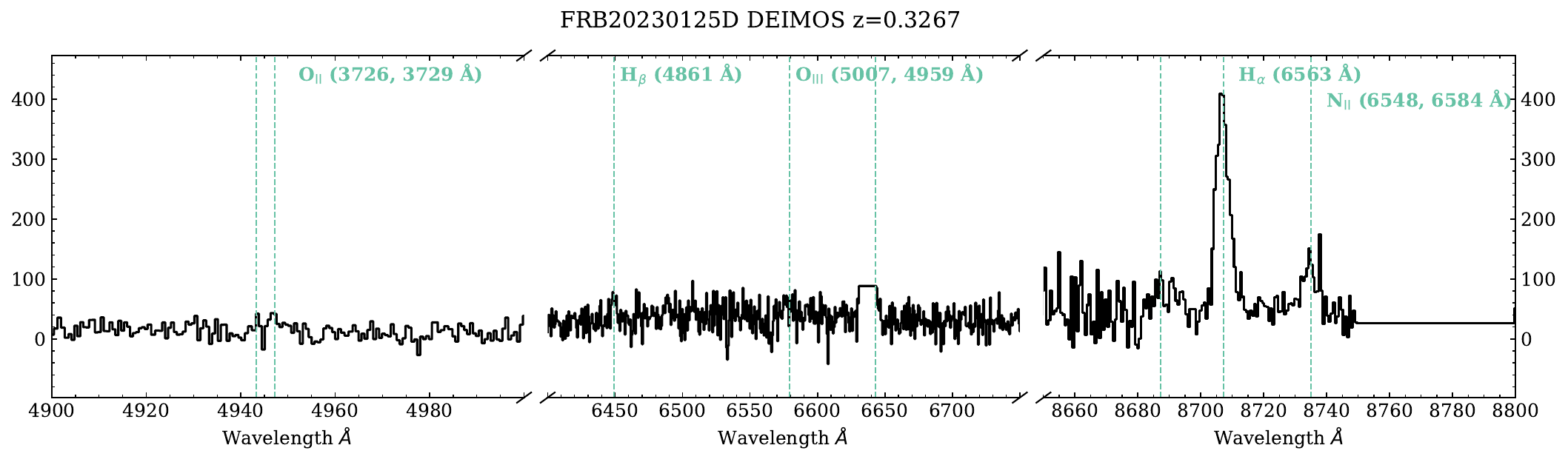}\\
    \includegraphics[width=0.99\textwidth]{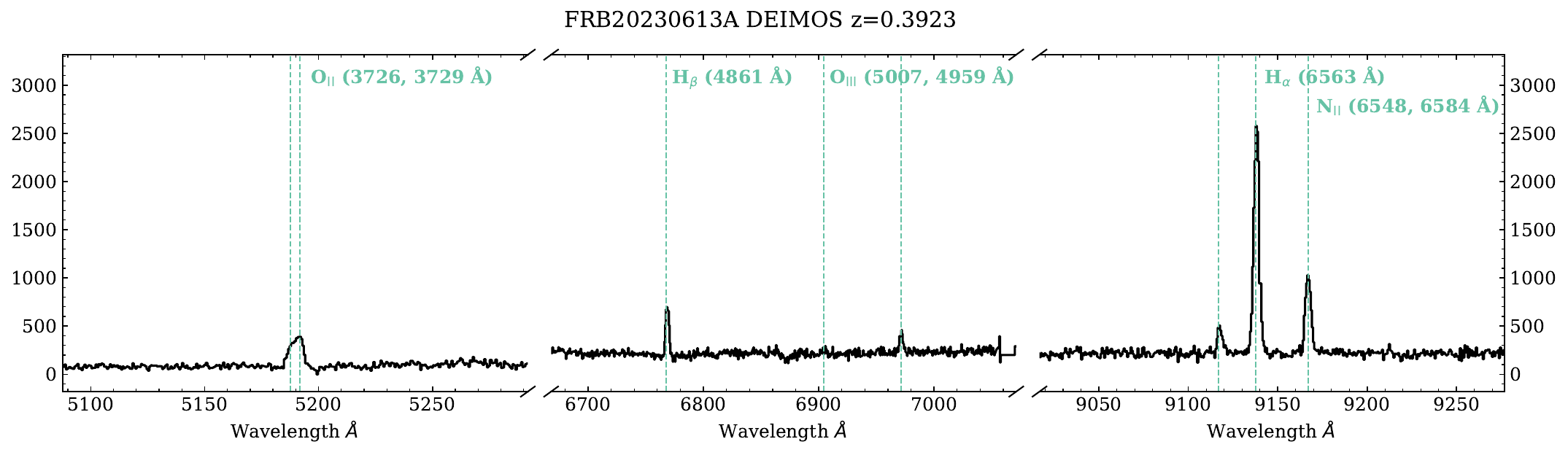}\\
    \includegraphics[width=0.99\textwidth]{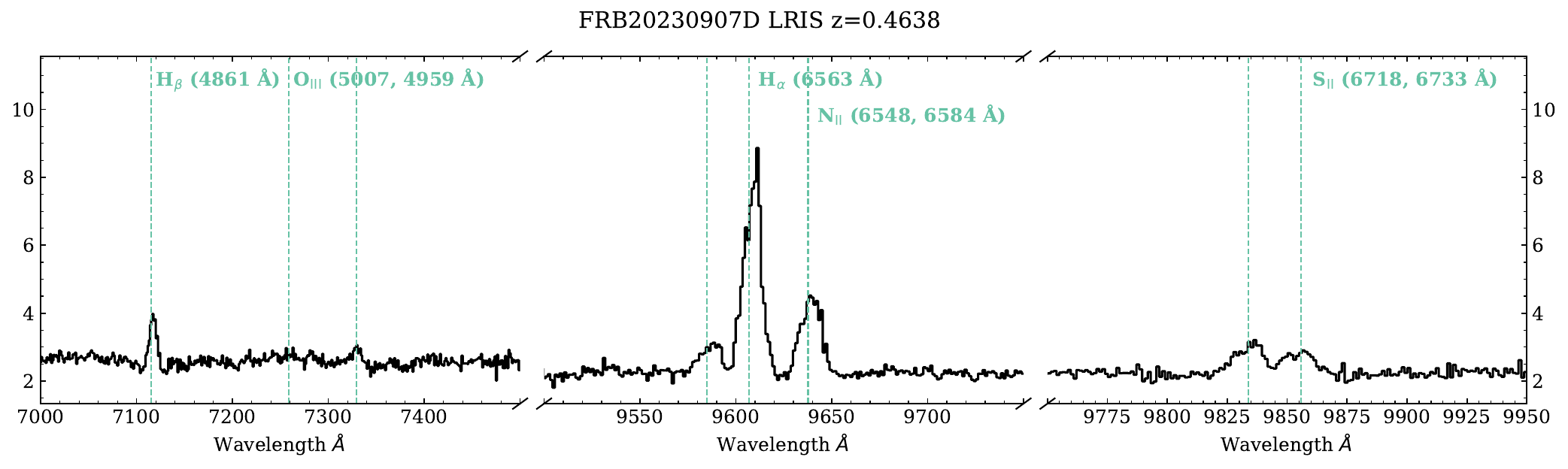}\\
    \caption{Spectra of FRB host galaxies. Each panel shows the counts as a function of wavelength in \r{A}. The black line shows the spectrum, while the green vertical dashed lines show the spectral line position at the given redshift. The text next to each line indicates which spectral line it is and its wavelength at rest. In the case of doublets, two wavelengths are indicated. For each FRB host, one or more panels are shown zooming into the spectral lines that were use to determine the redshift. The title of each panel group gives the FRB id, the instrument that was used to obtain the spectrum, and the redshift that was measured.}
    \label{fig:host_spectra}
\end{figure*}

\begin{figure*}
    \ContinuedFloat
    \includegraphics[width=0.66\textwidth]{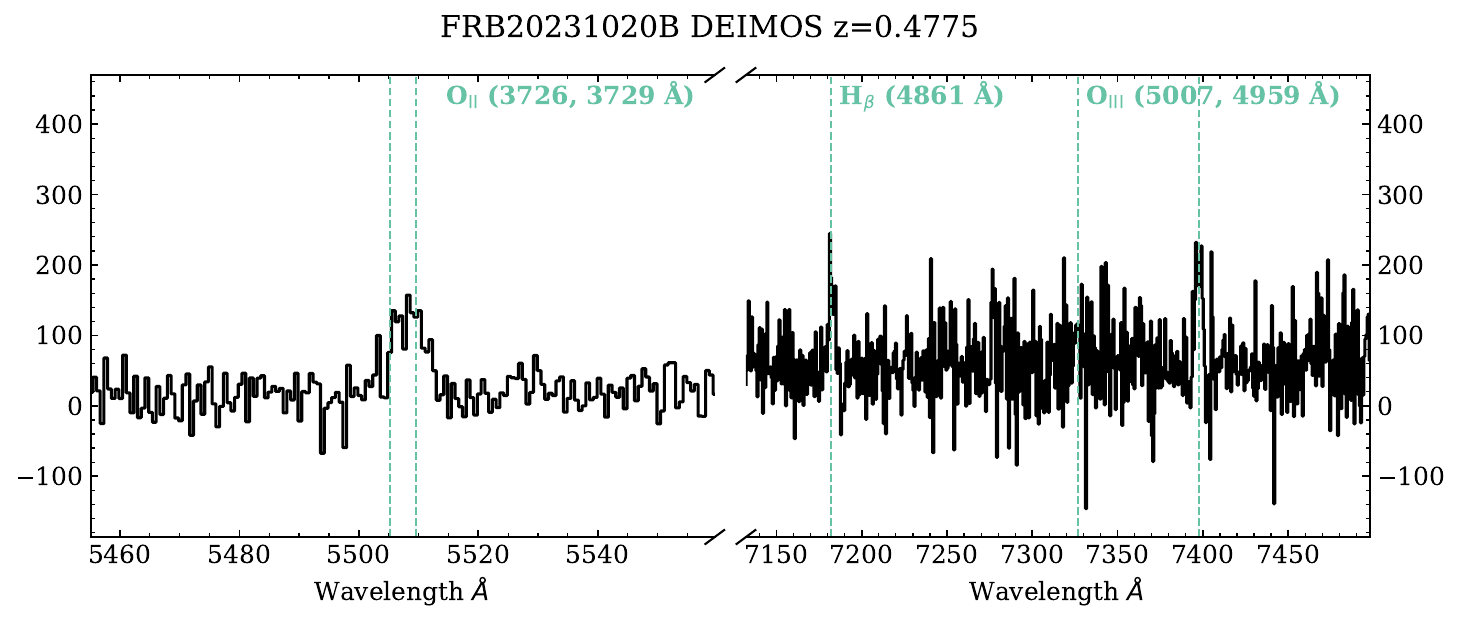}
    \caption{Continued.}
\end{figure*}

\begin{figure*}
    \includegraphics[width=0.48\textwidth]{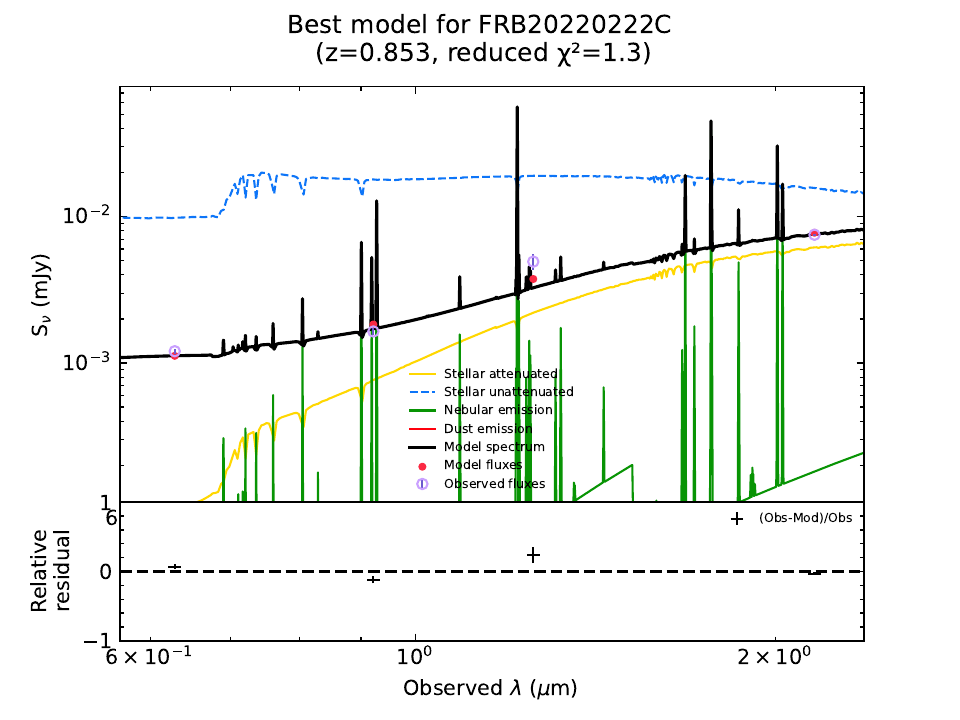}
    \includegraphics[width=0.48\textwidth]{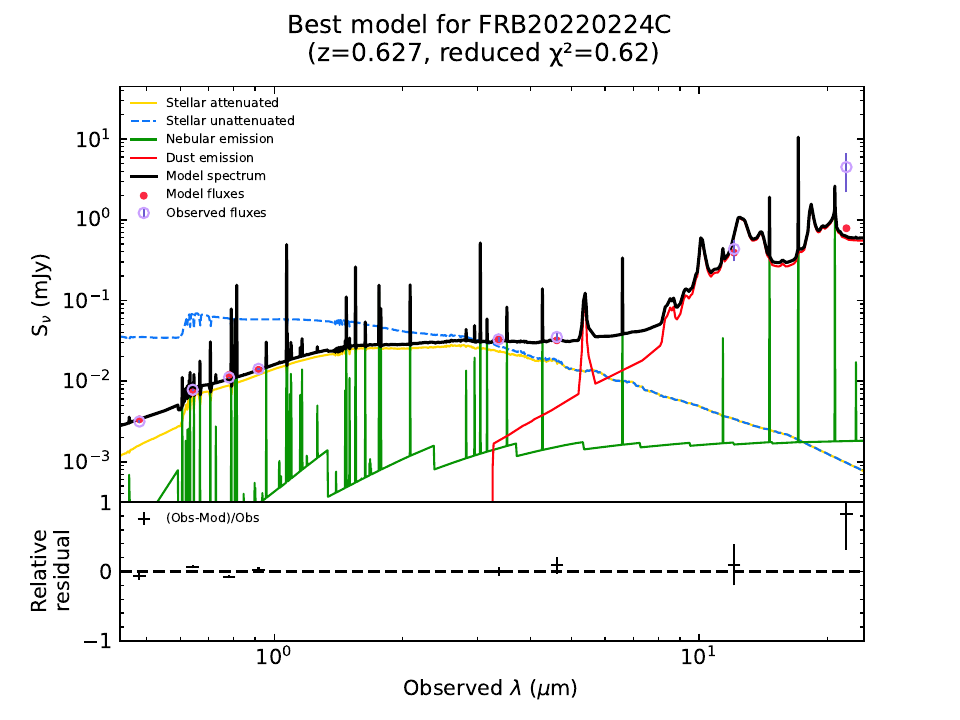}
    \includegraphics[width=0.48\textwidth]{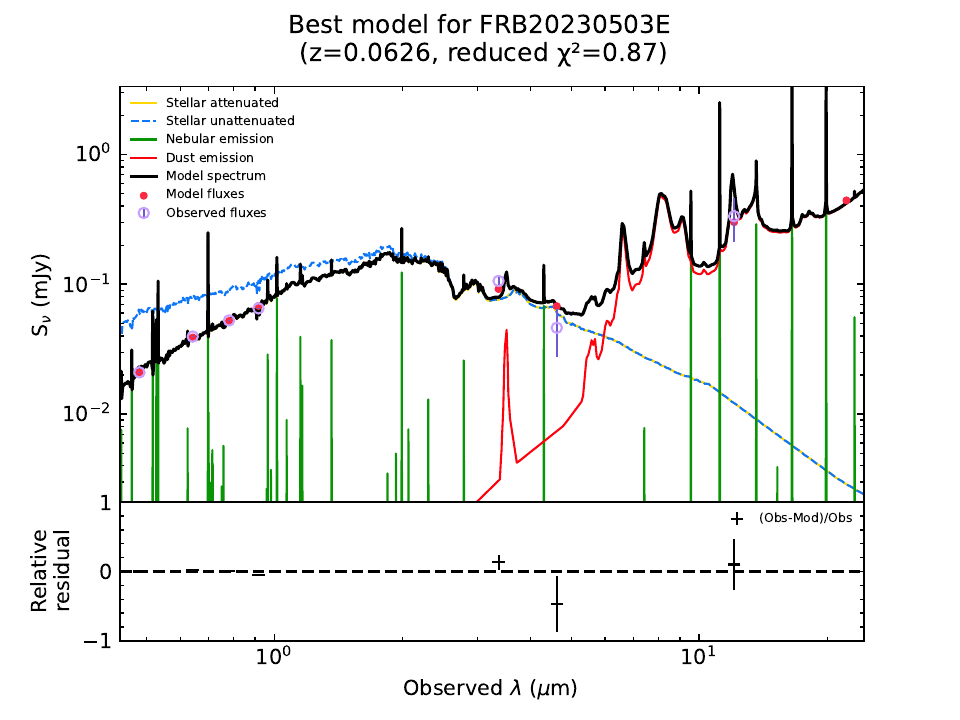}
    \includegraphics[width=0.48\textwidth]{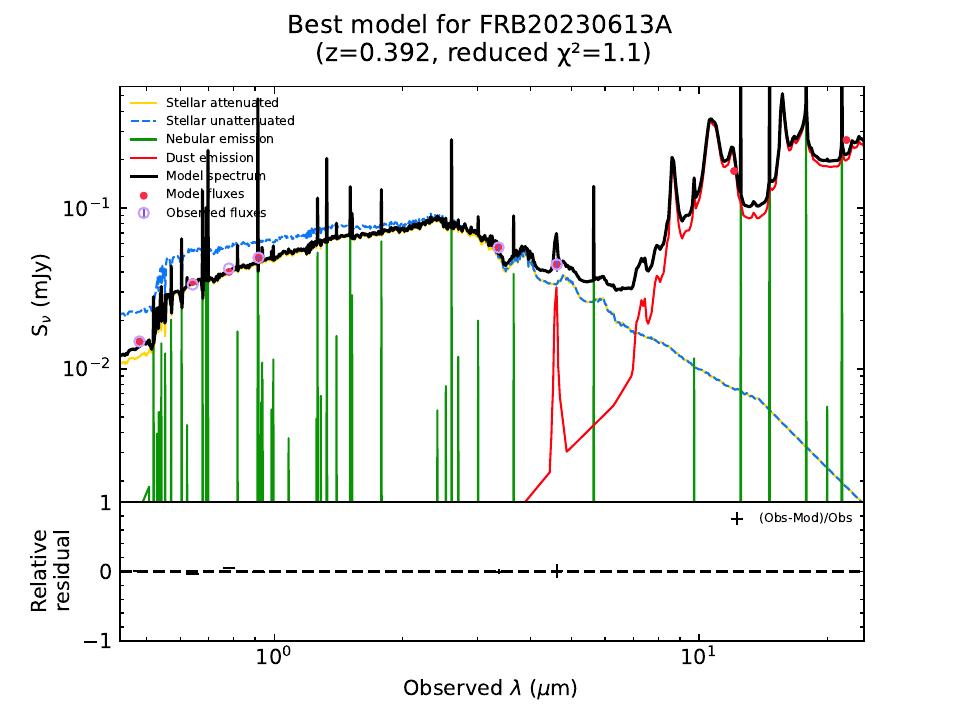}
    \caption{Spectral energy distribution fitting results with CIGALE.}
    \label{fig:host_sed}
\end{figure*}

\begin{figure*}
    \ContinuedFloat
    \raggedright
    \includegraphics[width=0.48\textwidth]{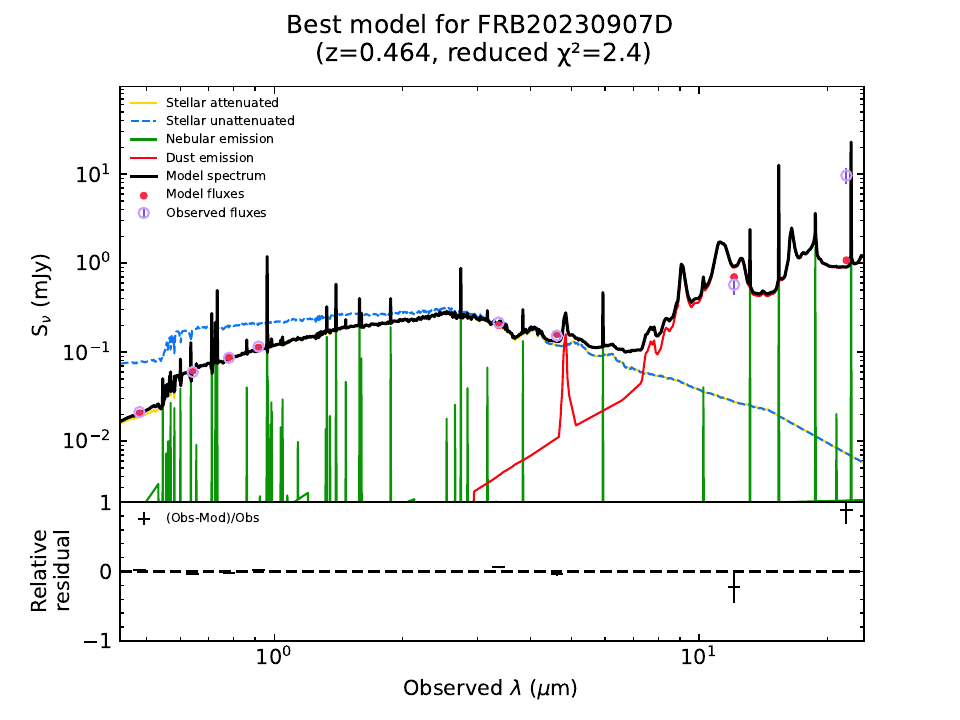}
    \includegraphics[width=0.48\textwidth]{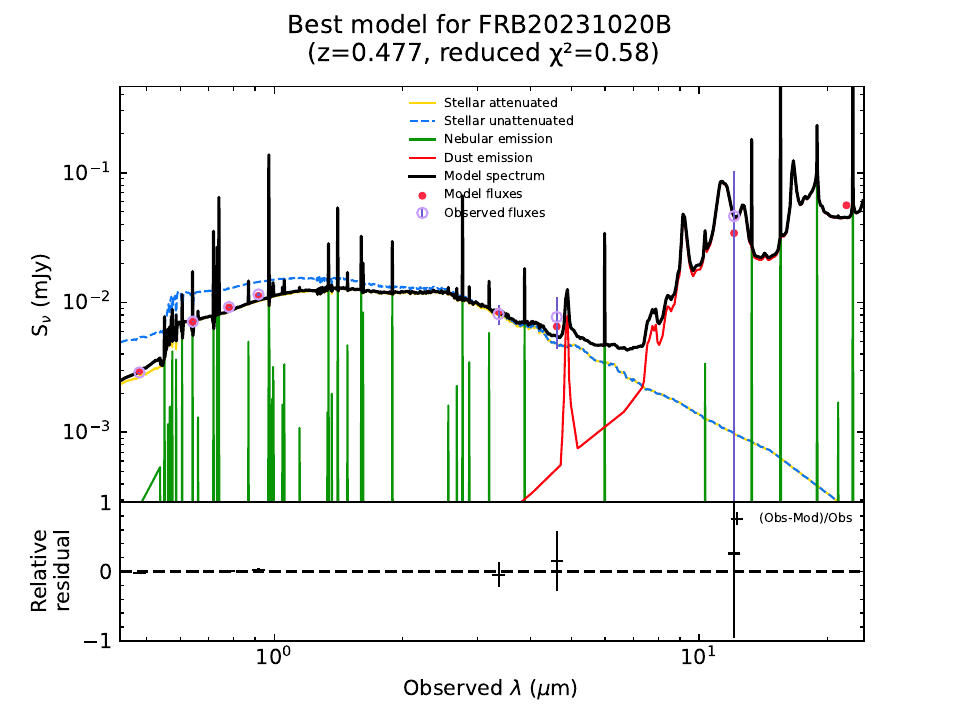}
    \includegraphics[width=0.48\textwidth]{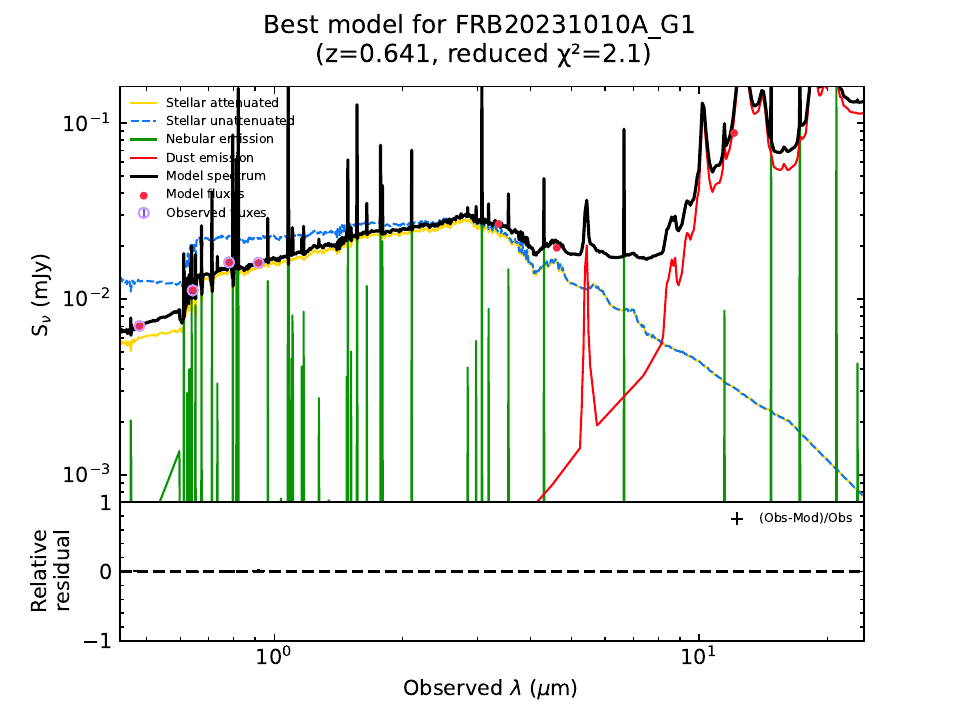}
    \includegraphics[width=0.48\textwidth]{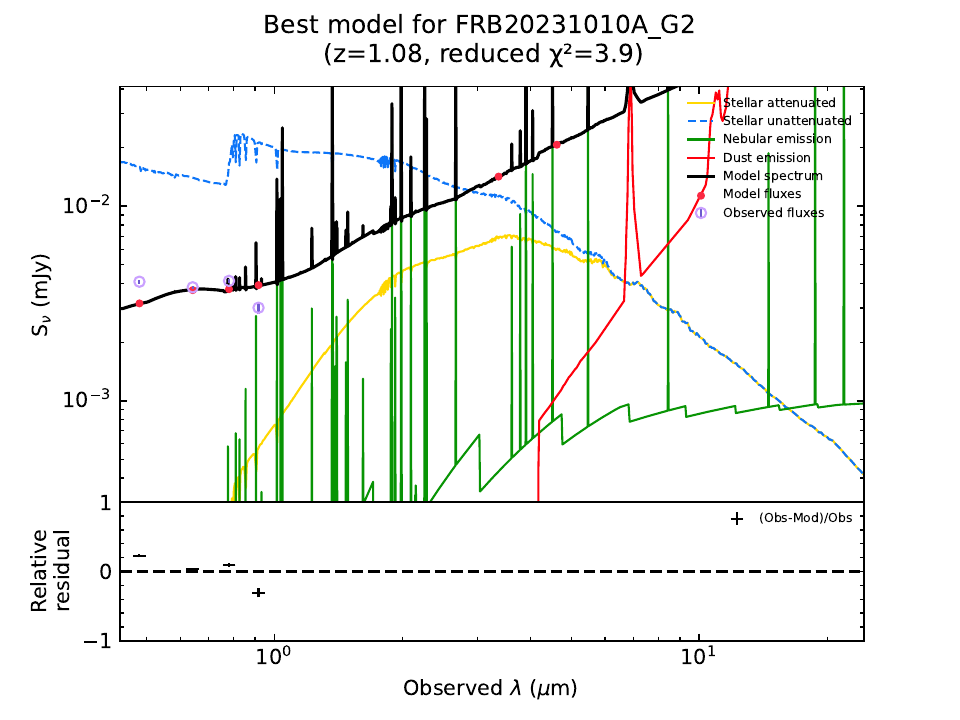}
    \includegraphics[width=0.48\textwidth]{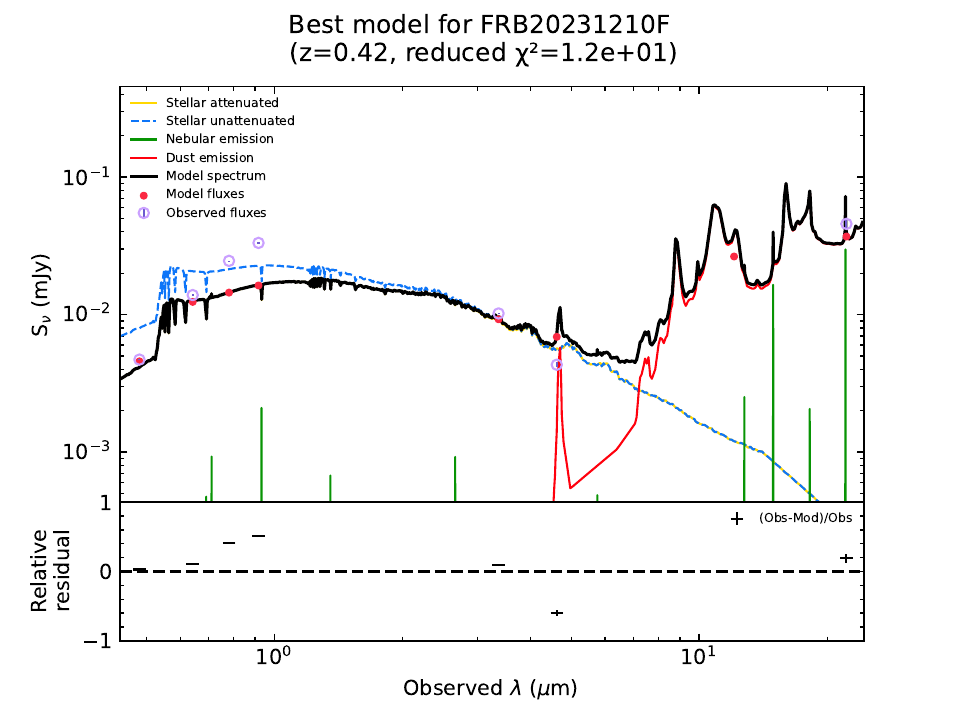}
    \caption{Continued.}
    \label{fig:host_sed}
\end{figure*}


\begin{table*}
    \centering
    \caption{Optical and near-infrared imaging and spectroscopic observation summary.}
    \label{tab:observations}
    \begin{tabular}{l|lllllllll}
    \hline\hline
        FRB ID & Facility & Instrument & Obs. date & Type & Configuration & Exp. time & Seeing* & Program ID & PI \\ \hline
        20220222C & Gemini-S & GMOS & 2022 May 25 & Imaging & r-band & $28\times100$\,s & 0.6\arcsec & GS-2022A-Q-143 & N. Tejos \\
        ~ & Gemini-S & GMOS & 2023 June 16 & Imaging & z-band & $25\times100$\,s & 0.6\arcsec & GS-2023A-Q-215 & A. Gordon \\
        ~ & Gemini-S & Flamingos2 & 2024 June 16 & Imaging & Ks-band & $60\times10$\,s & 0.5\arcsec & GS-2023A-Q-215 & A. Gordon \\
        ~ & Gemini-S & Flamingos2 & 2025 June 16 & Imaging & J-band & $30\times60$\,s & 0.5\arcsec & GS-2023A-Q-215 & A. Gordon\\
        ~ & Keck & LRIS & 2023 Apr 17 & Spec & R600/10000, 8082 Å & 6450\,s & 1.5\arcsec & U173 & X. Prochaska \\
        ~ & Keck & LRIS & 2023 Apr 17 & Spec & B300/5000 & 6600\,s & 1.5\arcsec & U173 & X. Prochaska \\ \hline
        20220224C & Gemini-S & GMOS & 2023 Jan 29 & Imaging & r-band & $15\times180$\,s & 0.8\arcsec &  GS-2022B-Q-123 & A. Gordon\\
        ~ & Gemini-S & GMOS & 2023 Jan 30 & Imaging & i-band & $15\times180$\,s & 0.7\arcsec & GS-2022B-Q-123 & A. Gordon\\
        ~ & Keck & LRIS & 2023 Jan 24 & Spec & R600/10000, 8927 Å & 2740\,s & 1.1\arcsec & U055 & J. Hennawi\\
        ~ & Keck & LRIS & 2023 Jan 24 & Spec & B300/5000 & 2780\,s & 1.1\arcsec & U055 & J. Hennawi\\ \hline
        20230125D & Gemini-S & GMOS & 2023 May 16 & Imaging & r-band & $15\times180$\,s & 0.7\arcsec & GS-2023A-Q-215 & A. Gordon\\
        ~ & Keck & DEIMOS & 2023 Dec 14 & Spec & 600ZD, 6500 Å & 2700\,s & 2.1\arcsec & U051 & X. Prochaska\\ \hline
        20230613A & Keck & DEIMOS & 2023 Aug 11 & Spec & 600ZD, 7000 Å & 2835\,s & 1.3\arcsec & U051 & X. Prochaska\\ \hline
        20230827E & Keck & DEIMOS & 2023 Sept 06 & Imaging & R-band & $9\times300$\,s & 0.9\arcsec & O438 & A. Gordon\\
        ~ & Keck & DEIMOS & 2023 Oct 07 & Imaging & Z-band & $9\times300$\,s & 0.9\arcsec & O439 & A. Gordon\\ \hline
        20230907D & Keck & LRIS & 2024 Jan 05 & Spec & R400/8500, 7830 Å & 2400\,s & 1.0\arcsec & O438 & A. Gordon\\
        ~ & Keck & LRIS & 2025 Jan 05 & Spec & B400/3400 & 2400\,s & 1.0\arcsec & O438 & A. Gordon\\ \hline
        20231020B & Keck & DEIMOS & 2023 Dec 14 & Spec & 600ZD, 6500 Å & 730\,s & 1.0\arcsec & U051 & X. Prochaska\\ \hline
        20231210F & SOAR & Goodman & 2024 Feb 02 & Imaging & r-band & $12\times300$\,s & 0.4\arcsec & SOAR2024A-002 & A. Gordon \\ \hline\hline
    \end{tabular}
\begin{minipage}{\linewidth}
    \textbf{Notes.} 
    * Seeing: for imaging observations, the seeing is estimated from the FWHM of the non-saturated stars contained in the image, fitted to a Moffat profile \citep{moffat_theoretical_1969}.  For spectroscopic observations, the seeing is obtained as a \texttt{PypeIt} output \citep{prochaska_pypeit_2020}. 
\end{minipage}
\end{table*}



\begin{table*} 
    \caption{Photometric data and \texttt{CIGALE} SED fitting results.}
    \label{tab:sed_results}
    \begin{tabular}{cccccccccc} 
    \hline\hline
Property & 220222 G1 & 220224 G1 & 230503 G1 & 230613 G1 & 230907 G1 & 231010 G1 & 231010 G2 & 231020 G1 & 231210 G1 \\ \hline
$z_{\text{spec}}$ & 0.853 & 0.6271 & -- & 0.39239 & 0.4637 & -- & -- & 0.4775 & -- \\
\zphot & -- & -- & 0.29$\pm$0.26 & -- & -- & 0.61$\pm$0.09 & 0.84$\pm$0.3 & -- & 0.43$\pm$0.01 \\
\hline
g & -- & 22.65$\pm$0.06 & 20.891$\pm$0.007 & 20.97$\pm$0.01 & 20.58$\pm$0.01 & 21.87$\pm$0.02 & 22.46$\pm$0.02 & 22.75$\pm$0.01 & 22.28$\pm$0.01 \\
r & 23.86$\pm$0.04 & 21.63$\pm$0.03 & 20.113$\pm$0.04 & 20.132$\pm$0.004 & 19.83$\pm$0.09 & 21.24$\pm$0.01 & 22.52$\pm$0.02 & 21.81$\pm$0.02 & 21.19$\pm$0.02 \\
i & -- & 21.29$\pm$0.02 & 19.749$\pm$0.05 & 19.84$\pm$0.01 & 19.06$\pm$0.01 & 20.93$\pm$0.01 & 22.41$\pm$0.03 & 21.49$\pm$0.01 & 20.45$\pm$0.01 \\
z & 23.45$\pm$0.06 & 21.03$\pm$0.04 & 19.481$\pm$0.008 & 19.67$\pm$0.02 & 18.74$\pm$0.01 & 20.92$\pm$0.03 & 22.74$\pm$0.07 & 21.25$\pm$0.03 & 20.12$\pm$0.02 \\
J & 22.22$\pm$0.13 & -- & -- & -- & -- & -- & -- & -- & -- \\
W1 & -- & 22.81$\pm$0.07 & 18.85$\pm$0.12 & 22.21$\pm$0.04 & 20.77$\pm$0.01 & -- & -- & 24.33$\pm$0.2 & 21.38$\pm$0.02 \\
W2 & -- & 23.38$\pm$0.13 & 19.75$\pm$0.44 & 23.12$\pm$0.11 & 21.77$\pm$0.03 & -- & -- & 25.02$\pm$0.47 & 22.31$\pm$0.05 \\
W3 & -- & 22.49$\pm$0.33 & 17.6$\pm$0.4 & -- & 22.18$\pm$0.26 & -- & -- & 24.91$\pm$1.32 & -- \\
W4 & -- & 21.39$\pm$0.56 & -- & -- & 20.56$\pm$0.22 & -- & -- & -- & -- \\
\hline
\logMstar & 10.11$\pm$0.17 & 10.32$\pm$0.12 & <10.56 & 10.13$\pm$0.18 & 10.88$\pm$0.2 & <10.54 & 9.11$\pm$0.25 & 10.75$\pm$0.3 & 9.75$\pm$0.04 \\
SFR* & 7.15 (0.29) & 18.88 (0.17) & <25.8  & 4.5 (0.28) & 14.91 (0.33) & 11.05 (0.48) & 2.51 (0.24) & 13.58 (0.99) & <0.2  \\
\met & <-0.13 & <0.70 & <0.21 & <-0.13 & <-0.11 & <0.18 & <-0.43 & <0.09 & <-2.26 \\
Age (Gyr) & 1.8$\pm$0.7 & 1.4$\pm$0.5 & 2.0$\pm$1.0 & 2.0$\pm$0.8 & 2.3$\pm$1.1 & 1.4$\pm$0.4 & 1.1$\pm$0.1 & 2.2$\pm$1.4 & 1.0$\pm$0.1 \\
\hline
    \end{tabular}
\begin{minipage}{\linewidth}
    \textbf{Notes.} 
    * SFR: star formation rate in \Msunyr. The uncertainties between parentheses are in dex.
\end{minipage}
\end{table*}



\bsp	
\label{lastpage}
\end{document}